\documentclass[a4paper,12pt]{article}
\pdfoutput=1

\usepackage{jheppub,fancyhdr,graphicx,epstopdf,color,amsmath,cases,slashed,hyperref}
\usepackage{makecell}
\usepackage{caption, subcaption}  
\usepackage{dcolumn}   
\usepackage{amsmath, amssymb}        
\usepackage{color,latexsym, array,multirow,  verbatim, enumerate,cancel}
\usepackage{slashed}
\usepackage{bigcenter}
\usepackage[utf8]{inputenc}
\usepackage{footnote}
\makesavenoteenv{tabular}
\makesavenoteenv{table}

\def\issue(#1,#2,#3){{\bf #1}, #2 (#3)}

\catcode`\@=11
\def\lsim{\mathrel{\mathpalette\@versim<}}
\def\gsim{\mathrel{\mathpalette\@versim>}}
\def\@versim#1#2{\vcenter{\offinterlineskip
\ialign{$\m@th#1\hfil##\hfil$\crcr#2\crcr\sim\crcr } }}
\catcode`\@=12

\catcode`@=12

\newcommand{\met}{$\cancel E_T$}

\newcommand{\newc}{\newcommand}
\newc{\wt}{\widetilde}
\newc{\ra}{\rightarrow}

\def\beq {\begin{equation}}
\def\eeq {\end{equation}}
\def\bi {\begin{itemize}}
\def\ei {\end{itemize}}
\def\bea {\begin{eqnarray}}
\def\eea {\end{eqnarray}}

\def \met{\slashed{E}_T}

\newcommand{\br}{\begin{eqnarray}}
\newcommand{\er}{\end{eqnarray}}
\newcommand{\be}{\begin{equation}}
\newcommand{\ee}{\end{equation}}

\newcommand{\ch}{\widetilde \chi^\pm}

\def \ch2p {{\wt\chi_2^+}}

\def \ch2m {{\wt\chi_2^-}}

\newc{\dmchi}{\Delta m_{\wt\chi}}

\def\issue(#1,#2,#3){{\bf #1}, #2 (#3)}

\hypersetup{
   colorlinks=true,       
   linkcolor=blue,        
   citecolor=blue,         
   filecolor=magenta,      
   }

\title{Revisiting the non-resonant Higgs pair production at the HL-LHC }

\preprint{LAPTH-062/17, IPPP/17/100}

\author[a]{Amit Adhikary}
\author[b]{Shankha Banerjee}
\author[a]{Rahool Kumar Barman}
\author[a]{Biplob Bhattacherjee}
\author[c]{Saurabh Niyogi}
\affiliation[a]{Centre for High Energy Physics, Indian Institute of Science, Bangalore 560012, India}
\affiliation[b]{Institute for Particle Physics Phenomenology, Department of Physics, Durham University,
Durham DH1 3LE, United Kingdom}
\affiliation[b]{Universit\'{e} Grenoble Alpes, USMB, CNRS, LAPTh, F-74000 Annecy, France}
\affiliation[c]{Gokhale Memorial Girls' College, 1/1, Harish Mukherjee Road, Kolkata 700020, India}
\emailAdd{amitadhikary@iisc.ac.in}
\emailAdd{shankha.banerjee@durham.ac.uk}
\emailAdd{rahoolbarman@iisc.ac.in}
\emailAdd{biplob@iisc.ac.in}
\emailAdd{saurabhphys@gmail.com}
\date{\today}

\abstract
{We study the prospects of observing the non-resonant di-Higgs pair production in the Standard Model (SM) at the high luminosity run of
the 14 TeV LHC (HL-LHC), upon combining multiple final states chosen on the basis of their yield and cleanliness. In particular, we 
consider the $b\bar{b}\gamma \gamma, b\bar{b} \tau^+ \tau^-, b\bar{b} WW^*, WW^*\gamma \gamma$ and $4W$ channels mostly focusing on final states
with photons and/or leptons and study 11 final states. We employ multivariate analyses to optimise the discrimination between signal 
and backgrounds and find it performing better than simple cut-based analyses. The various differential distributions for the Higgs pair production have non-trivial dependencies on the Higgs self-coupling ($\lambda_{hhh}$). We thus explore the implications of varying $\lambda_{hhh}$ for the most sensitive search channel for the double Higgs production, \textit{viz.}, $b\bar{b}\gamma\gamma$. 
The number of signal events originating from SM di-Higgs 
production in each final state is small and for this reason measurement of differential distributions may not be possible. In order to 
extract the Higgs quartic coupling, we have to rely on the total number of events in each final state and these channels can be 
contaminated by various new physics scenarios. Furthermore, we consider various physics beyond the standard model scenarios to quantify
the effects of contamination while trying to measure the SM di-Higgs signals in detail. In particular, we study generic resonant heavy 
Higgs decays to a pair of SM-like Higgs bosons or to a pair of top quarks, heavy pseudoscalar decaying to an SM-like Higgs and a 
$Z$-boson, charged Higgs production in association with a top and a bottom quark and also various well-motivated supersymmetric 
channels. We set limits on the cross-sections for the aforementioned new physics scenarios, above which these can be seen as excesses 
over the SM background and affect the measurement of Higgs quartic coupling. We also discuss the correlations among various channels which 
can be useful to identify the new physics model.}

\begin{document}

\maketitle

\section{Introduction}
\label{intro}

The existence of a scalar boson with a mass around 125 GeV has been unambiguously confirmed by both the ATLAS and CMS collaborations at 
the Large Hadron Collider (LHC). It is, however, still early to conclude whether this discovered scalar is the Higgs boson as conjectured 
in the Standard Model of particle physics (SM)~\footnote{We will call this the Higgs from now on, for convenience.}. Therefore, it is 
of paramount importance to precisely measure its couplings to the various SM particles, its width, spin and $CP$ properties. As already 
seen from the Run I data and gradually being reiterated by the Run II data, the Higgs couplings to the SM electroweak gauge bosons are 
in excellent agreement with the SM expectations~\cite{Khachatryan:2016vau,CMS-PAS-HIG-16-020,ATLAS-CONF-2016-067,ATLAS-CONF-2016-079,
ATLAS-CONF-2016-081,CMS:2017jkd,ATLAS-CONF-2016-112,CMS-PAS-HIG-15-003}. The Yukawa couplings to the first two generation fermions are 
extremely difficult to measure owing to their smallness~\cite{ATLAS-CONF-2017-014}. However, the couplings to the third generation 
quarks and lepton are gradually gaining in significance~\cite{Aaboud:2017xsd,Sirunyan:2017khh,CMS-PAS-HIG-17-004,ATLAS-CONF-2016-080,
CMS-PAS-HIG-16-003,ATLAS-CONF-2016-091,CMS-PAS-HIG-16-006}. The only other measurable coupling (the first generation Yukawa couplings 
being extremely small, is considerably challenging to getting measured in the near future) which also describes the scalar potential of 
the theory is the elusive Higgs self-coupling ($\lambda_{hhh}$). The focus of the present work is to study in considerable details, 
various possible final states of the Higgs pair production and to study the effects of contamination due to the presence of several new 
physics effects. The only direct probe to this coupling is via a pair 
production of Higgs bosons which further decay to various SM final states. However, it has been shown in Refs.~\cite{Goertz:2013kp,
McCullough:2013rea,Degrassi:2016wml,Bizon:2016wgr,Gorbahn:2016uoy,DiVita:2017eyz,Maltoni:2017ims} that an indirect measurement of the Higgs trilinear 
coupling is possible through radiative corrections of single Higgs processes both at the HL-LHC and at future $e^+ e^-$ colliders. 
Ref.~\cite{DiVita:2017eyz} has shown that this coupling can be constrained in the range of [0.1,2.3] times that of its SM expectation 
at 68\% confidence level. It has also been shown in Ref.~\cite{Kribs:2017znd} that it is possible to constrain $\lambda_{hhh}$ from the electroweak 
oblique parameters. The triumph of the experiments in having already probed most of the standard Higgs couplings, urges the community to constrain the 
self-coupling in a plethora of channels. Such measurements have received considerable attention in recent times both from theoretical 
and experimental communities~\cite{Glover:1987nx,Eboli:1987dy,Plehn:1996wb,Dawson:1998py,Baglio:2014nea,Hespel:2014sla,Lu:2015qqa,
Bian:2016awe,Kribs:2012kz,Dawson:2012mk,Pierce:2006dh,Kanemura:2008ub,Nishiwaki:2013cma,Ellwanger:2013ova,Chen:2014xra,Liu:2014rba,
Slawinska:2014vpa,Goertz:2014qta,Azatov:2015oxa,Lu:2015jza,Carvalho:2015ttv,Gorbahn:2016uoy,Carvalho:2016rys,Banerjee:2016nzb,
Cao:2016zob,Dolan:2012ac,Contino:2010mh,Grober:2010yv,Contino:2012xk,Grober:2016wmf,Liu:2004pv,Dib:2005re,Wang:2007zx,Barger:2014taa,
Nakamura:2017irk,Crivellin:2016ihg,Cao:2013si,Sun:2012zzm,Costa:2015llh,No:2013wsa,Kotwal:2016tex,deFlorian:2017qfk,Cao:2015oaa,DiLuzio:2017tfn}. 
However, a precise direct measurement of the self-coupling is extremely challenging at the LHC because the SM 
production cross-section is small even at $\sqrt{s}=14$ TeV. The dominant di-Higgs production process proceeds through top quark loop 
diagrams in the gluon fusion channel. An interesting aspect of this process lies in the fact that there is a fine cancellation owing to 
a destructive interference between the box and the triangle diagrams. This results in an extremely small cross section, \textit{viz.}, 
$39.56^{+7.32\%}_{-8.38\%}$ fb at the NNLO+NNLL level~\cite{deFlorian:2013jea,deFlorian:2015moa,hhtwiki} (with full top mass effects at 
NLO~\cite{Borowka:2016ypz}) for the 14 TeV run of the LHC. 
However, various decay channels of the Higgs provide phenomenologically rich final states and appropriate combinations might help in 
improving the discovery potential at the high luminosity run of the LHC (HL-LHC), provided we identify optimised sets of selection cuts 
to reduce backgrounds ($B$) and improve the signal ($S$) over background ratio ($S/B$) and the statistical significance ($S/\sqrt{B}$). 
Searches for both resonant and non-resonant Higgs pair production have been performed in various channels by both the ATLAS and CMS 
experiments~\cite{Aad:2015xja,ATLAS-CONF-2016-049,ATLAS-CONF-2016-071,CMS-PAS-HIG-16-002,CMS-PAS-B2G-16-008,CMS-PAS-HIG-16-032,
CMS-PAS-HIG-16-011,CMS-PAS-HIG-17-002,CMS-PAS-HIG-17-006,CMS-PAS-HIG-16-028,ATLAS-CONF-2016-004}. At present, one of the strongest 
bounds on the non-resonant Higgs pair production comes from the $4b$ search performed by ATLAS~\cite{ATLAS-CONF-2016-049} with an 
integrated luminosity of 13.3 fb$^{-1}$, putting an upper bound of 29 times that of the SM expectation. Very recently, the $b b \tau 
\tau$ search by CMS~\cite{CMS-PAS-HIG-17-002,Sirunyan:2017djm} has put a strong observed limit at 30 times the SM number, with an integrated
luminosity of 35.9 fb$^{-1}$. The strongest (second strongest) constraint, at 13 (19.2) times 
that of the SM expectation, comes from the $b\bar{b}b\bar{b}$ ($b\bar{b}\gamma\gamma$) search 
by ATLAS~\cite{Aaboud:2018knk} (CMS~\cite{CMS-PAS-HIG-17-008}) with an integrated luminosity of 36.1 (35.9) fb$^{-1}$. As for the 
resonant searches, at present, the strongest limits are obtained from the $hh \rightarrow b\bar{b}\gamma\gamma$~\cite{CMS-PAS-HIG-17-008}, 
$hh \rightarrow b\bar{b}b\bar{b}$~\cite{CMS-PAS-HIG-17-009,Aaboud:2018knk} and $b\bar{b}\tau^+ \tau^-$~\cite{Sirunyan:2017djm} modes, 
competing in the mass range [$\sim 250$ GeV; 3 TeV]. However, the $b\bar{b}WW^*$ channel is also predicted to be a competitive probe 
in the future runs of the LHC~\cite{Papaefstathiou:2012qe,Huang:2017jws}.

The di-Higgs production rate can be enhanced in various beyond the Standard Model (BSM) scenarios. Some such new physics 
scenarios involve new heavy coloured states propagating in both the box and triangle loops, \textit{e.g.}, supersymmetric and extra-dimensional theories, theories 
with heavy resonance(s) decaying into a pair of SM-like Higgs, \textit{viz.}, a multitude of models with an extended Higgs sector, 
strongly interacting theories, composite Higgs models and also various effective field theories (EFTs) modifying the $t\bar{t}h$ 
coupling~\cite{Baglio:2014nea,Hespel:2014sla,Lu:2015qqa,Bian:2016awe,Kribs:2012kz,Dawson:2012mk,Pierce:2006dh,Kanemura:2008ub,
Nishiwaki:2013cma,Ellwanger:2013ova,Chen:2014xra,Liu:2014rba,Slawinska:2014vpa,Goertz:2014qta,Azatov:2015oxa,Lu:2015jza,
Carvalho:2015ttv,Gorbahn:2016uoy,Carvalho:2016rys,Banerjee:2016nzb,Cao:2016zob,Dolan:2012ac,Contino:2010mh,Grober:2010yv,Contino:2012xk,
Grober:2016wmf,Liu:2004pv,Dib:2005re,Wang:2007zx,Barger:2014taa,Nakamura:2017irk,Crivellin:2016ihg,Cao:2013si,Sun:2012zzm,
Costa:2015llh,No:2013wsa,Kotwal:2016tex,deFlorian:2017qfk,Cao:2015oaa,DiLuzio:2017tfn}. Since the Higgs discovery, many of the models 
exhibiting new coloured states, have been severely constrained owing to the near-precise measurements in the single Higgs channels. Many of these 
extensions are responsible not only for an enhancement in the di-Higgs production cross-section, but also for certain distinct kinematic 
distributions, often having minimal overlap with their SM counterparts. We must, however, remember that even the enhanced cross-sections 
might not be entirely sufficient to obtain an adequate significance because large SM backgrounds, primarily ensuing from $t\bar{t}$, 
$ZZ$, $ZH$, pure QCD and also fakes, may swamp the signal completely. In this regard, modified kinematics, especially the presence of 
resonances might be somewhat helpful. In the quest to reduce backgrounds to the best of one's abilities, one has to envision a combination 
of optimal final states. In addition, for each such final state, one has to identify the most suited set of selection cuts in order to 
enhance signal-to-background ratio. A thorough literature survey points us to studies which show that the trilinear coupling can be 
best probed when studied in multiple channels with a combination of the numerous final states of the Higgses. These final states are 
chosen owing to the largeness of the Higgs branching ratios and their cleanliness with respect to the backgrounds. A more inclusive 
search procedure takes a closer look into various kinematic regions of di-Higgs processes. In particular, studies utilising variables 
reconstructed from boosted objects, jet substructure techniques, stransverse mass ($m_{T_2}$) and other novel variables, are also shown 
to have potential importance in the future runs of the LHC~\cite{Dolan:2012rv,Barr:2013tda}. Multivariate analyses also turn out to be 
very efficient in segregating the signal from the backgrounds, thus offering encouraging results~\cite{Barger:2013jfa,Kling:2016lay,Alves:2017ued}. 
Nevertheless, an exhaustive study in the di-Higgs sector, involving detector simulations and also alongside an inclusion of the 
effects of new physics effects (as we shall discuss below) on such measurements, is by and large missing from the literature, 
since some of the aforementioned studies claiming very optimistic results have been performed at the parton level or with minimal 
detector effects. Hence, one of the primary goals in this work is to optimise the di-Higgs search strategy by systematically studying a 
number of final states taking into account detector effects and conservative systematic uncertainties. 

In the first part of our study, we focus on the non-resonant di-Higgs production in the familiar $b \bar{b} \gamma \gamma$, 
$b \bar{b} \tau^+ \tau^-$ and $b \bar{b} WW^*$ channels and try to estimate the statistical significances at the
HL-LHC. Being mostly agnostic to the previous studies, we try to identify the sets of optimised cuts which show the greatest 
sensitivities in these channels. The $b \bar{b} \gamma \gamma$ and $b \bar{b} W W^*$ have been shown to be the most promising channels in 
this regard~\cite{Baglio:2012np,Baur:2003gpa,Huang:2017jws}. The $b \bar{b} \tau^+ \tau^-$ channel, however, suffers from large 
$t \bar{t}$ backgrounds. The reconstruction of $\tau$s, which is always accompanied by missing transverse energy ($\slashed{E}_T$), 
is a complicated process at the colliders and involves identifying optimal $\tau$-tagging and mistagging efficiencies. However, 
improvements in the reconstruction of invariant mass of the di-tau system using the missing mass algorithm~\cite{Elagin:2010aw}, 
dynamical likelihood techniques~\cite{Bianchini:2014vza} or the modified $m_{T_2}$ algorithm~\cite{Barr:2011he,Barr:2013tda} may 
provide encouraging results in this channel. Before performing these 
studies, we stress that the analyses involving these channels are not novel and hence we will be more cautious in our claims. CMS 
predicts final significances of 1.6$\sigma$, 0.39$\sigma$, 0.45$\sigma$ and 0.39$\sigma$ respectively in the $b\bar{b}\gamma \gamma, 
b \bar{b} \tau^+ \tau^-, b \bar{b} V V^*$ and $b\bar{b}b\bar{b}$ channels for the non-resonant di-Higgs production, at the end of 
HL-LHC run with an integrated luminosity of 3 ab$^{-1}$~\cite{CMS-DP-2016-064}. ATLAS on the other hand predicts their best-case 
significance at 1.05$\sigma$ for the $b\bar{b} \gamma \gamma$ non-resonant channel at the HL-LHC~\cite{ATL-PHYS-PUB-2017-001}. Moreover, 
for the $b\bar{b}WW^*$ channel, we study both the semi-leptonic and di-leptonic modes. Besides, we look into the $\gamma \gamma WW^*$ 
channel with both the semi-leptonic and di-leptonic final states. Finally, we also look for the $4W$ channel in the same-sign di-lepton 
($SS2\ell$), tri-lepton ($3\ell$) and four lepton ($4\ell$) final states. We compare the numbers obtained from the experimental 
projections with our study by including detailed detector effects and conservative background systematics.

In this work, we will not concern ourselves with dedicated analyses for resonant di-Higgs searches. Neither will we focus on scenarios 
where the rescaling or the modification in the $t\bar{t}h$ Yukawa coupling ($y_t$)may alter the nature of interference 
between the triangle and box diagrams. However, we will briefly discuss the case where 
one can have $\lambda_{hhh}$ different from the SM expectations. These, in principle, can have drastic ramifications in the production cross-sections as well as 
the kinematics of the di-Higgs system. New physics contributions may also show up in the $\textrm{BR}(h \to XX)$, modifying the total
rate. These will be considered as a separate future study. In the present work, we will however consider various BSM signatures which
have the potential to contaminate the non-resonant SM di-Higgs production and affect the measurement of $\lambda_{hhh}$. Observing 
any significant difference in the number of events for a particular channel, with respect to its SM expectation, may be interpreted 
as a modification in the value of $\lambda_{hhh}$. This is one of the main aims of this present work. We want
to quantify the degree to which we can discard such contamination after having established a robust set of cuts which optimises the
SM signal. We will be using multivariate analyses for this purpose. We classify these contaminating scenarios into three broad 
categories, \textit{viz.}, $h h (+ X)$, $h + X$ and $X$, where $X$ denotes an object or a group of objects not coming from an SM Higgs 
decay. The $h h (+ X)$ mode is one of the most studied scenarios. Di-Higgs production from the decays of heavy scalar particles is the classic
case considered in the literature~\cite{Bhattacherjee:2014bca,Ren:2017jbg,Nakamura:2017irk,vanBeekveld:2015tka}. A heavy scalar particle
arises naturally in many extensions of the SM, for instance, in the minimal supersymmetric standard model (MSSM) or in further extended
scenarios~\cite{Dolan:2012ac,Cao:2013si,Biswas:2016ffy}, general two-Higgs doublet models (2HDMs)~\cite{Hespel:2014sla,Baglio:2014nea}, extra-dimensional 
models~\cite{Sun:2012zzm}, models with an extra $U(1)$ gauge group~\cite{Costa:2015llh,No:2013wsa,Kotwal:2016tex}, to name a 
few. In the present work, we do not focus on any particular model and consider a generic heavy resonance decaying to a pair of SM Higgses 
which further decay to various final states. We vary the mass of the heavy resonance but do not optimise the selection cuts for each 
benchmark and keep them fixed at the optimisation obtained for the corresponding SM non-resonant Higgs pair production channel. 
Delving a bit more into well-motivated models, we consider certain different channels in the MSSM from which we can obtain 
a pair of SM-like Higgs bosons. For generic supersymmetric (SUSY) scenarios, we will encounter high effective masses ($m_{\textrm{eff}}$)
and high missing transverse momentum ($\met$). This will lead to a minimal or no overlap of kinematic variables with their SM di-Higgs 
counterparts. For a degenerate SUSY spectrum, however, we will obtain low $m_{\textrm{eff}}$ and low $\met$ and this may potentially 
contaminate several di-Higgs final states. The $hh(+X)$ state may come from a squark pair production, \textit{i.e.}, $p p \to \tilde{q_i} 
\tilde{q_j} \to q_i q_j + h h + \chi_1^0 \chi_1^0$, where $\tilde{q_i}$ refers to squarks (anti-squarks), $q_i$ refers to quarks (anti-quarks) 
with $i$ being the flavour index and $\chi_1^0$ to the lightest supersymmetric particle (LSP), here the lightest neutralino. Thus, we obtain a $hh + \; \textrm{jets} \; + 
\met$ state which has the potential to contaminate the SM di-Higgs signal unless specific cuts are designed to subdue its effect. For 
the second category, we consider a mono-Higgs production in association with other objects and this can specifically mimic some of the
Higgs pair production final states. We consider few such scenarios, \textit{viz.}, $A \to Z h$, \textit{i.e.}, a pseudoscalar decaying 
to the $Z$ boson along with the SM-like Higgs: this scenario is particularly interesting in the MSSM and also in classes of generic 
2HDMs. We will encounter the $b \bar{b} \gamma \gamma$, $b \bar{b} W W^*$, $b \bar{b} \tau^+ \tau^-$ final states from this channel. Besides, we will even
have some contamination to the $SS2\ell$, $3\ell$ and $4\ell$ final states. Furthermore, an electroweakino pair production may also 
exhibit a mono-Higgs final state with a significant rate. Processes like $p p \to \chi_2^0 \chi_1^{\pm} \to h W^{\pm} + \chi_1^0 
\chi_1^0$, where the lightest chargino and the second-lightest neutralino are wino-like can contribute significantly. For such a 
scenario, $\textrm{BR}(\chi_2^0 \to h \chi_1^0)$ can be dominant and $\textrm{BR}(\chi_1^{\pm} \to W^{\pm} + \chi_1^0)$ is close to
unity. From such channels, we can have possible contaminations to the semi-leptonic $b\bar{b}W^+ W^-$, $\gamma \gamma W^+ W^-$ and 
$b\bar{b} \tau^+ \tau^-$ channels and also to the $SS2\ell$ and $3\ell$ modes. The final category of BSM scenarios having potential
contaminating effects to the SM di-Higgs production are processes with no SM-like Higgs bosons. In this paper, we study three such examples.
We may have the production of a pair of top quarks emanating from a heavy (pseudo-)scalar resonance, displaying prowess for resonant
masses above the $t\bar{t}$ threshold. Besides, in various classes of models we have an associated production of a charged Higgs boson 
with a top and a bottom quark ($H^{\pm}tb$). For $m_{H^{\pm}} > m_t$, we have the $tbtb$ production. Another potential contamination 
can come from the stop-anti-stop ($\tilde{t_i} \tilde{t_i^*}$, where $i=1,2$ ) pair production which can lead to the $t\bar{t} + 
\slashed{E}_T$ or the $bWbW + \met$ final states. All the above three channels can mimic the $hh \to b \bar{b} W W^*$ and $b \bar{b} 
\tau^+ \tau^-$ modes. In the following, we make an attempt to study these contamination effects as functions of the neutral/charged
heavy Higgs masses for certain well-chosen benchmark points. In the following sections, we will see the importance of multivariate 
analyses in discriminating the SM di-Higgs signal from the SM backgrounds and later also from possible new physics contaminations. 
Hence, the backbone of the analyses techniques used in this work, are the boosted-decision tree (BDT) algorithms.

Having described the various aspects studied in this work, we dissect our paper into the following sections. In section~\ref{sec2}, we
study the SM non-resonant di-Higgs final states in considerable details and present the reach of the HL-LHC in observing various 
channels. We discuss the variation of the Higgs self-coupling and the effects 
one obtains on the signal sensitivity, in section~\ref{sec2-2}. In section~\ref{sec3}, we consider the contamination effects ensuing from the aforementioned three categories with the help
of benchmark points. Finally, in section~\ref{sec4}, we summarise our results, conclude and present a future outlook for the vast 
field of di-Higgs searches.


\section{Non-resonant di-Higgs production}
\label{sec2}

As discussed in the introduction, the objective of this present work is two-fold, \textit{viz.}, estimating the observability of SM 
di-Higgs production in multifarious channels at the HL-LHC and also to decipher the contamination to such SM processes from various new 
physics scenarios as we will discuss at length in section~\ref{sec3}. In this section, we will focus on several possible final states 
of the SM Higgs pair production. Our guiding principles in choosing these final states are cleanliness and substantial production rates. 
Hence, we choose states containing either photons or leptons ($e$, $\mu$ and $\tau$) or both. Thus, we consider the $b\bar{b} \gamma 
\gamma$, $b \bar{b} \tau^+ \tau^-$, $b \bar{b} W W^*$, $W W^* \gamma \gamma$ and $4W$ channels for the present work. We do not 
consider the $4 \tau$, $W W^* \tau^+ \tau^-$, $Z Z^* \tau^+ \tau^-$, $4 \gamma$, $Z Z^* \gamma \gamma$ and $4 Z$ states on account of 
their negligible rates. We must mention however that some of these neglected channels at the 14 TeV study may have important 
ramifications for 100 TeV collider studies~\cite{Papaefstathiou:2015iba}. At this point, it is important to mention that we closely follow the 
ATLAS and CMS analyses whenever available. For channels where we are unable to find such studies, we optimise the cuts to maximise the 
significance.

As we have emphasised in the introduction, the gluon fusion mode prevails as the dominant contribution to the SM di-Higgs production 
when compared with the remaining modes, such as vector boson fusion, associated production with a vector boson~\cite{Cao:2015oxx}, or 
double Higgs production in association with a pair of top quarks~\cite{Baglio:2012np}. Hence, for the present study, we concern ourselves only with 
the former production mode. On the simulation front, we generate the di-Higgs signal samples at leading order (LO) upon using 
{\tt MG5\_aMC@NLO}~\cite{Alwall:2014hca}. To attain the final states discussed above, we decay these samples with {\tt Pythia-6}~\cite
{Sjostrand:2001yu,Sjostrand:2014zea}. We generate the background event samples also at LO using {\tt MG5\_aMC@NLO}~\footnote{We must 
clarify here that even though we generate our signal and background samples at LO, we use the higher order cross-sections~\cite{signal_twiki_cs,
bkg_twiki_cs} throughout our analysis, whenever available.}. Unless the decays are done at the {\tt MG5\_aMC@NLO} level, we decay these 
with {\tt Pythia-6}. The generation level cuts for the various processes are listed in Appendix~\ref{sec:appendixA}. For all our simulations, the NN23LO parton distribution function (PDF)~\cite{Ball:2012cx} has been employed. Also 
for all our sample generations, we use the default factorisation and renormalisation scales as defined in {\tt MG5\_aMC@NLO}~\cite{renfac}. 
Next, we shower and hadronise the signal and background samples with {\tt Pythia-6}. Following this, the final state jets are reconstructed 
with the anti-$kT$~\cite{Cacciari:2008gp} algorithm with a minimum $p_T$ of 20 GeV and a jet parameter of $R = 0.4$ in the {\tt 
FastJet}~\cite{Cacciari:2011ma} framework. In order to simulate detector effects, we use {\tt Delphes-3.4.1}~\cite{deFavereau:2013fsa}. 
Unless otherwise stated, we demand the electrons, muons and photons to be isolated as follows: the total energy activity within a cone 
of $\Delta R = 0.5$ around each such object, is required to be smaller than 12\%, 25\% and 12\% respectively of its $p_T$. Besides, we 
consider the default identification efficiencies of the electrons, muons and photons as specified in the ATLAS detector card in {\tt 
Delphes-3.4.1}. For channels with $b$-jets as final state objects, we consider a flat $b$-tagging efficiency of 70\%~\cite{ATL-PHYS-PUB-2017-001}. 
We also consider flat $j \to b$ and $c \to b$ mistag rates of 1\% and 30\% respectively. Here we would also like to clarify that 
whenever in the following sub-sections, we mention a lepton ($\ell$) as a final state, we always refer to an electron or a muon.

In almost all the channels which follow, we perform a cut-based analysis (whenever an equivalent analysis has been performed by CMS or
ATLAS) for the signal optimisation. For these channels and also for the rest where we do not perform a cut-based analysis, we perform a
multivariate analysis in order to capture the full machinery of an optimised search. For such studies, we choose numerous discriminatory
variables, depending on the analysis and use the TMVA framework~\cite{2007physics3039H} to discriminate between the signal and background
samples. For the following analyses, we use the decorrelated boosted decision tree (BDTD) algorithm. We must admit here that, it is 
possible to have a further improved algorithm but here we stick to a standard discriminator. In all cases, we train the signal and
background samples, carefully avoiding overtraining of the samples at each step. For this purpose, we demand that the Kolmogorov-Smirnov
test results are always greater than 0.1. It is, however, mentioned in Ref.~\cite{KS} that a non-oscillatory critical test value of 0.01
may also suffice as a test for overtraining. We systematically modulate the BDT optimisation procedure with sufficiently large number 
of signal and background samples and always ensure a KS test value greater than 0.1 for both signal and background. 

With this machinery in hand, we outline and detail the prospects of the non-resonant di-Higgs process in various  final states in the
following sections. We also note that all our generated samples are at a centre of mass energy of 14 TeV and the final analyses are 
performed for an integrated luminosity of 3 ab$^{-1}$.

\subsection{The $b\bar{b}\gamma\gamma$ channel }
\label{sec2.1:2b2gamma}

Having set the stage, we begin by studying one of the most promising non-resonant di-Higgs search channels at the HL-LHC, \textit{viz.}, 
the $b\bar{b} \gamma \gamma$ final state. Even though this channel is somewhat at a disadvantage from the point of view of the total 
rate, because of the extremely small branching ratio of $h \to \gamma \gamma$, the cleanliness of this channel makes way for an 
adequate compensation, as we will gather at the end of this section. Numerous studies in the literature~\cite{Baur:2003gp,Baglio:2012np,
Azatov:2015oxa,Barger:2013jfa,ATL-PHYS-PUB-2017-001,ATL-PHYS-PUB-2014-019} have attempted to constrain the Higgs 
self-coupling ($\lambda$) by focusing on this particular final state. In performing this study, we closely follow the analysis presented 
in Ref.~\cite{ATL-PHYS-PUB-2017-001}.

The most dominant background stems from the QCD-QED $b \bar{b}\gamma \gamma$ process. We generate this background upon merging 
with an additional jet by employing the MLM merging scheme~\cite{Mangano:2006rw}. We must also mention here that the pure QED contribution 
(not involving the Higgs) to $b \bar{b} \gamma \gamma$ is $\mathcal{O}(1\%)$ that of its QCD-QED counterpart. Other significant 
backgrounds arise from the associated production of the Higgs with a pair of bottom (top) quarks, $b \bar{b} h$ ($t \bar{t} h$) and the associated production of 
Higgs with a $Z$-boson ($Z h$). In addition to these backgrounds, contributions also arise from numerous fakes, having event yields 
comparable to the QCD-QED $b \bar{b}\gamma \gamma$ process. Although, the list of such relevant fake backgrounds is exhaustive, 
\textit{viz.}, $c \bar{c}\gamma \gamma$, $j j \gamma \gamma$, $b \bar{b} j j$, $b \bar{b} j \gamma$ and $c \bar{c} j \gamma$, it is 
considerably difficult to simulate them. Thus, for the $c \bar{c}\gamma \gamma$ and $j j \gamma \gamma$ channels, which bear a similar 
topology to the QCD-QED $b \bar{b} \gamma \gamma$ process, we estimate the fake event yields upon employing a simple scaling: 
$N^{c \bar{c}\gamma \gamma \; (j j \gamma \gamma)} = (N^{c \bar{c}\gamma \gamma \; (j j \gamma \gamma)}_{\textrm{ATLAS}} / N^{b \bar{b} 
\gamma \gamma}_{\textrm{ATLAS}}) \cdot N^{b \bar{b}\gamma \gamma}$, where the subscript ATLAS denotes the event yields as listed 
in Ref.~\cite{ATL-PHYS-PUB-2017-001}, while, $N^{b \bar{b}\gamma \gamma}$ is our simulated estimation. In an analogous manner, we simulate 
the $b \bar{b} j \gamma$ and $b \bar{b} j j$ backgrounds and scale $N^{c \bar{c} j \gamma} = (N^{c \bar {c} j \gamma}_{\textrm{ATLAS}}/
N^{b \bar{b} j \gamma}_{\textrm{ATLAS}}) \cdot N^{b \bar{b}j \gamma}$. Following Ref.~\cite{ATL-PHYS-PUB-2017-001}, we consider a $j 
\to \gamma$ fake probability of $\sim 0.1\%$. Also, at this point, we would like to mention that the fake rates are $p_T/\eta$ dependent 
functions and for precise analyses, these must be dealt with more care.

Upon generating the samples, for every event we require exactly two $b$-tagged jets and two photons in the final state. The leading 
(sub-leading) $b$-jet is required to have $p_{T,b_1 (b_2)} > \; 40 \; (30)~{\rm GeV}$ and must lie within a pseudo-rapidity 
range of $|\eta_{b_1,b_2}| < 2.4$. The two photons are required to have $p_{T,\gamma} > 30~{\rm GeV}$ and are required to lie within 
$|\eta_{\gamma}| < 1.37$ (barrel) or $1.52 < |\eta_{\gamma}|< 2.37$ (endcap). Additionally, we also veto events having one or more 
isolated leptons with $p_T > 25~{\rm GeV}$ and within $|\eta| < 2.5$. The following selection cuts are implemented and are also 
tabulated in Table~\ref{tab:bbgamgam_sel_cut}. We demand that the jet multiplicity, $N_{j}$ must be less than 6 in order to reduce the 
large $t\bar{t}h$ background when either or both the top-quarks decay hadronically via the decays of the $W$-bosons. We also 
find that the $\Delta{R}$ cuts are highly effective in tackling the QCD-QED $b \bar{b} \gamma 
\gamma$ background. Here, $\Delta{R}_{ab}$ refers to the distance between the final state particles $a$ and $b$ in the $\eta$-$\phi$ 
plane. In addition, we also impose an upper and lower limits on the invariant masses of the two $b$-jets (100 GeV $< m_{b b} < 150$ 
GeV) and the two photons (122 GeV $< m_{\gamma\gamma} < 128$ GeV), which impressively reduces the QCD-QED $b\bar{b}\gamma\gamma$ 
background and sufficiently affects all the other backgrounds as well. Lastly, we also impose a lower bound on the transverse momenta 
of the $b$-jet pair ($p_{T, b b} > 80$ GeV) and the transverse momenta of the di-photon pair ($p_{T,\gamma \gamma} > 80$ GeV).

We tabulate the signal and background yields for each selection cut in Table~\ref{tab:bbgamgam_yield}. We also quote the statistical 
significance $S/\sqrt{B}$, where $S$ represents the signal yield and $B$ refers to the sum of all relevant backgrounds. Upon applying
all the aforementioned cuts, we obtain a final significance of $1.46$, assuming zero systematic uncertainty. Because this first part 
of our paper somewhat serves as a validation of the studies performed by the ATLAS and CMS collaborations, we would like to confirm 
that our statistical significance is consistent with the results obtained by ATLAS~\cite{ATL-PHYS-PUB-2017-001}.

\begin{table}
\begin{center}
\begin{tabular}{||c||}\hline \hline
Selection cuts \\ \hline
$N_j < 6$ \\
$0.4 < \Delta R_{\gamma \gamma} < 2.0$, $0.4 < \Delta R_{bb} < 2.0$, $\Delta R_{\gamma b} > 0.4$\\
$100~{\rm GeV} < m_{bb} < 150~{\rm GeV}$ \\
$122~{\rm GeV} < m_{\gamma \gamma} < 128~{\rm GeV}$ \\
$p_{T, b b} > 80~{\rm GeV},~ p_{T, \gamma \gamma} > 80~{\rm GeV}$\\ \hline
\end{tabular}
\caption{Selection cuts for the cut-based analysis in the $b\bar{b}\gamma\gamma$ channel following Ref.~\cite{ATL-PHYS-PUB-2017-001}.}
\label{tab:bbgamgam_sel_cut}
\end{center}
\end{table}

\begin{table}[htb!]
\begin{bigcenter}
\scalebox{0.7}{%
\begin{tabular}{||c||c|c|c|c|c|c|c||c||}
\hline
 & \multicolumn{7}{c||}{Event rates with $3000 \; \textrm{fb}^{-1}$ of integrated luminosity} & \\ \cline{2-8}
 Cut flow   & Signal & \multicolumn{6}{c||}{SM Backgrounds} &$\frac{S}{\sqrt{B}}$\\
\cline{3-8} 
 & $hh \to 2b2\gamma$ & $hb\bar{b}$ & $t\bar{t}h$ & $Zh$ & $b\bar{b}\gamma\gamma *$~\footnote{$b \bar{b} \gamma \gamma + c \bar{c} \gamma \gamma + j j \gamma \gamma $.}  & Fake 1~\footnote{$b \bar{b} j \gamma + c \bar{c} j \gamma$.} & Fake 2~\footnote{$b \bar{b} j j$.}   & \\
\cline{1-8}  
Order                                & NNLO~\cite{hhtwiki}         & NNLO (5FS) +                   & NLO~\cite{bkg_twiki_cs}       & NNLO (QCD) +                & LO   & LO & LO & \\ 
                                     &                             & NLO (4FS)~\cite{bkg_twiki_cs}  &                               & NLO EW~\cite{bkg_twiki_cs}  &      &    &    & \\
\cline{1-8} 
\hline
\hline 
$2b+2\gamma$                         & $31.63$     & $21.20$ &  $324.91$ & $39.32$ & $25890.31$ & $1141.18$ & $393.79$  & $0.19$\\

\hline
lepton veto                          & $31.63$     & $21.20$ &  $255.66$ & $39.32$ & $25889.94$ & $1141.18$ & $393.79$  & $0.19$ \\

\hline
$N_j < 6$                            & $31.04$     & $21$    &  $192.05$ & $39.23$ & $25352.78$ & $1064.64$ & $167.32$  & $0.19$\\

\hline
$\Delta R$ cuts                      & $22.19$     & $7.75$  &  $38.71$  & $23.48$ & $4715.21$  & $130.10$  & $28.81$   & $0.31$ \\

\hline
${m_{bb}}$                           & $12.71$     & $1.53$  &  $13.80$  & $1.09$  & $862.37$   & $22.11$   & $6.88$    & $0.42$ \\

\hline
${m_{\gamma\gamma}}$                 & $12.36$     & $1.5$   &  $13.16$  & $1.06$  & $26.54$    & $22.11$   & $6.88$    & $1.46$\\
 
\hline
${p_{T,bb}}$,${p_{T,\gamma \gamma}}$ & $12.32$     & $1.48$  &  $13.03$  & $1.06$  & $26.54$    & $21.82$   & $6.88$    & $1.46$\\
 
\hline

\end{tabular}}
\end{bigcenter}
\caption{The cut-flow and significance table for the $b\bar{b} \gamma \gamma$ mode.}

\label{tab:bbgamgam_yield}
\end{table}

Before moving on to discussing the multivariate analyses, we slightly digress in discussing the effects of certain possible cuts in
improving the significance when compared to the one we derived just above. One of the largest background yields even after imposing
all the aforementioned cuts is $t\bar{t}h$. However, it is interesting to note that this channel is associated with missing
transverse energy even at the parton level when at least one of the $W$-bosons decays leptonically. Our signal, on the other 
hand, other than $\met$ emanating from experimental noise, does
not have any missing energy. Hence, we demand an upper limit of $\met < 50$ GeV and show in Table~\ref{tab:bbgamgam_etmiss} (a) that
the $t\bar{t} h$ background reduces to almost half its previous value. The $b\bar{b} \gamma \gamma$ and Fake 1 backgrounds also incur
modest reductions. The signal on the other hand reduces marginally. This improves the $S/B$ from 0.17 to 0.19. Accordingly, the signal 
significance with zero systematics, acquires a slight increase at $1.51$. 

On a slightly different note, the ATLAS analysis~\cite{ATL-PHYS-PUB-2017-001} that we follow has considered jet energy corrections, to 
account for the parton radiation sourced from outside the jet cone. This results in the invariant mass distribution of the $b\bar{b}$ 
pair coming from the Higgs boson to peak at a value less than that of the Higgs mass. In the present study, we have however, only 
implemented the default jet energy correction considered in Delphes. As a result, we attempt to study the consequence of modifying the range of the selection 
cut on $m_{bb}$ to 90 GeV $<m_{bb}<130$ GeV. We present the new results in Table.~\ref{tab:bbgamgam_etmiss} (b). This modified 
selection cut results in an increase in the $Zh$ background but the signal also receives a relatively large increase, resulting in an 
$S/B$ of 0.19 and a significance of $1.64$. We left these last two modified cuts at the discussion level as issues concerning both 
$\met$ and jet-energy correction are primarily experimental and it is non-trivial to predict if our modified cuts can be incorporated 
seamlessly in an experimental setup.


\begin{center}
\begin{table}[htb!]
\centering
\scalebox{0.9}{%
\begin{tabular}{|c|c|c|}\hline
(a) & Process & Events \\ \hline \hline

\multirow{4}{*}{Background}  
 & $hb\bar{b}$              & $1.31$ \\ 
 & $t\bar{t}h$              & $7.87$ \\  
 & $Zh$                     & $1.03$\\  
 & $b\bar{b}\gamma\gamma *$ & $23.18$ \\ 
 & Fake 1                   & $20.69$ \\  
 & Fake 2                   & $6.52$ \\ \cline{2-3}  
 & Total                    & $60.60$ \\ \hline
\multicolumn{2}{|c|}{Signal ($hh \to 2b2\gamma$)}  & $11.75$ \\\hline 
\multicolumn{2}{|c|}{Significance ($S/\sqrt{B}$)} & $1.51$ \\ \hline \hline  
 
\end{tabular}}
\quad
\scalebox{0.9}{%
\begin{tabular}{|c|c|c|}\hline
(b) & Process & Events \\ \hline \hline

\multirow{4}{*}{Background}  
 & $hb\bar{b}$              & $1.55$ \\ 
 & $t\bar{t}h$              & $11.91$ \\  
 & $Zh$                     & $4.43$\\  
 & $b\bar{b}\gamma\gamma *$ & $28.41$ \\ 
 & Fake 1                   & $22.39$ \\  
 & Fake 2                   & $7.25$ \\ \cline{2-3}  
 & Total                    & $75.94$ \\ \hline
\multicolumn{2}{|c|}{Signal ($hh \to 2b2\gamma$)}  & $14.27$ \\\hline 
\multicolumn{2}{|c|}{Significance ($S/\sqrt{B}$)} & $1.64$ \\ \hline \hline  
 
\end{tabular}}
\caption{Signal, background yields and statistical significance after applying (a) $\met < 50~{\rm GeV}$ on top of the selection 
cuts and (b) modifying $m_{bb}$ to 90 GeV $< m_{bb}<130$  GeV.}
\label{tab:bbgamgam_etmiss}
\end{table}
\end{center}

\begin{figure}
\includegraphics[scale=0.2]{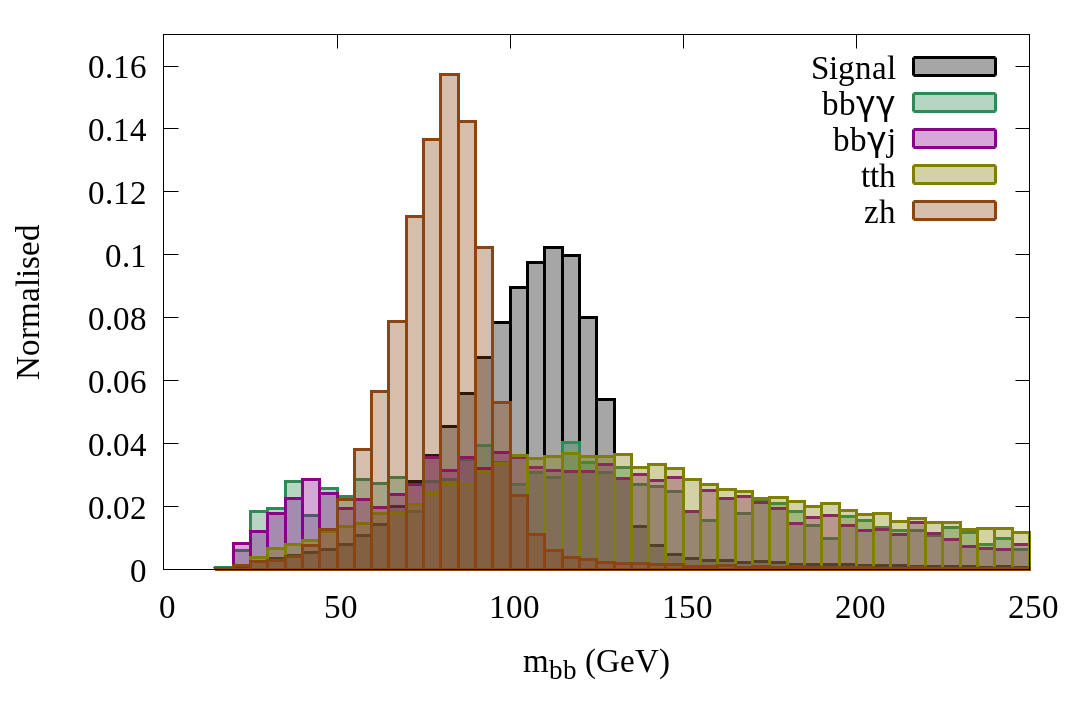}\includegraphics[scale=0.2]{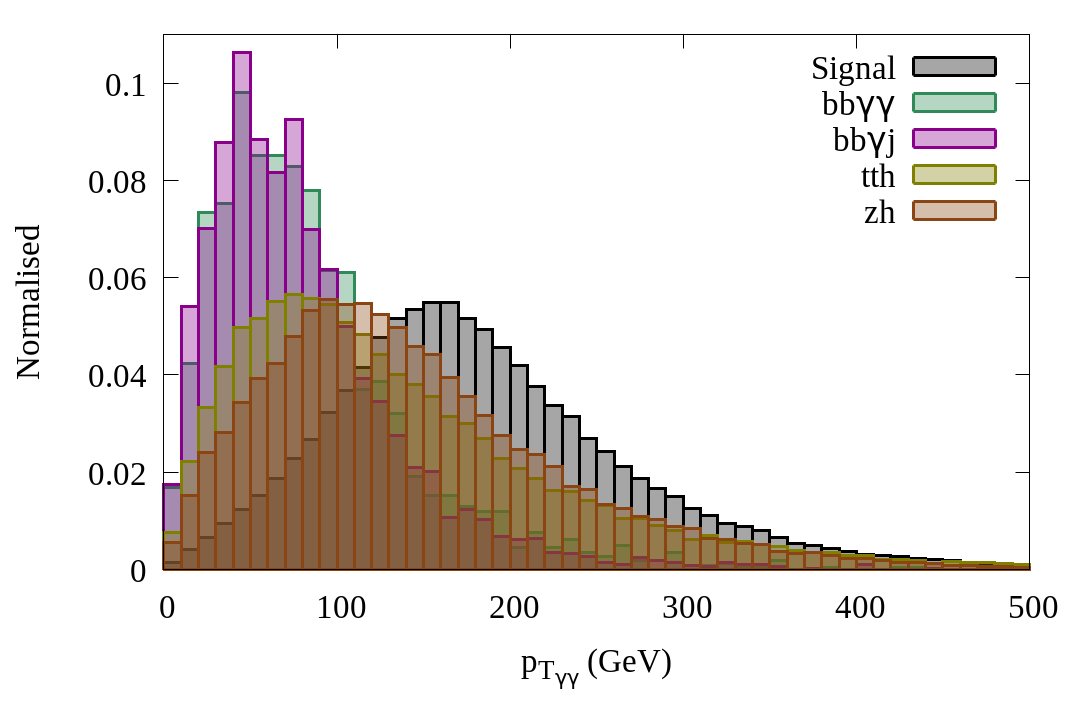}\\
\includegraphics[scale=0.2]{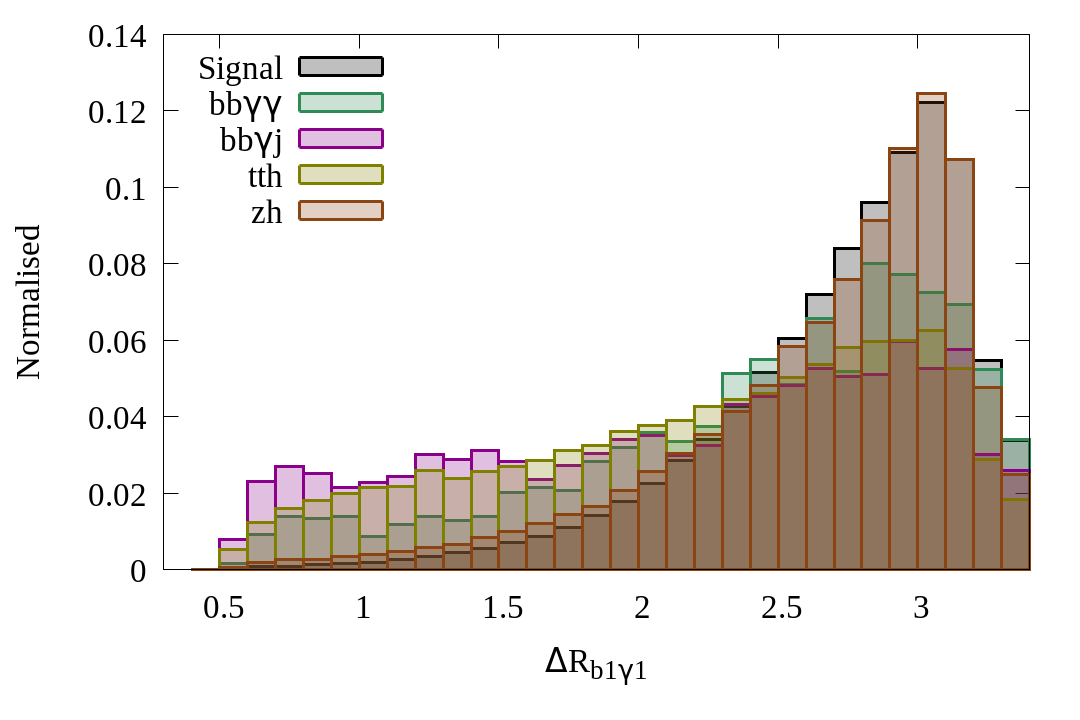}\includegraphics[scale=0.2]{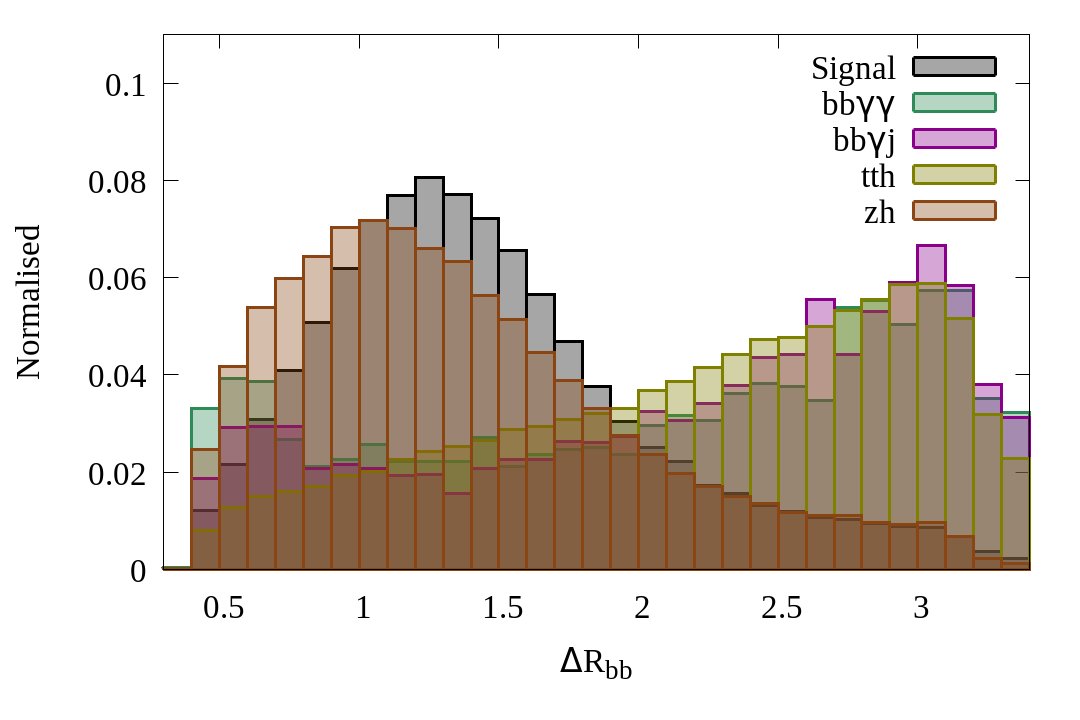}
\caption{Normalised distributions of $m_{bb}$, $p_{T,\gamma\gamma}$, $\Delta R_{b_1 \gamma_1}$ and $\Delta R_{b b}$ for the signal and 
the relevant backgrounds in the $b\bar{b} \gamma \gamma$ channel after the basic selection cuts.}
\label{fig:bbgamgam_tmva}
\end{figure}

In the last leg of this subsection, we perform a multivariate analysis of the $b\bar{b}\gamma \gamma$ final state by utilising the BDT 
algorithm in an attempt to isolate the signal and backgrounds more efficiently and improve upon the signal significance. The BDT 
optimisation procedure is performed upon using the following kinematic variables: 
\begin{eqnarray}
 m_{bb},~p_{T,\gamma\gamma},~\Delta R_{\gamma \gamma},~p_{T,bb},~\Delta R_{b_1\gamma_1},~p_{T,\gamma_1},~\Delta R_{bb},\nonumber \\
 ~p_{T,\gamma_2}, ~\Delta R_{b_2\gamma_1},~\Delta R_{b_2\gamma_2},~p_{T,b_1},~\Delta R_{b_1\gamma_2},~p_{T,b_2},\met \nonumber,
\end{eqnarray}
where the numerical subscripts signify the $p_T$ ordering of an object with the subscript 1 corresponding to the hardest object.
In the course of training the BDT, the kinematic variables $m_{bb}$, $p_{T,\gamma\gamma}$, $\Delta{R}_{b_1\gamma_1}$ and 
$\Delta R_{bb}$ showed the maximal prowess in discriminating the signal from the background. We present the normalised distributions 
of these variables for the signal and the dominant backgrounds in Fig.~\ref{fig:bbgamgam_tmva} after the basic selection cuts. The 
corresponding signal and background yields along with the final significance are tabulated in Table~\ref{tab:bbgamgam_tmva}. We 
observe that the multivariate analysis features a $\sim20\%$ improvement in the significance ($S/\sqrt{B}=1.76$) over its cut-based 
counterpart.

\begin{center}
\begin{table}[htb!]
\centering
\scalebox{0.7}{
\begin{tabular}{|c|c|c|}\hline
Sl. No. & Process & Events \\ \hline \hline

\multirow{4}{*}{Background}  
 & $hb\bar{b}$              & $2.75$ \\ 
 & $t\bar{t}h$              & $14.85$ \\  
 & $Zh$                     & $12.28$\\  
 & $b\bar{b}\gamma\gamma *$ & $34.46$ \\ 
 & Fake 1                   & $14.25$ \\  
 & Fake 2                   & $8.46$ \\ \cline{2-3}  
 & \multicolumn{1}{|c|}{Total}                    & $87.05$ \\ \hline
\multicolumn{2}{|c|}{Signal ($hh \to 2b2\gamma$)}  & $16.46$ \\\hline 
\multicolumn{2}{|c|}{Significance ($S/\sqrt{B}$)} & $1.76$ \\ \hline \hline  
 
\end{tabular}}
\caption{Signal and background yields after the BDT analysis along with the significance.}
\label{tab:bbgamgam_tmva}
\end{table}
\end{center}

\subsection{The $b\bar{b}\tau\tau$ channel}
\label{sec2.2:2b2tau}

Having studied the cleanest di-Higgs channel, we now turn our focus towards the channel which at present imposes one of the stronger 
limits on the di-Higgs cross-section. The $b \bar{b} \tau^+ \tau^-$ channel has a considerably larger rate compared to $b \bar{b} \gamma 
\gamma$ and has the advantage of three different final states as we shall discuss in details below. The $\tau$-lepton can decay either
leptonically with a $\sim 34\%$ branching ratio or hadronically. This yields us rich final states, \textit{viz.}, $b b \ell \ell$, 
$b b \ell j$ and $b b j j$, all accompanied with $\met$. The jets are formed from the hadronic $\tau$-decays and we will tag them in 
order to discriminate more from the backgrounds.

The major backgrounds for these channels stem from the fully hadronic, semi-leptonic and fully leptonic decays of pair produced 
$t \bar{t}$. The QCD-QED background, $g g \to b \bar{b} Z^{(*)}/\gamma^* \to b \bar{b} \tau^+ \tau^-$ is also substantial. As we will 
see, demanding a large invariant mass in the $\tau^+ \tau^-$ system, eradicates the $\gamma^*$ contribution almost completely. Other 
backgrounds include $b\bar{b} h$, $Zh$, $t\bar{t}W$, $t\bar{t}Z$ and $t\bar{t}h$. Besides, we also have the $b\bar{b} j j$ background, 
with jets faking hadronic $\tau$s. In context of the $Zh$ channel, we once decay the $Z$-boson to a pair of bottom quarks while forcing 
the Higgs to decay to a pair of $\tau$-leptons and then interchange these decay modes in order to have all possible $b\bar{b} \tau^+ 
\tau^-$ final states. The cross-sections of the backgrounds are large and hence in order to improve statistics in our final analyses, 
we generate the samples with hard generation level cuts (see Appendix~\ref{sec:appendixA}). We neglect $W (\to \tau \nu) + \; 
\textrm{jets}$, $Wh$, $WZ$, $h \to ZZ^*$ and single top production owing to their very small production rate. 
 
On the one hand the $t\bar{t}$ backgrounds are significantly large when compared to the small signal rate. However, 
boosted techniques and several kinematic variables do provide us some handle over the situation~\cite{Dolan:2012rv}. On the other hand, reconstruction of 
invariant mass of the $\tau$-pair is a delicate issue at the LHC since it is always accompanied by missing transverse energy. Several 
$m_{\tau \tau}$ reconstruction techniques have been discussed in the literature~\cite{Ellis:1987xu,Elagin:2010aw,Hagiwara:2016zqz} and 
extensively used in various previous analyses. In this work, we will considerably focus on the collinear mass approximation 
technique~\cite{Elagin:2010aw}. This approximation is based on two important assumptions, \textit{viz.}, the visible decay products of a $\tau$
lepton along with the neutrinos coming from it are all nearly collinear (\textit{i.e.}, $\theta_{vis}=\theta_{\nu}$ and 
$\phi_{vis}=\phi_{\nu}$) and the total missing energy in the event is solely due to these neutrinos. Upon utilising these two 
assumptions, the $x$- and the $y$-components of $\met$ can be easily expressed in terms of the momenta of the neutrinos. Solving
this, one obtains the individual momentum of each neutrino. The above method has a drawback because only in the cases where 
the $\tau\tau$ system is boosted against a hard object (examples being energetic jet, boosted objects), do we recover a 
reasonable mass. In our present scenario, the $\tau\tau$ system ($h\to\tau\tau$) is boosted against the other Higgs which decays
to a pair of $b$-quarks. The reason for this drawback is that this technique is extremely sensitive to the $\met$ resolution 
and may overestimate the reconstructed mass, $M_{\tau\tau}$. Another drawback of this assumption is that, the solutions of 
the $\met$ equation diverge when the visible $\tau$ decay products are produced back to back in the transverse plane. 
We discuss another $\tau\tau$ reconstruction technique, \textit{viz.}, the Higgs-bound technique~\cite{Lester:1999tx,Barr:2003rg},
in Appendix~\ref{sec:appendixB}. We are aware of the fact that the ATLAS~\cite{Elagin:2010aw}~\footnote{This code is however neither available publicly nor 
upon request.} and CMS~\cite{Bianchini:2014vza} collaborations use different algorithms to reconstruct a resonance decaying 
into a pair of $\tau$-leptons.

In the following sub-subsections, we present the analyses with sets of optimised cuts aimed for the HL-LHC. For the major part, we closely 
follow the predicted performance of an upgraded ATLAS detector~\cite{ATL-PHYS-PUB-2013-004} to model the detector effects and tagging 
efficiencies. For this part of the study, we use a different isolation criteria for the leptons ($e, \mu$) upon following this ATLAS 
reference~\cite{ATL-PHYS-PUB-2015-046}. We demand the total energy activity around the lepton and within a cone of radius $\Delta R=0.2$ 
must be less than 10 GeV. Following Ref.~\cite{ATL-PHYS-PUB-2013-004}, we fix the medium-level $\tau$ selection efficiencies for 
candidates with $p_T > 20$ GeV and $|\eta| < 2.3$ at 55\% and 50\% respectively for the  one-pronged and three-pronged $\tau$ 
candidates. We also allow for QCD-jets faking $\tau$-jets with mistag rates of 5\% and 2\% respectively for one and three tracks passing
the medium level $\tau$ identification.

We dissect the analysis into three independent parts corresponding to the decay mode of the $\tau$-lepton, \textit{viz.}, the 
$b b \tau_h \tau_h$, $b b \tau_h \tau_{\ell}$ and $b b \tau_{\ell} \tau_{\ell}$ final states, where the subscript $h (\ell)$ denotes 
the hadronic (leptonic) decay mode of the $\tau$. For the following three sub-analyses, we demand some common sets of cuts. We select events 
with exactly two reconstructed $b$-tagged jets with a minimum $p_T$ requirement of 40 (30) GeV for the leading (subleading) jet. We also 
require these $b$-tagged jets to be within a pseudorapidity range of $|\eta| < 2.5$. We require ${m}_{bb}>50$ GeV in order to bring the 
signal and backgrounds on the same footing because the backgrounds have been generated with this cut at the generation level. In case of the Higgs decaying to $\tau$ pair, we take $\Delta R_{b\tau} > 0.4$, $\Delta R_{\tau\tau} > 0.4$ and $m_{\tau \tau}^{\textrm{vis}} > 30$ GeV, which signifies the minimum invariant mass on the visible products from the $\tau$-pair. We also 
apply a common set of selection cuts as follows:

\begin{itemize}
\item $0.4~ < \Delta R_{bb} < ~2.0$
\item $100~\text{GeV} < {m}_{bb} < 150~\text{GeV}$
\end{itemize} 

\subsubsection{The $b\bar{b}\tau_h\tau_h$ channel}
\label{sec2.2.2:2b2tauhh}

In addition to the aforementioned common cuts, we require exactly two $\tau$-tagged jets having a minimum $p_T$ of 30 GeV and a 
maximal pseudorapidity range of $|\eta| < 2.5$. In each of these sub-analyses, we first consider the variable $m_{\tau \tau}^{\textrm{vis}}$, 
constructed out of the visible $\tau$ objects and afterwards we consider the collinear mass variable, $M_{\tau\tau}$. For the
first case, we further optimise $p_{T,bb}, m_{T2}$ and $m_{\tau \tau}^{\textrm{vis}}$ in order to have 
the best possible signal over background ratio.

\begin{itemize}
\item $p_{T,bb} > 110$ GeV
\item $m_{T2} > 105$ GeV
\item $55 \; \textrm{GeV} < m_{\tau \tau}^{\textrm{vis}} < \; 140$ GeV 
\end{itemize}

Upon performing the optimised cut-based analysis, we obtain a final significance of 0.44 for the HL-LHC. The cut-flow
and the final significance are tabulated in Table~\ref{tab1:2b2tauhh}. In contrast to the $b\bar{b} \gamma \gamma$ channel, the $S/B$ 
ratio here is $\sim 0.67\%$ and hence one needs data-driven background techniques and a drastic reduction in systematic uncertainties 
in order for this channel to be relevant in the future.

\begin{table}[htb!]
\begin{bigcenter}\
\scalebox{0.7}{%
\begin{tabular}{||c||c|c|c|c|c|c|c|c|c||c||}
\hline
 & \multicolumn{9}{c||}{Number of events at $3000$ fb$^{-1}$}   & \\ \cline{2-10}
 Cut flow  & Signal & \multicolumn{8}{c||}{Backgrounds} & $\frac{S}{\sqrt{B}}$\\
\cline{2-10}
                   & $hh\to 2b2\tau$     & $t\bar{t}$ had         & $t\bar{t}$ semi-lep    & $t\bar{t}$ lep         &  $\ell\ell b\bar{b}$                   & $hb\bar{b}$                   & $Zh$                                                   & Others~\footnote{``Others'' include $t\bar{t}h$, $t\bar{t}W$ and $t\bar{t}Z$.} & $b\bar{b}jj$ & \\
\cline{1-10}  
Order              & NNLO~\cite{hhtwiki} & NNLO +                 & NNLO +                 & NNLO +                 &  LO                            & NNLO (5FS) +                  & NNLO (QCD) +                                           & NLO                                                                            & LO & \\ 
                   &                     & NNLL~\cite{ttbarNNLO}  & NNLL~\cite{ttbarNNLO}  & NNLL~\cite{ttbarNNLO}  &                                & NLO (4FS)~\cite{bkg_twiki_cs} & NLO (EW)~\cite{bkg_twiki_cs}                           & ~\cite{bkg_twiki_cs,Campbell:2012dh,Lazopoulos:2008de}                         &  &\\
\hline
event selection          & 75.67 & 3405.26   & 37092.00 & 103073.95 & 16561.12 & 13.72  & 273.92 & 5278.22 & 52377.27 & 0.16 \\
\hline
${\Delta R}_{b\bar{b}}$  & 62.00 & 1196.24   & 11288.87 & 25190.00  & 3857.81  & 2.41  & 184.72 & 1837.20 & 23106.23 & 0.24\\
\hline
$m_{b\bar{b}}$           & 40.90 & 433.00    & 4188.53  & 7672.70   & 973.82   & 0.64  & 97.12  & 678.52  & 4586.82  & 0.30\\
\hline
$p_{T,h}^{b\bar{b}}$     & 37.42 & 330.25    & 2934.21  & 4485.89   & 742.85   & 0.44  & 82.43  & 549.84  & 3290.74  & 0.33\\
\hline
$m_{T2}$                 & 33.32 & 124.76    & 1791.88   & 2598.16  & 611.76   & 0.33  & 74.23  & 309.74  & 2418.24  & 0.37\\
\hline
$m_{\tau\tau}^{vis}$     & 30.09  & 80.72    & 1254.32   & 1928.32  & 474.42   & 0.31  & 56.24  & 189.80  & 688.80   & 0.44\\ 
\hline  
\end{tabular}}
\end{bigcenter}
\caption{The cut-flow and significance table for the $b\bar{b} \tau_h \tau_h$ mode.}

\label{tab1:2b2tauhh}
\end{table}

Next, we use the collinear approximation technique, discussed above, to reconstruct the invariant mass of the Higgs decaying 
to a pair of $\tau$ leptons. To overcome the limitations as discussed above, we select events by putting an additional cut, 
$\Delta \phi_{\tau\tau}<3.0~\text{radian}$. For the BDT analysis, we impose an upper cut on the collinear mass, 
$M_{\tau\tau} < 200$ GeV. The cut-flow and the statistical significance are tabulated in Table~\ref{tab1:2b2tauhhc} with the 
following optimised cuts on top of the other variables. We obtain a significance of $0.65$, which shows a small improvement 
over the previous analysis with the $m_{\tau\tau}^{vis}$ variable. 
\begin{itemize}
\item $p_{T,bb} > 125$ GeV
\item $m_{T2} > 110$ GeV
\item $80 \; \textrm{GeV} < M_{\tau\tau} < \; 170$ GeV 
\end{itemize}
%

\begin{table}[htb!]
\begin{bigcenter}\
\scalebox{0.7}{%
\begin{tabular}{||c||c|c|c|c|c|c|c|c|c||c||}
\hline
 & \multicolumn{9}{c||}{Number of events at $3000$ fb$^{-1}$}   & \\ \cline{2-10}
 Cut flow  & Signal & \multicolumn{8}{c||}{Backgrounds} & $\frac{S}{\sqrt{B}}$\\
\cline{2-10}
                   & $hh\to 2b2\tau$     & $t\bar{t}$ had         & $t\bar{t}$ semi-lep    & $t\bar{t}$ lep         &  $\ell\ell b\bar{b}$                   & $hb\bar{b}$                   & $Zh$                                                   & Others~\footnote{``Others'' include $t\bar{t}h$, $t\bar{t}W$ and $t\bar{t}Z$.} & $b\bar{b}jj$ & \\
\cline{1-10}  
Order              & NNLO~\cite{hhtwiki} & NNLO +                 & NNLO +                 & NNLO +                 &  LO                            & NNLO (5FS) +                  & NNLO (QCD) +                                           & NLO                                                                            & LO & \\ 
                   &                     & NNLL~\cite{ttbarNNLO}  & NNLL~\cite{ttbarNNLO}  & NNLL~\cite{ttbarNNLO}  &                                & NLO (4FS)~\cite{bkg_twiki_cs} & NLO (EW)~\cite{bkg_twiki_cs}                           & ~\cite{bkg_twiki_cs,Campbell:2012dh,Lazopoulos:2008de}                         &  &\\
\hline
event selection          & 75.00 & 3061.53   & 34670.67 & 93679.19  & 15968.09 & 12.93 & 270.97 & 1832.58 & 51997.54 & 0.17 \\
\hline
${\Delta R}_{b\bar{b}}$  & 61.87 & 1133.90   & 11556.89 & 23462.09  & 4288.54  & 2.32  & 183.41 & 620.46  & 23509.51 & 0.24\\
\hline
$m_{b\bar{b}}$           & 41.10 & 340.17    & 4430.58  & 7392.16   & 1154.85  & 0.63  & 97.71  & 230.21  & 4585.01  & 0.30\\
\hline
$p_{T,h}^{b\bar{b}}$     & 34.21 & 166.30    & 2455.50  & 2588.10   & 580.54   & 0.35  & 70.71  & 146.27  & 2550.85  & 0.37\\
\hline
$m_{T2}$                 & 30.65 & 120.95    & 1467.96  & 1184.10   & 518.12   & 0.27  & 65.25  & 98.21   & 2005.53  & 0.41\\
\hline
$M_{\tau\tau}$           & 26.19 & 83.15     & 400.35   & 186.07    & 355.82   & 0.23  & 48.37  & 42.60   & 480.10   & 0.65\\ 
\hline  
\end{tabular}}
\end{bigcenter}
\caption{The cut-flow and significance table for the $b\bar{b} \tau_h \tau_h$ mode with collinear mass variable.}

\label{tab1:2b2tauhhc}
\end{table}

\begin{figure}
\includegraphics[scale=0.2]{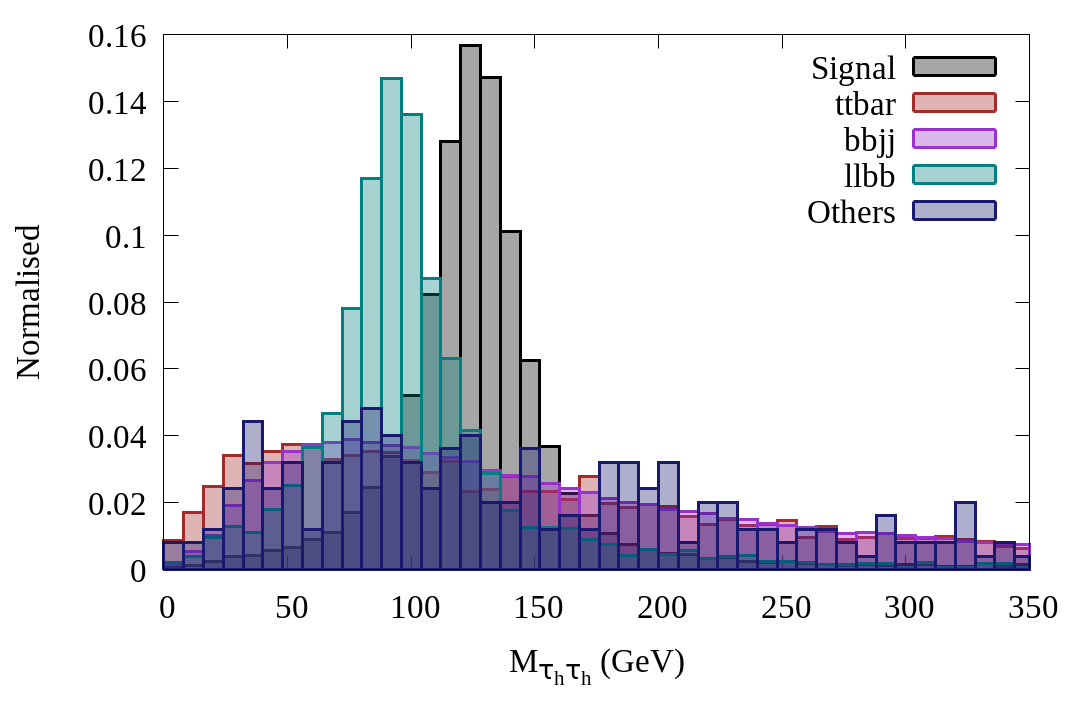}\includegraphics[scale=0.2]{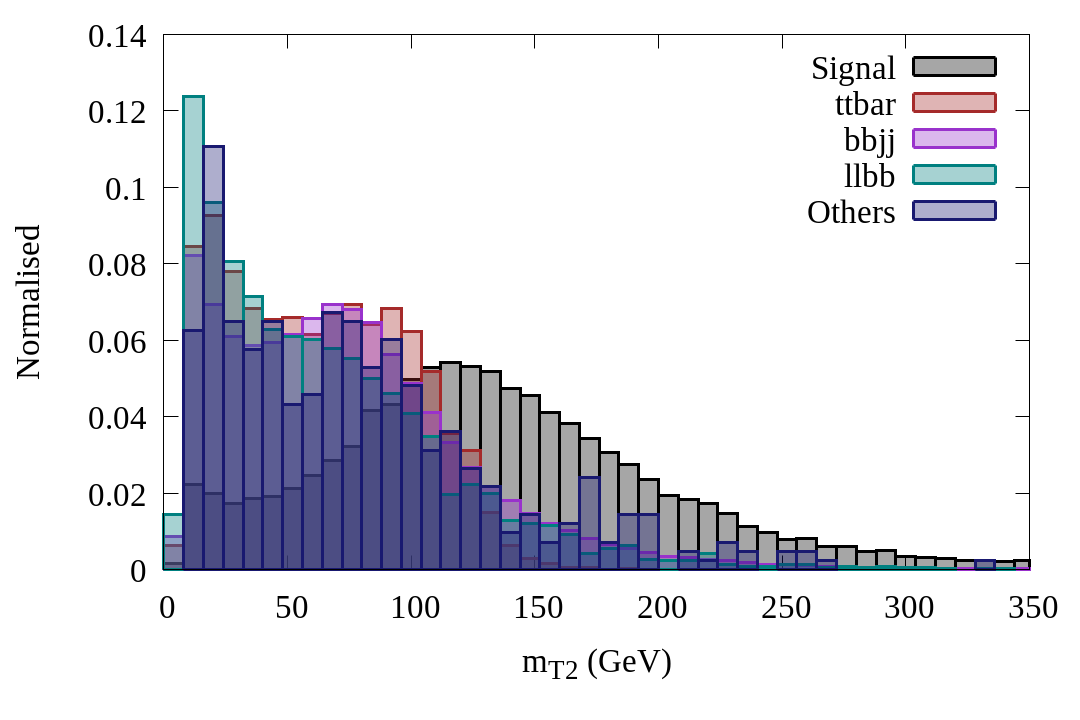}\\
\includegraphics[scale=0.2]{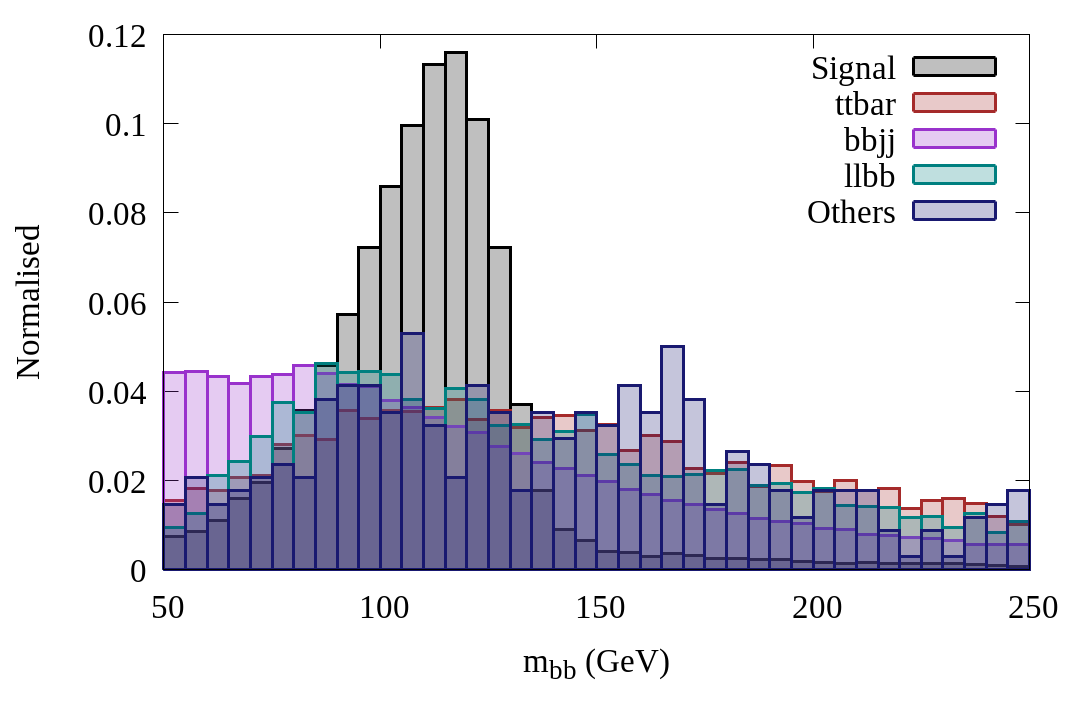}\includegraphics[scale=0.2]{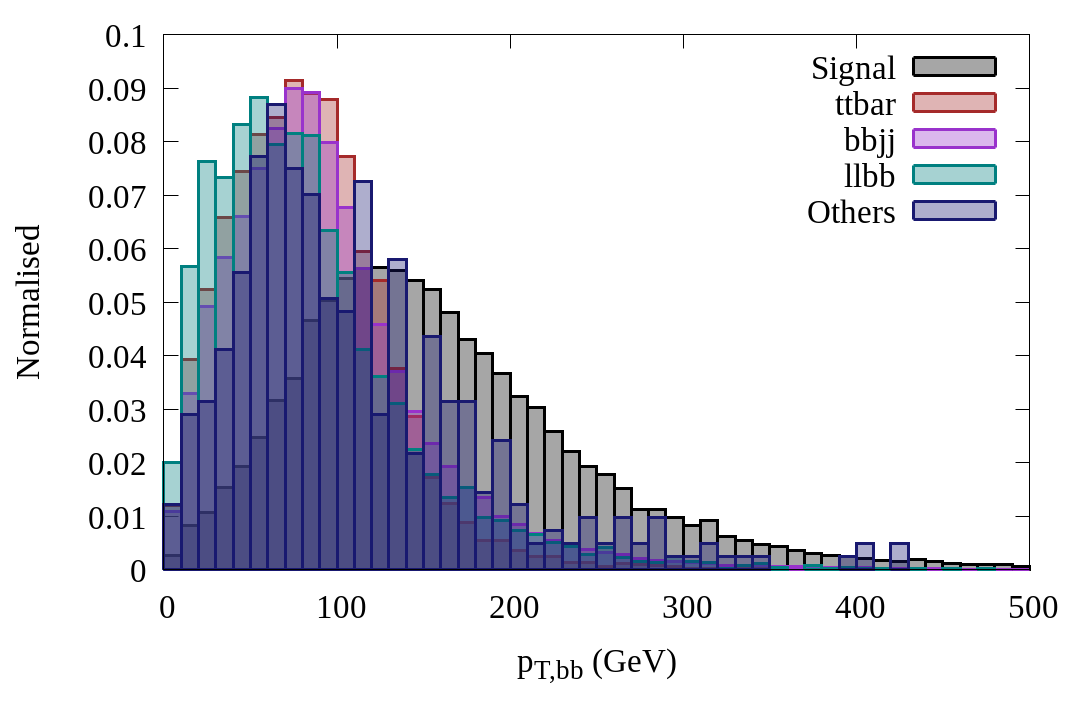}
\caption{Normalised distributions of $M_{\tau_h\tau_h}$, $m_{T2}$, $m_{bb}$ and $p_{T,bb}$ for the signal 
and dominant backgrounds in $b\bar{b}\tau_h\tau_h$ channel after the basic selection cuts.}
\label{fig1:2b2tauhh}
\end{figure}

In order to be certain if our optimised cuts can be improved further, we employ a multivariate analysis using the BDT algorithm after 
the basic selection cuts. We train our signal and background samples with the following 12 kinematic variables for the case
with the $m_{\tau_h\tau_h}^{\textrm{vis}}$ variable:

\begin{equation}
\begin{split}
p_{T,bb},~m_{bb},~\Delta R_{bb},~m_{\tau_h\tau_h}^{\textrm{vis}},~\Delta R_{\tau_h\tau_h},~\Delta\phi_{\tau_{h_1}\slashed{E}_T},
\nonumber \\~\Delta\phi_{\tau_{h_2}\slashed{E}_T},~m_{hh}^{\textrm{vis}},~p_{T,hh}^{\textrm{vis}},~\Delta R_{b_1\tau_{h_1}},
~\Delta R_{hh}^{\textrm{vis}},~m_{T2} 
\end{split}
\end{equation}

For the other case, with the $M_{\tau\tau}$ variable, we train our signal and background samples with the following 9 kinematic variables:

\begin{equation}
\begin{split}
p_{T,bb},~m_{bb},~\Delta R_{bb},~M_{\tau_h\tau_h},~m_{T2},~\Delta\phi_{\tau_{h_1}\slashed{E}_T}, ~m_{hh}^{\textrm{vis}},~p_{T,hh}^{\textrm{vis}},~\Delta R_{hh}^{\textrm{vis}} \nonumber
\end{split}
\end{equation}
where the symbols have their usual meaning. $\Delta\phi_{ab}$ is the azimuthal angle separation for the $ab$ system. $M_{\tau_h\tau_h}$ is the collinear mass of Higgs from hadronic $\tau$ decays. The signal and 
background yields after this multivariate analysis are shown in Table~\ref{tab2:2b2tauhh}. The normalised distributions of the four 
best discriminating kinematic variables, \textit{viz.}, $M_{\tau_h\tau_h}$, $m_{T2}$, $m_{bb}$ and $p_{T,bb}$ are shown in Fig.~\ref{fig1:2b2tauhh}. We find that the $S/B$ ratio increases slightly and we also have a non-negligible 
increase in the significance at 0.74, assuming zero systematic uncertainty.

\begin{center}
\begin{table}[htb!]
\centering
\scalebox{0.8}{%
\begin{tabular}{|c|c|c|}\hline
(a) & Process & Events \\ \hline \hline

\multirow{4}{*}{Background}  
 & $t\bar{t}$ had      & $315.57$ \\ 
 & $t\bar{t}$ semi-lep & $3673.36$ \\  
 & $t\bar{t}$ lep      & $2456.07$\\  
 & $b\bar{b}jj$        & $2906.30$ \\ 
 & $\ell\ell b\bar{b}$ & $2078.72$ \\  
 & $b\bar{b}h$         & $1.00$ \\
 & $Zh$                & $139.14$ \\ 
 & $t\bar{t}h$         & $300.58$ \\ 
 & $t\bar{t}Z$         & $270.26$ \\ 
 & $t\bar{t}W$         & $110.26$ \\  \cline{2-3}
 & \multicolumn{1}{c|}{Total}                    & $12251.26$ \\ \hline
\multicolumn{2}{|c|}{Signal ($hh\to 2b2\tau$)}    & $51.85$ \\\hline 
\multicolumn{2}{|c|}{Significance ($S/\sqrt{B}$)} & $0.47$ \\ \hline \hline  
\end{tabular}}
\quad
\scalebox{0.8}{%
\begin{tabular}{|c|c|c|}\hline
(b) & Process & Events \\ \hline \hline

\multirow{4}{*}{Background}  
 & $t\bar{t}$ had      & $109.09$ \\ 
 & $t\bar{t}$ semi-lep & $800.71$ \\  
 & $t\bar{t}$ lep      & $642.80$\\  
 & $b\bar{b}jj$        & $879.39$ \\ 
 & $\ell\ell b\bar{b}$ & $605.51$ \\  
 & $b\bar{b}h$         & $0.70$ \\
 & $Zh$                & $69.61$ \\ 
 & $t\bar{t}h$         & $96.05$ \\ 
 & $t\bar{t}Z$         & $42.27$ \\ 
 & $t\bar{t}W$         & $9.38$ \\  \cline{2-3}
 & \multicolumn{1}{c|}{Total}                    & $3255.51$ \\ \hline
\multicolumn{2}{|c|}{Signal ($hh\to 2b2\tau$)}    & $42.09$ \\\hline 
\multicolumn{2}{|c|}{Significance ($S/\sqrt{B}$)} & $0.74$ \\ \hline \hline  
 
\end{tabular}}
\caption{Signal, background yields and final significance for the $b\bar{b} \tau_h \tau_h$ channel after the BDT analysis with (a) $m_{\tau \tau}^{\textrm{vis}}$ (b) $M_{\tau\tau}$ variable.}
\label{tab2:2b2tauhh}
\end{table}
\end{center}

\subsubsection{The $b\bar{b}\tau_h\tau_{\ell}$ channel}
\label{sec2.2:2b2tauhl}

In the present instalment, we choose events containing exactly one isolated lepton and one reconstructed $\tau$-tagged jet 
over and above the common requirements. We also require the isolated lepton to have a $p_T > 20~\text{GeV}$ and $|\eta| < 2.5$. The 
additional optimised selection cuts for this present mode, involving the $m_{\tau \tau}^{\textrm{vis}}$, are:

\begin{itemize}
\item $p_{T,bb} > 150$ GeV
\item $m_{T2} > 145$ GeV
\item $50 \; \textrm{GeV} < m_{\tau \tau}^{\textrm{vis}} < 105$ GeV
\end{itemize}

After imposing the various cuts, we obtain a signal significance of 0.26 for the HL-LHC. The event yields along with the significance 
are shown in Table~\ref{tab1:2b2tauhl}.

\begin{table}[htb!]
\begin{bigcenter}
\scalebox{0.7}{%
\begin{tabular}{||c||c|c|c|c|c|c|c|c||c||}
\hline
 & \multicolumn{8}{c||}{Event yield at $3000 fb^{-1}$}   & \\ \cline{2-9}
 Cut flow  & Signal & \multicolumn{7}{c||}{Backgrounds} & $\frac{S}{\sqrt{B}}$\\
\cline{2-9}
 & $hh\to 2b2\tau$ & $t\bar{t}$ had & $t\bar{t}$ semi-lep  & $t\bar{t}$ lep &  $\ell\ell b\bar{b}$ & $b\bar{b}h$  & $Zh$ & Others &  \\
\hline
event selection          & 114.47 & 52032.93 & 746566.27 & 2056850.98 & 28983.51 & 19.47 & 387.26 & 34951.83 & 0.07\\
\hline
${\Delta R}_{b\bar{b}}$  & 94.30  & 16042.87 & 212114.33 & 520586.36  & 6854.17  & 3.61  & 271.52 & 12000.51 & 0.11\\
\hline
$m_{b\bar{b}}$           & 56.00  & 5467.49  & 74094.42  & 168799.32  & 1910.18  & 1.01  & 131.36 & 4282.50  & 0.11\\
\hline
$p_{T,h}^{b\bar{b}}$     & 38.73  & 2164.98  & 24683.21  & 24621.64   & 618.00   & 0.33  & 76.34  & 1989.73  & 0.17\\
\hline
$m_{T2}$                 & 30.11  & 447.67   & 12587.99  & 1847.13    & 412.00   & 0.18  & 61.78  & 840.09   & 0.24\\ 
\hline 
$m_{\tau\tau}^{vis}$     & 22.34  & 205.49   & 5980.41  & 629.24      & 218.48   & 0.14  & 32.24  & 320.23   & 0.26\\ 
\hline
\end{tabular}}
\end{bigcenter}
\caption{Same as in Table~\ref{tab1:2b2tauhh} for the $b\bar{b} \tau_h \tau_{\ell}$ mode. The various orders of the signal and
backgrounds are same as in Table~\ref{tab1:2b2tauhh}.}
\label{tab1:2b2tauhl}
\end{table}

 We get the following optimised cuts upon the other variables with $M_{\tau\tau}$ variable. The event yields at HL-LHC are shown in Table~\ref{tab1:2b2tauhlc} with a significance of $0.44$
\begin{itemize}
\item $p_{T,bb} > 60$ GeV
\item $m_{T2} > 135$ GeV
\item $105 \; \textrm{GeV} < M_{\tau\tau} < 150$ GeV
\end{itemize}
%

\begin{table}[htb!]
\begin{bigcenter}
\scalebox{0.7}{%
\begin{tabular}{||c||c|c|c|c|c|c|c|c||c||}
\hline
 & \multicolumn{8}{c||}{Event yield at $3000 fb^{-1}$}   & \\ \cline{2-9}
 Cut flow  & Signal & \multicolumn{7}{c||}{Backgrounds} & $\frac{S}{\sqrt{B}}$\\
\cline{2-9}
 & $hh\to 2b2\tau$ & $t\bar{t}$ had & $t\bar{t}$ semi-lep  & $t\bar{t}$ lep &  $\ell\ell b\bar{b}$ & $b\bar{b}h$  & $Zh$ & Others &  \\
\hline
event selection          & 111.53 & 49309.49 & 690610.94 & 1916786.62 & 27922.31 & 18.11 & 372.74 & 16902.24 & 0.07\\
\hline
${\Delta R}_{b\bar{b}}$  & 92.38  & 15065.74 & 196814.09 & 486072.84  & 6766.77  & 3.37  & 262.37 & 5798.86  & 0.11\\
\hline
$m_{b\bar{b}}$           & 55.66  & 4739.70  & 67873.37  & 151446.33  & 1897.69  & 0.89  & 126.70 & 1997.44  & 0.12\\
\hline
$p_{T,h}^{b\bar{b}}$     & 55.66  & 4739.70  & 67873.37  & 151446.33  & 1897.69  & 0.89  & 126.70 & 1997.44  & 0.12\\
\hline
$m_{T2}$                 & 38.54  & 846.64   & 18656.50  & 10758.39   & 692.91   & 0.27  & 79.94  & 674.45   & 0.22\\ 
\hline 
$M_{\tau\tau}$           & 29.98  & 136.07   & 3096.07   & 1031.86    & 255.94   & 0.20  & 23.61  & 103.94   & 0.44\\ 
\hline
\end{tabular}}
\end{bigcenter}
\caption{Same as in Table~\ref{tab1:2b2tauhh} for the $b\bar{b} \tau_h \tau_{\ell}$ mode with collinear mass variable. The various orders of the signal and
backgrounds are same as in Table~\ref{tab1:2b2tauhh}.}
\label{tab1:2b2tauhlc}
\end{table}

\begin{figure}
\includegraphics[scale=0.2]{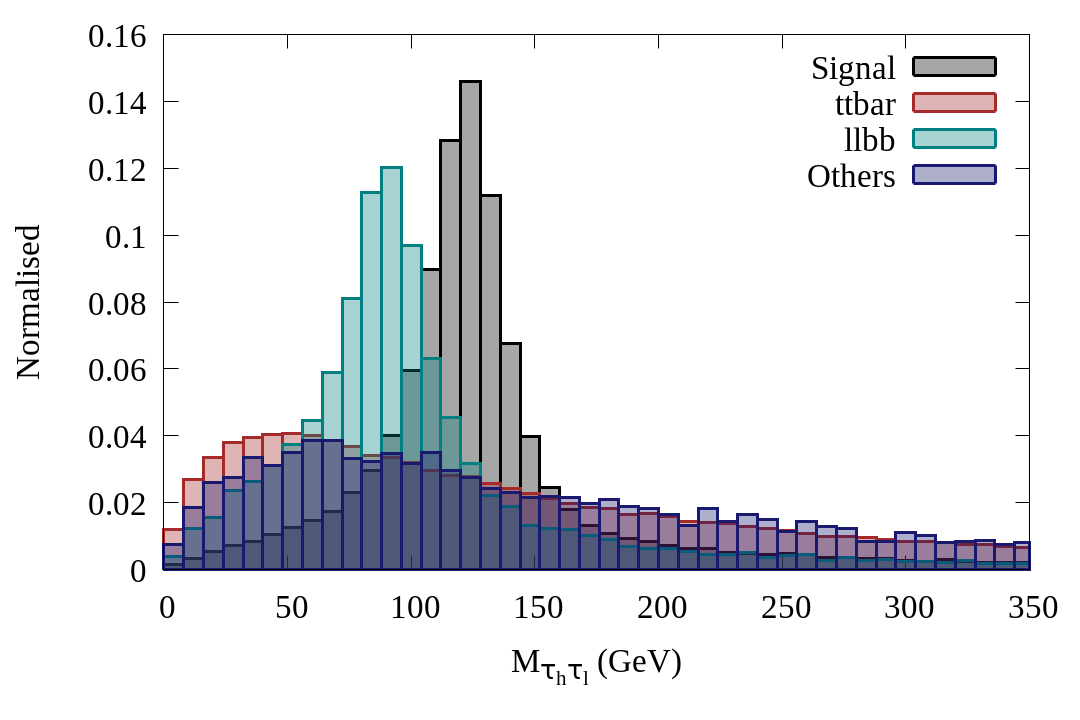}\includegraphics[scale=0.2]{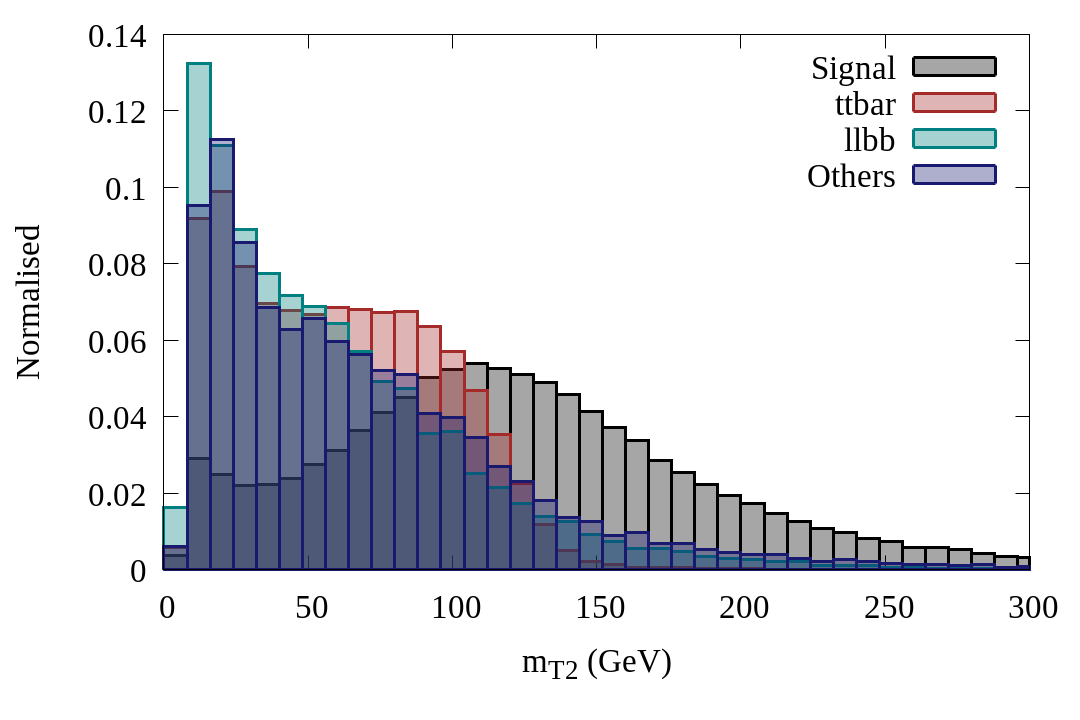}\\
\includegraphics[scale=0.2]{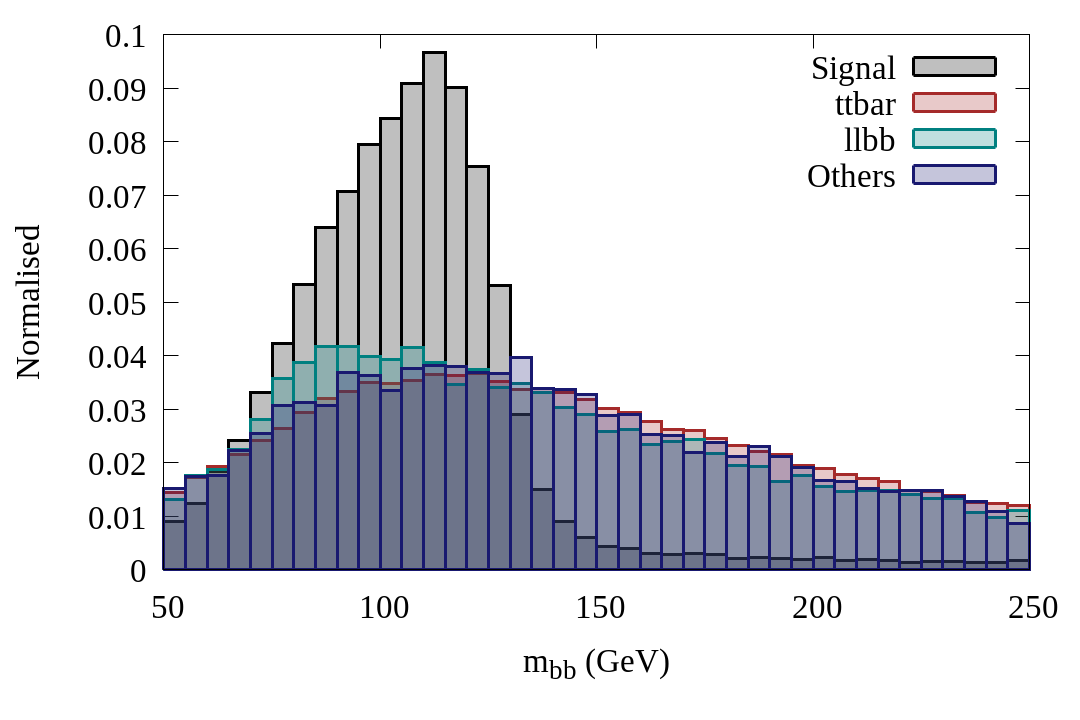}\includegraphics[scale=0.2]{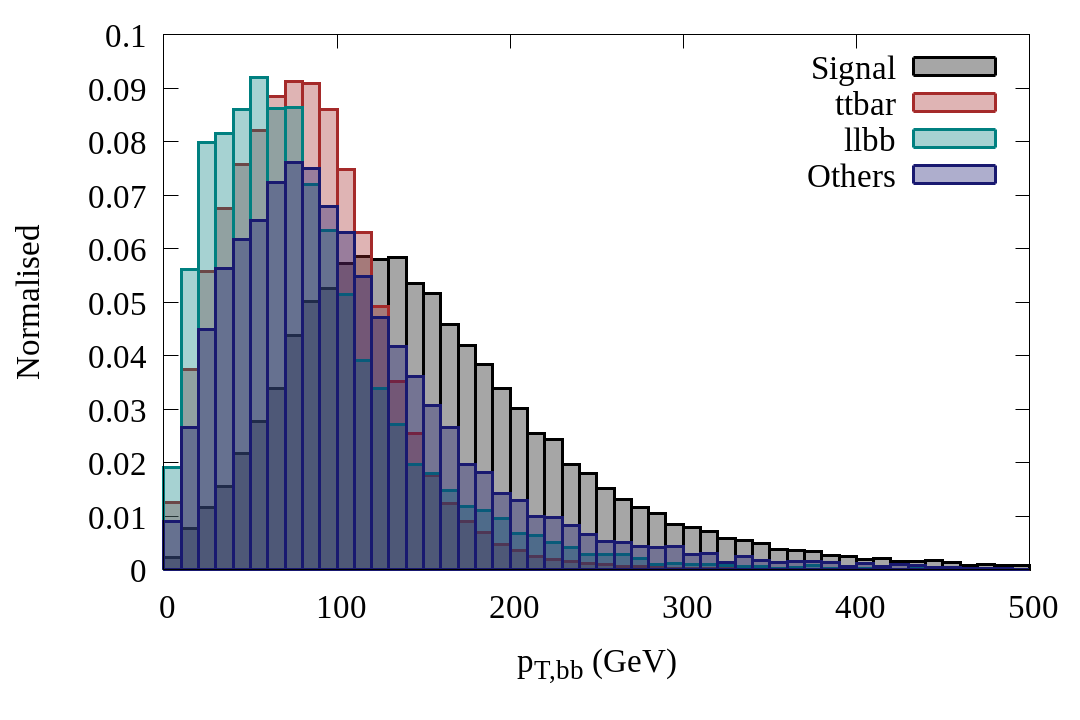}
\caption{Normalised distributions of $M_{\tau_h\tau_l}$, $m_{T2}$, $m_{bb}$ and $p_{T,bb}$ for the signal and dominant backgrounds in $b\bar{b}\tau_h\tau_{\ell}$ channel before the basic selection 
cuts.}
\label{fig1:2b2tauhl}
\end{figure}

Here also we perform a BDT analysis to see its potential. We choose the following 13 kinematic variables to train our signal 
and background event samples with the $m_{\tau_h\tau_l}^{\textrm{vis}}$ variable: 

\begin{equation}
\begin{split}
p_{T,bb},~m_{bb},~\Delta R_{bb},~m_{\tau_h\tau_{\ell}}^{\textrm{vis}},~\Delta\phi_{\tau_h \met},~\Delta\phi_{\tau_\ell \met},
~p_{T,hh}^{\textrm{vis}}, \nonumber \\~\Delta R_{b_1\tau_h},~\Delta R_{b_1\tau_{\ell}},~\Delta R_{hh}^{\textrm{vis}},~m_{T2},p_{T,\ell} 
\end{split}
\end{equation}

Furthermore, we consider the following 9 kinematic variables to train our signal and 
background event samples while having the $M_{\tau_h\tau_l}$ variable: 

\begin{equation}
\begin{split}
p_{T,bb},~m_{bb},~\Delta R_{bb},~M_{\tau_h\tau_l},~m_{T2},~\Delta\phi_{\tau_h \met},~\Delta\phi_{\tau_\ell \met}, ~m_{hh}^{\textrm{vis}},~\Delta R_{hh}^{\textrm{vis}} \nonumber
\end{split}
\end{equation}

We ensure a proper training of the event samples. In Table~\ref{tab2:2b2tauhl}, the signal, background yields and the significance 
after the multivariate analysis, are presented. The normalised distribution of the four maximal discriminating kinematic variables, 
\textit{viz.}, $M_{\tau_h\tau_l}$, $m_{T2}$, $m_{bb}$ and $p_{T,bb}$ are shown 
in Fig.~\ref{fig1:2b2tauhl}. Upon imposing a suitable cut on the BDT variable, we find that the zero-systematics significance 
is 0.49 for the case involving the collinear mass variable.

\begin{center}
\begin{table}[htb!]
\centering
\scalebox{0.8}{%
\begin{tabular}{|c|c|c|}\hline
(a) & Process & Events \\ \hline \hline

\multirow{4}{*}{Background}  
 & $t\bar{t}$ had      & $2267.72$ \\ 
 & $t\bar{t}$ semi-lep & $24078.45$ \\  
 & $t\bar{t}$ lep      & $11752.62$\\  
 & $\ell\ell b\bar{b}$ & $1566.84$ \\  
 & $b\bar{b}h$         & $0.92$ \\
 & $Zh$                & $142.57$ \\ 
 & $t\bar{t}h$         & $558.51$ \\ 
 & $t\bar{t}Z$         & $516.18$ \\ 
 & $t\bar{t}W$         & $304.19$ \\  \cline{2-3}
 & \multicolumn{1}{c|}{Total}                 & $41188.00$ \\ \hline
\multicolumn{2}{|c|}{Signal ($hh\to 2b2\tau$)} & $54.87$ \\\hline 
\multicolumn{2}{|c|}{Significance ($S/\sqrt{B}$)} & $0.27$ \\ \hline \hline  
 
\end{tabular}}
\quad
\scalebox{0.8}{%
\begin{tabular}{|c|c|c|}\hline
(b) & Process & Events \\ \hline \hline

\multirow{4}{*}{Background}  
 & $t\bar{t}$ had      & $166.30$ \\ 
 & $t\bar{t}$ semi-lep & $3816.71$ \\  
 & $t\bar{t}$ lep      & $1454.75$\\  
 & $\ell\ell b\bar{b}$ & $255.94$ \\  
 & $b\bar{b}h$         & $0.67$ \\
 & $Zh$                & $34.57$ \\ 
 & $t\bar{t}h$         & $94.40$ \\ 
 & $t\bar{t}Z$         & $35.86$ \\ 
 & $t\bar{t}W$         & $14.86$ \\  \cline{2-3}
 & \multicolumn{1}{c|}{Total}                 & $5874.06$ \\ \hline
\multicolumn{2}{|c|}{Signal ($hh\to 2b2\tau$)} & $37.33$ \\\hline 
\multicolumn{2}{|c|}{Significance ($S/\sqrt{B}$)} & $0.49$ \\ \hline \hline  
 \end{tabular}}
\caption{Same as in Table~\ref{tab2:2b2tauhh} for the $b\bar{b} \tau_h \tau_{\ell}$ mode with (a) $m_{\tau \tau}^{\textrm{vis}}$ (b) $M_{\tau\tau}$ variable.}
\label{tab2:2b2tauhl}
\end{table}
\end{center}

\subsubsection{The $b\bar{b}\tau_{\ell}\tau_{\ell}$ channel}
\label{sec2.2:2b2taull}

The last segment of the $b\bar{b}\tau^+\tau^-$ channel consists of two leptonically decaying $\tau$s. We demand events containing 
exactly two oppositely charged isolated leptons with $p_T > 20$ GeV, over and above the requirements stated above. We impose 
the following optimised cuts on top of the other variables for the scenario where we consider the invariant mass from the visible
products of the $\tau$-leptons.

\begin{itemize}
\item $p_{T,bb} > 105$ GeV
\item $m_{T2} > 140$ GeV
\item $30 \; \textrm{GeV} < m_{\tau \tau}^{\textrm{vis}} < 85$ GeV
\end{itemize}

A final signal significance, $S/\sqrt{B}$, of 0.044 is obtained, upon assuming zero systematic uncertainties. We show the 
event yields and the significance in Table~\ref{tab1:2b2taull}.

\begin{table}[htb!]
\begin{bigcenter}
\scalebox{0.7}{%
\begin{tabular}{||c||c|c|c|c|c|c|c|c||c||}
\hline
 & \multicolumn{8}{c||}{Number of events at $3000 fb^{-1}$}   & \\ \cline{2-9}
 Cut flow  & Signal & \multicolumn{7}{c||}{Backgrounds} & $\frac{S}{\sqrt{B}}$\\
\cline{2-9}
 & $hh\to 2b2\tau$ & $t\bar{t}$ had & $t\bar{t}$ semi-lep & $t\bar{t}$ lep &  $\ell\ell b\bar{b}$ & $b\bar{b}h$  & $Zh$ & Others &  \\
\hline
event selection          & 33.60 & 39197.16 & 1568324.50 & 10671096.85 & 731173.68 & 5.50  & 111.09  & 69821.95 & 0.009\\
\hline
${\Delta R}_{b\bar{b}}$  & 26.84 & 13767.81 & 592173.03  & 2665084.71  & 144168.50   & 1.11 & 77.40   & 24366.86 & 0.014\\
\hline
$m_{b\bar{b}}$           & 17.69 & 4462.06  & 223291.21  & 843895.11   & 33378.17   & 0.31  & 39.86   & 8756.58  & 0.017\\
\hline
$p_{T,h}^{b\bar{b}}$     & 16.65 & 3860.27  & 185258.46  & 587286.04   & 24776.13   & 0.26 & 34.45   & 7432.93  & 0.018\\
\hline
$m_{T2}$                 & 10.99 & 579.77   & 56489.16   & 16279.11    & 6404.71    & 0.07 & 20.02   & 2188.44  & 0.038\\
\hline 
$m_{\tau\tau}^{vis}$     & 10.30 & 499.05   & 46645.12   & 6109.74     & 1098.66    & 0.06 & 19.93   & 863.14  & 0.044\\ 
\hline
\end{tabular}}
\end{bigcenter}
\caption{Same as in Table~\ref{tab1:2b2tauhh} for the $b\bar{b} \tau_{\ell} \tau_{\ell}$ mode. The various orders of the signal 
and backgrounds are same as in Table~\ref{tab1:2b2tauhh}.}
\label{tab1:2b2taull}
\end{table}

For the second category involving the collinear mass variable, we choose the following optimised cuts on top of the other 
variables. The results are tabulated in Table~\ref{tab1:2b2taullc}. 
\begin{itemize}
\item $p_{T,bb} > 60$ GeV
\item $m_{T2} > 140$ GeV
\item $85 \; \textrm{GeV} < M_{\tau\tau} < 165$ GeV
\end{itemize}
%

\begin{table}[htb!]
\begin{bigcenter}
\scalebox{0.7}{%
\begin{tabular}{||c||c|c|c|c|c|c|c|c||c||}
\hline
 & \multicolumn{8}{c||}{Number of events at $3000 fb^{-1}$}   & \\ \cline{2-9}
 Cut flow  & Signal & \multicolumn{7}{c||}{Backgrounds} & $\frac{S}{\sqrt{B}}$\\
\cline{2-9}
 & $hh\to 2b2\tau$ & $t\bar{t}$ had & $t\bar{t}$ semi-lep & $t\bar{t}$ lep &  $\ell\ell b\bar{b}$ & $b\bar{b}h$  & $Zh$ & Others &  \\
\hline
event selection          & 32.96 & 33185.44 & 1439433.25 & 9931026.00  & 688219.62  & 5.14  & 105.50  & 69963.93 & 0.009\\
\hline
${\Delta R}_{b\bar{b}}$  & 26.78 & 11973.97 & 564045.62  & 2543006.00  & 141746.44  & 1.00  & 76.68   & 24644.38 & 0.015\\
\hline
$m_{b\bar{b}}$           & 17.64 & 4134.95  & 217579.14  & 786056.06   & 32341.93   & 0.27  & 39.99   & 8774.64  & 0.017\\
\hline
$p_{T,h}^{b\bar{b}}$     & 17.64 & 4134.95  & 217579.14  & 786056.06   & 32341.93   & 0.27  & 39.99   & 8774.64  & 0.017\\
\hline
$m_{T2}$                 & 11.01 & 521.59   & 56876.98   & 15680.86    & 6130.05    & 0.05  & 21.62   & 2232.01  & 0.038\\
\hline 
$M_{\tau\tau}$           & 9.95  & 83.15    & 14012.39   & 2368.20     & 3033.81    & 0.05  & 12.02   & 528.57   & 0.070\\ 
\hline
\end{tabular}}
\end{bigcenter}
\caption{Same as in Table~\ref{tab1:2b2tauhh} for the $b\bar{b} \tau_{\ell} \tau_{\ell}$ mode with collinear mass variable. The various orders of the signal 
and backgrounds are same as in Table~\ref{tab1:2b2tauhh}.}
\label{tab1:2b2taullc}
\end{table}

\begin{figure}
\includegraphics[scale=0.2]{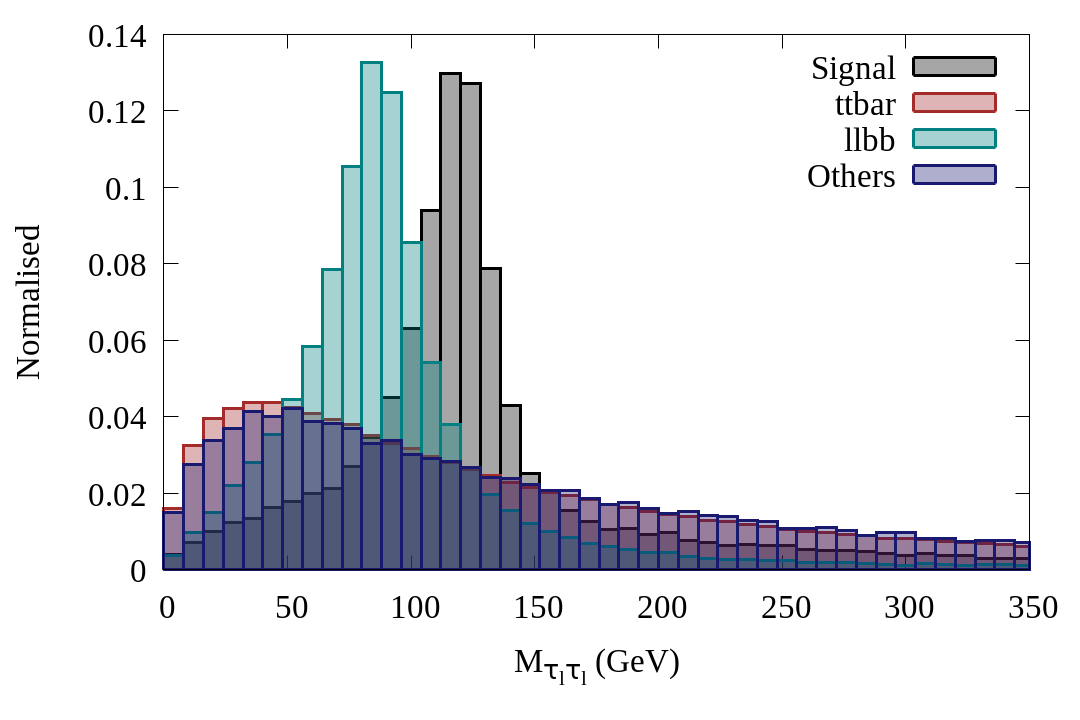}\includegraphics[scale=0.2]{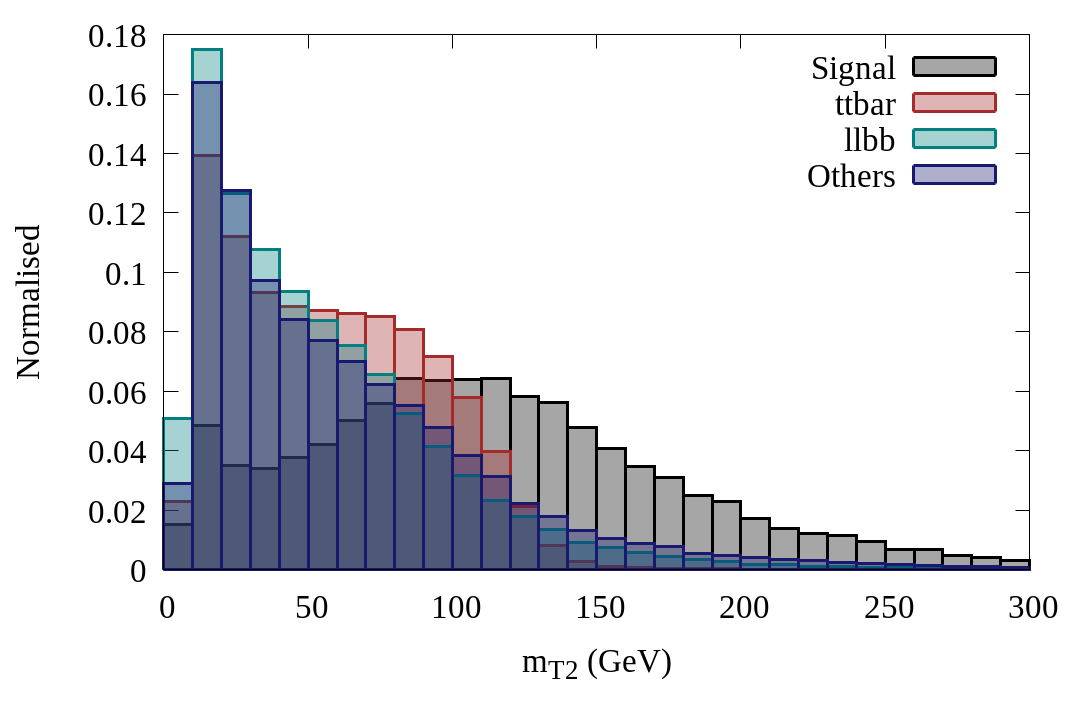}\\
\includegraphics[scale=0.2]{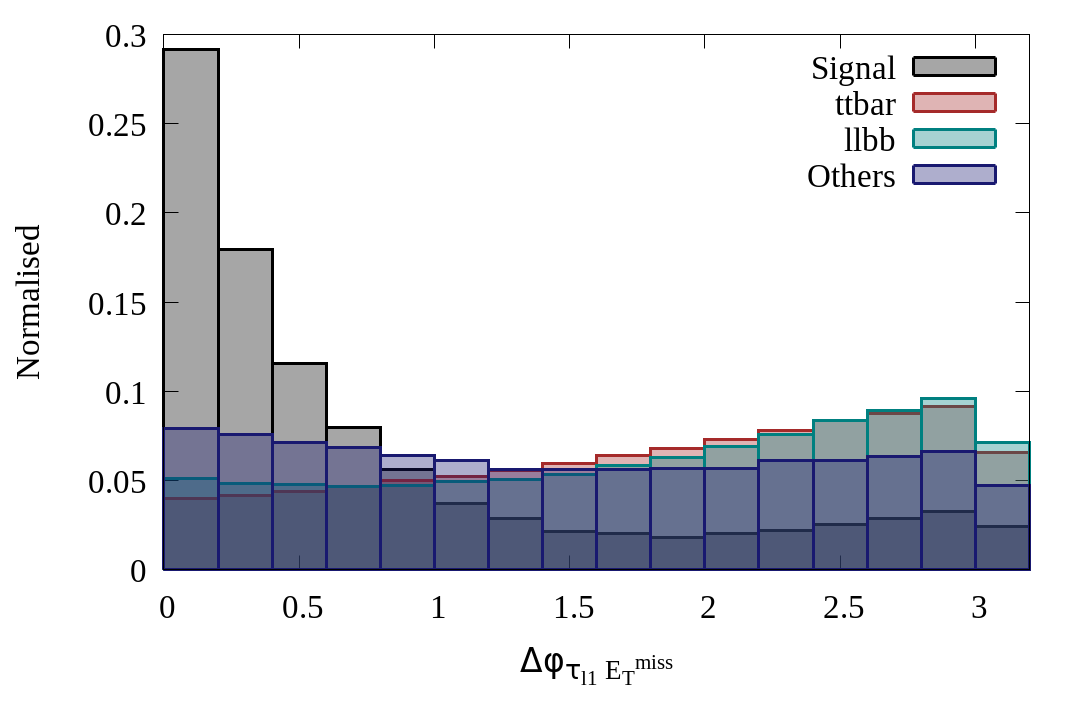}\includegraphics[scale=0.2]{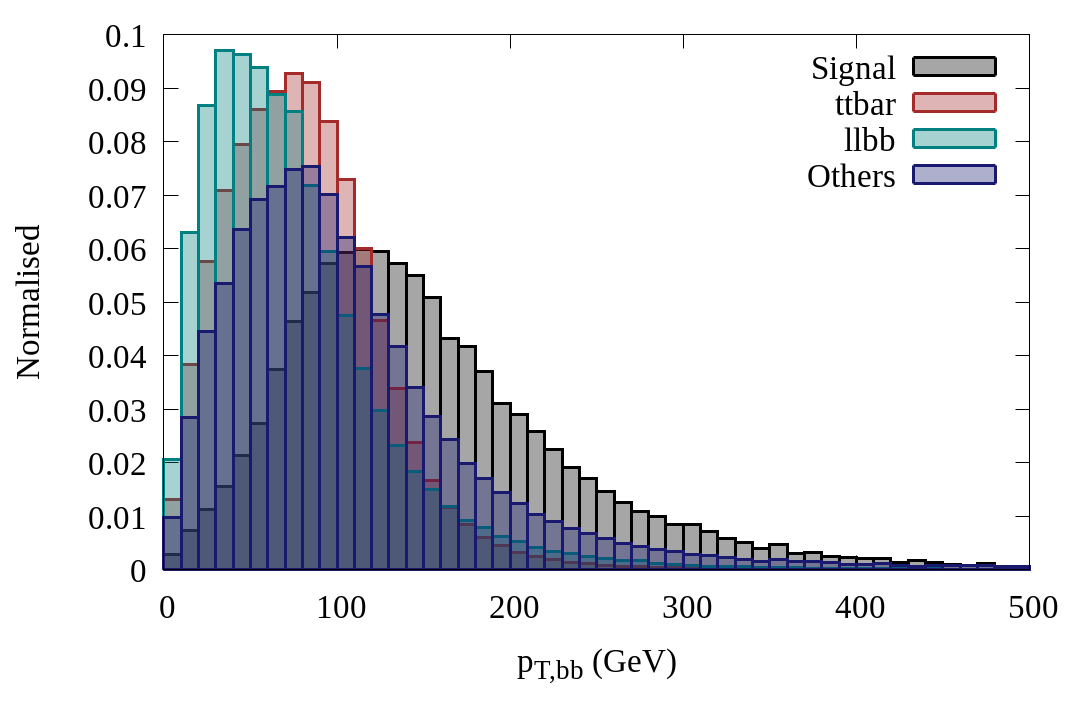}
\caption{Normalised distributions of $M_{\tau_l\tau_l}$, $m_{T2}$, $\Delta\phi_{\tau_{\ell_1} \met}$ and $p_{T,bb}$
 for the signal and dominant backgrounds in $b\bar{b}\tau_{\ell}\tau_{\ell}$ channel before applying basic selection cuts.}
\label{fig1:2b2taull}
\end{figure}

In an analogous manner to the previous two cases, we perform a multivariate analysis with the following 11 kinematic variables 
for the first case: 

\begin{equation}
\begin{split}
p_{T,bb},~m_{bb},~\Delta R_{bb},~m_{\tau_{\ell} \tau_{\ell}}^{\textrm{vis}},~\Delta\phi_{\tau_{\ell} \tau_{\ell}},
~\Delta\phi_{\tau_{\ell_1} \met},\nonumber \\~\Delta\phi_{\tau_{\ell_2} \met},~m_{hh}^{\textrm{vis}},~p_{T,hh}^{\textrm{vis}},
~\Delta R_{b_1\tau_{\ell_2}},~m_{T2} 
\end{split}
\end{equation}

Following this, we perform another multivariate analysis with the following 8 kinematic variables for the case involving the
collinear mass: 

\begin{equation}
\begin{split}
p_{T,bb},~m_{bb},~\Delta R_{bb},~M_{\tau_l\tau_l},~m_{T2},~\Delta\phi_{\tau_{\ell_1} \met},~\Delta\phi_{\tau_{\ell_2} \met}, ~m_{hh}^{\textrm{vis}} \nonumber
\end{split}
\end{equation}

In Table~\ref{tab2:2b2taull}, the signal, background yields and the significance after the BDT analysis are presented. We also show the 
normalised distributions of the four kinematic variables \textit{viz.}, $M_{\tau_l\tau_l}$, $m_{T2}$, $\Delta\phi_{\tau_{\ell_1} \met}$ and $p_{T,bb}$ in Fig.~\ref{fig1:2b2taull}. 
The BDT optimisation yields a statistical significance of 0.077 for the latter scenario where we use the collinear mass 
observable.

\begin{center}
\begin{table}[htb!]
\centering
\scalebox{0.8}{%
\begin{tabular}{|c|c|c|}\hline
(a) & Process & Events \\ \hline \hline

\multirow{4}{*}{Background}  
 & $t\bar{t}$ had      & $1181.56$ \\ 
 & $t\bar{t}$ semi-lep & $60632.89$ \\  
 & $t\bar{t}$ lep      & $34425.64$\\  
 & $\ell\ell b\bar{b}$ & $7684.41$ \\  
 & $b\bar{b}h$         & $0.38$ \\
 & $Zh$                & $39.78$ \\ 
 & $t\bar{t}h$         & $539.47$ \\ 
 & $t\bar{t}Z$         & $672.44$ \\ 
 & $t\bar{t}W$         & $353.46$ \\  \cline{2-3}
 & 
\multicolumn{1}{c|}{Total}                       & $105530.03$ \\ \hline
\multicolumn{2}{|c|}{Signal ($hh\to 2b2\tau$)}    & $14.62$ \\\hline 
\multicolumn{2}{|c|}{Significance ($S/\sqrt{B}$)} & $0.045$ \\ \hline \hline  
 
\end{tabular}}
\quad
\scalebox{0.8}{%
\begin{tabular}{|c|c|c|}\hline
(b) & Process & Events \\ \hline \hline

\multirow{4}{*}{Background}  
 & $t\bar{t}$ had      & $196.54$ \\ 
 & $t\bar{t}$ semi-lep & $18843.34$ \\  
 & $t\bar{t}$ lep      & $12230.06$\\  
 & $\ell\ell b\bar{b}$ & $1516.91$ \\  
 & $b\bar{b}h$         & $0.28$ \\
 & $Zh$                & $19.55$ \\ 
 & $t\bar{t}h$         & $199.97$ \\ 
 & $t\bar{t}Z$         & $199.81$ \\ 
 & $t\bar{t}W$         & $110.26$ \\  \cline{2-3}
 & 
\multicolumn{1}{c|}{Total}                       & $33316.72$ \\ \hline
\multicolumn{2}{|c|}{Signal ($hh\to 2b2\tau$)}    & $14.02$ \\\hline 
\multicolumn{2}{|c|}{Significance ($S/\sqrt{B}$)} & $0.077$ \\ \hline \hline  
 
\end{tabular}}
\caption{Same as in Table~\ref{tab2:2b2tauhh} for the $b\bar{b} \tau_{\ell} \tau_{\ell}$ mode with (a) $m_{\tau \tau}^{\textrm{vis}}$ (b) $M_{\tau\tau}$ variable.}
\label{tab2:2b2taull}
\end{table}
\end{center}

\subsection{The $b\bar{b}WW^*$ channel}
\label{sec2.3:2b2W}

A channel often neglected in terms of rigour and clarity is the $b \bar{b} WW^*$ final state, having three markedly different sub-states, 
\textit{viz.}, the fully leptonic ($b \bar{b} \ell \ell + \slashed{E}_T$), the semi-leptonic ($b \bar{b} \ell + \; \textrm{jets} \; + 
\slashed{E}_T$) and the fully hadronic ($b \bar{b} + \; \textrm{jets}$), where $\ell$ denotes an electron, muon or a tau lepton. Out of 
these three possible final states, the fully leptonic one (which has an overlapping final state from $bb \tau \tau$; see section~\ref{sec2.2:2b2taull}) 
is the cleanest owing to lesser backgrounds. The semi-leptonic channel has a larger background as compared to the former. The fully 
hadronic final state, on the other hand, will be swamped, mostly by QCD backgrounds and hence is omitted from any further discussion in 
this study. For both the leptonic and semi-leptonic channels, the major background comes in the form of $t\bar{t}$. The fully leptonic 
$t\bar{t}$ scenario contributes to being the dominant background for the leptonic signal and both the fully leptonic and semi-leptonic 
decays of $t\bar{t}$ act as the dominant backgrounds to the semi-leptonic signal. For the semi-leptonic channel, the second-most 
dominant background arises in the form of $Wb\bar{b}+\textrm{jets}$. The much less dominant backgrounds are comprised of 
$b\bar{b}h$, $t\bar{t}h$, $t\bar{t}V$, $Vh$, $Vb\bar{b}$ and $VVV$, where $V$ denotes a $W$ or a $Z$ boson. For both the analyses, we 
implement a common set of trigger cuts, \textit{viz.}, $p_{T,b/j} > 30$ GeV, $p_{T,e \; (\mu)} > 25 \; (20)$ GeV, $|\eta_{b,\ell}| < 2.5$ 
and $|\eta_j| < 4.7$. Furthermore, in order to deal with the large $t\bar{t}$ backgrounds, we apply, at the generator level a hard cut 
of $m_{bb} > 50$ GeV. We apply the same for the $\ell \ell b \bar{b}$ background. Hence, in order to be consistent, we implement this 
same cut for all the samples at the analysis level. In the following two sub-subsections, we focus only on multivariate analyses. We 
pass the signal and background samples to the BDTD algorithm upon implementing the aforementioned cuts.


\subsubsection{The $2b2\ell+\met$ channel}
\label{sec2.3:2b2l2nu}

\begin{figure}
\includegraphics[scale=0.2]{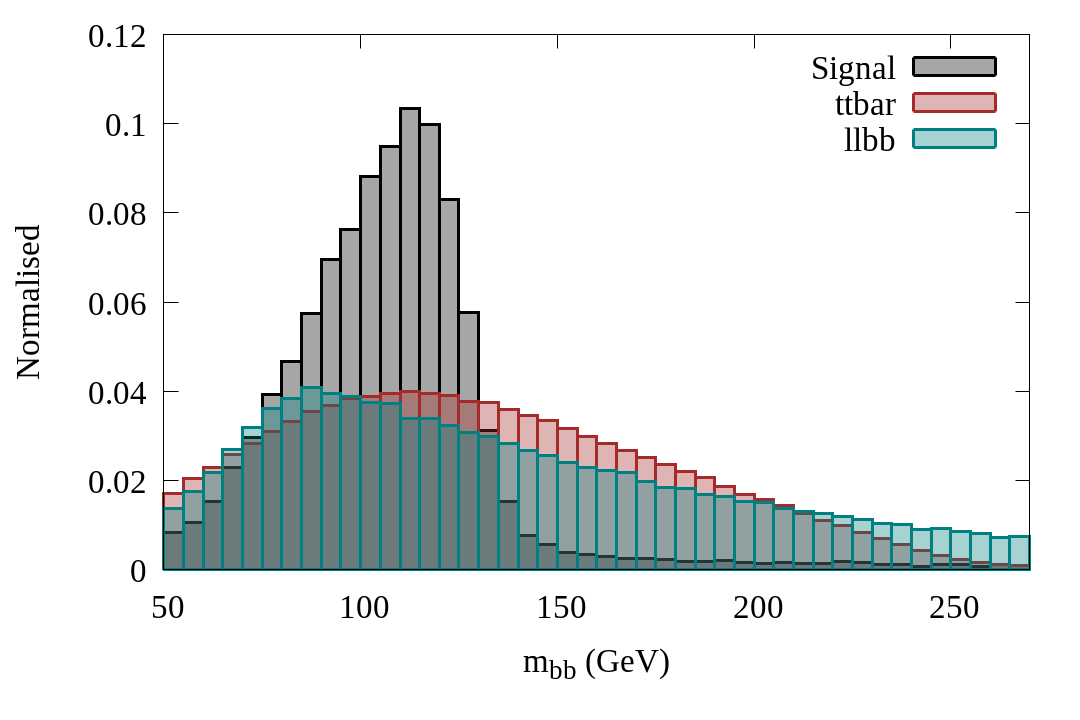}\includegraphics[scale=0.2]{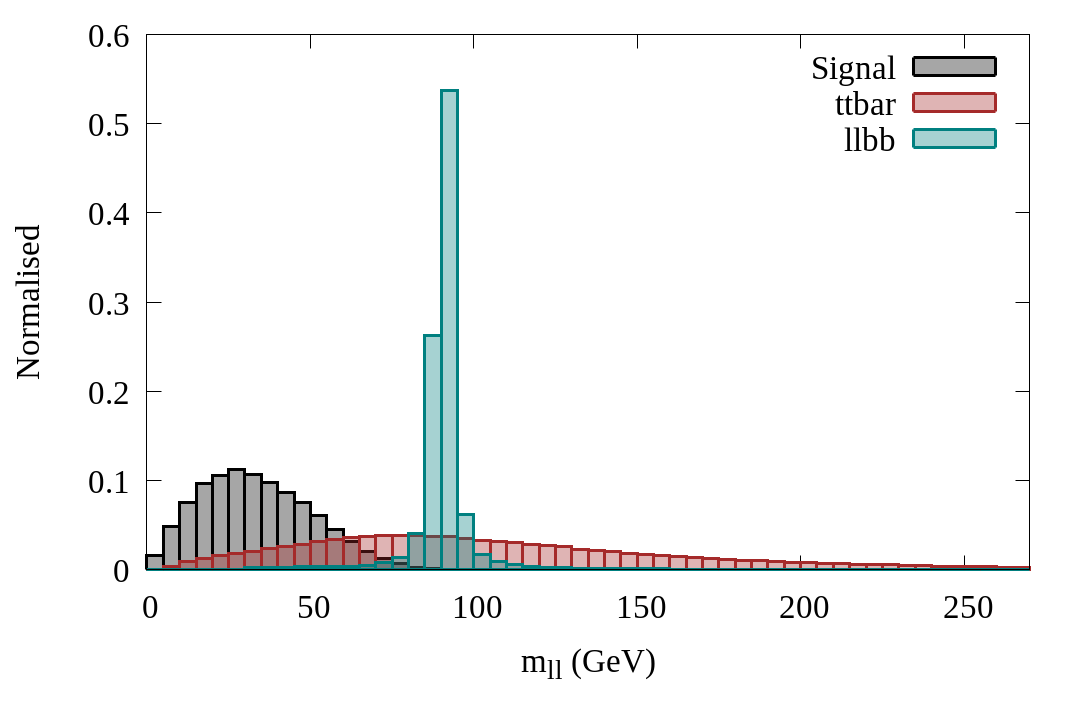}\\
\includegraphics[scale=0.2]{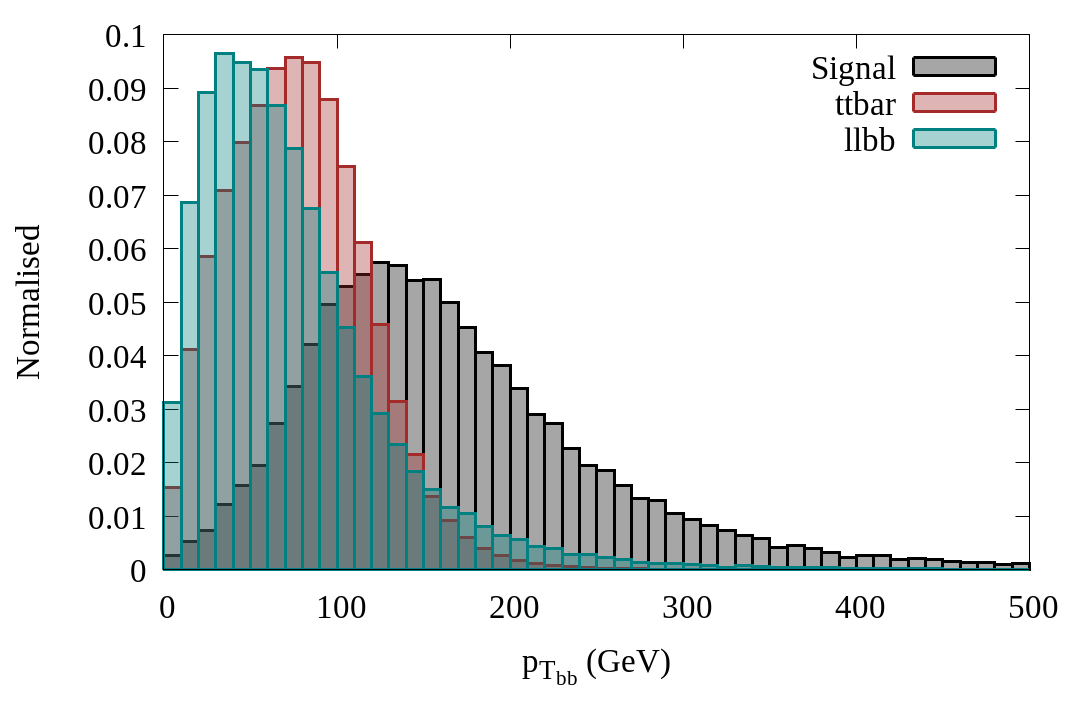}\includegraphics[scale=0.2]{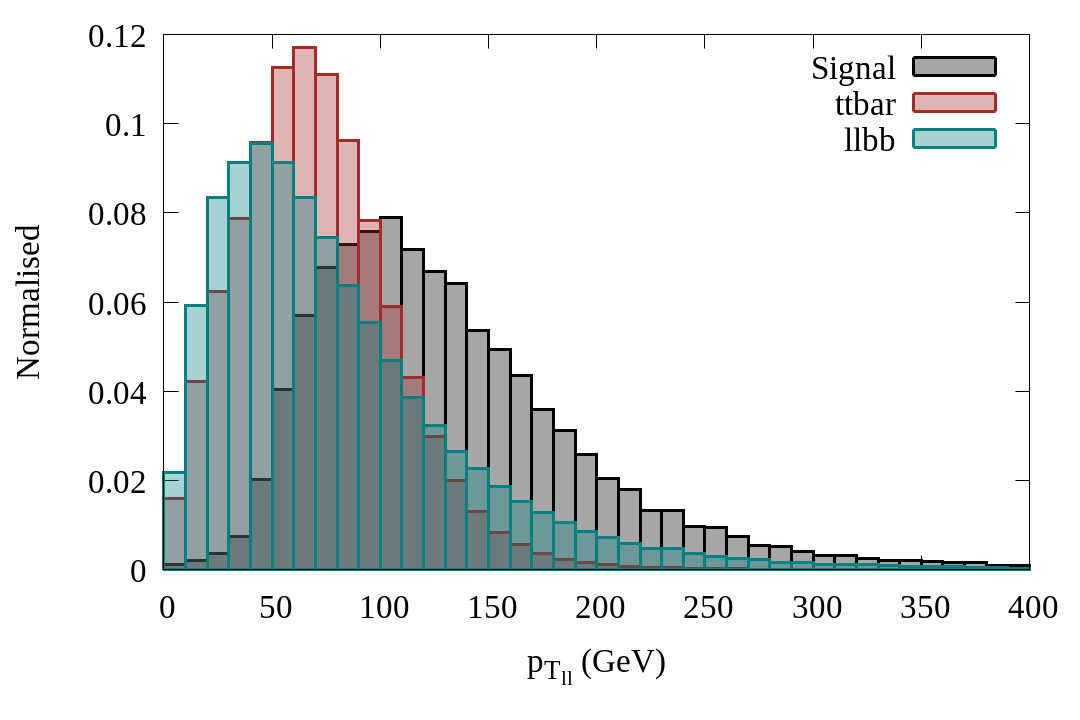}
\caption{Normalised distributions of $m_{bb}, \; m_{\ell \ell}, \; p_{T, bb}$ and $p_{T, \ell \ell}$ for the signal and dominant 
backgrounds in the $2b2\ell+\met$ channel after the basic selection cuts.}
\label{fig1:2b2l2nu}
\end{figure}

Inspired by the CMS HL-LHC studies~\cite{CMS-PAS-FTR-15-002}, we focus on the dileptonic mode of the $b \bar{b} WW^*$ channel in this
part. Differing slightly from CMS, we do not impose cuts on $m_{\ell \ell}$, $\Delta R_{\ell \ell}$ and $\Delta \phi_{bb \; \ell \ell}$.
Moreover, instead of using their neural network discriminator, we consider the BDTD algorithm. Besides, in addition to their analysis,
we include various subdominant backgrounds on top of the dominant $t\bar{t}$ backgrounds, as has been listed above. For this study, we
select events with exactly two $b$-tagged jets and two isolated leptons with opposite charges. Upon inspecting various kinematic 
distributions, we choose the following ten for our multivariate analysis:
\begin{equation}
p_{T,\ell_{1/2}},~\met,~m_{\ell \ell},~m_{b b},~\Delta R_{\ell \ell},~\Delta R_{b b},~p_{T,bb},~p_{T,\ell \ell},~\Delta \phi_{bb\; \ell \ell} \nonumber,
\end{equation}
where the last term implies the azimuthal angle separation between the reconstructed di $b$-tagged jet and di-lepton systems. Having 
$t\bar{t}$ as the dominant background by far, \textit{i.e.}, the weight of this background being several orders of magnitude larger than 
the rest, we train our BDTD algorithm with the signal sample along with this background only. We analyse the other backgrounds upon 
using this training. The final number of signal and background events along with the significance are listed in Table~\ref{tab1:2b2l2nu}. 
The distributions of the four best discriminatory variables, \textit{viz.}, $m_{bb}, \; m_{\ell \ell}, \; p_{T, bb}$ and $p_{T, \ell 
\ell}$, after the basic cuts as listed above, are shown in Fig.~\ref{fig1:2b2l2nu}.

\begin{center}
\begin{table}[htb!]
\centering
\scalebox{0.7}{%
\begin{tabular}{|c|c|c|c|}\hline
Sl. No. & Process & Order                                      & Events \\ \hline \hline

\multirow{5}{*}{Background}   
 & $t\bar{t}$ lep                                                               & NNLO~\cite{ttbarNNLO}                      & $2080.52$ \\  
 & $t\bar{t}h$                                                                  & NLO~\cite{bkg_twiki_cs}                    & $131.66$ \\  
 & $t\bar{t}Z$                                                                  & NLO~\cite{Lazopoulos:2008de}               & $106.31$\\  
 & $t\bar{t}W$                                                                  & NLO~\cite{Campbell:2012dh}                 & $35.97$ \\ 
 & $hb\bar{b}$                                                                  & NNLO (5FS) + NLO (4FS)~\cite{bkg_twiki_cs} & $\sim 0$ \\  
 & $\ell\ell b\bar{b}$                                                          & LO                                         & $842.72$ \\ \cline{2-4} 
 & Total                                                                        &                                            & $3197.18$ \\ \hline
\multicolumn{2}{|c|}{Signal ($hh \to b\bar{b}WW\to b\bar{b} \ell \ell + \met$)} & NNLO~\cite{hhtwiki}                        & $35.20$ \\\hline 
\multicolumn{3}{|c|}{Significance ($S/\sqrt{B}$)} & $0.62$ \\ \hline \hline  
 
\end{tabular}}
\caption{Signal, background yields and final significance for the $b\bar{b} \ell \ell + \met$ channel after the BDT analysis.}
\label{tab1:2b2l2nu}
\end{table}
\end{center}

Finally, with a judicious cut on the BDTD observable, we find $\sim 35$ signal and $\sim 3197$ background events, yielding a significance 
of $\sim 0.62$ upon neglecting systematic uncertainties. The numbers are in excellent agreement to the ones obtained by 
CMS~\cite{CMS-PAS-FTR-15-002}. This channel can thus act as an important combining channel to enhance the total SM di-Higgs 
significance at the HL-LHC and also serves as an important search for a resonant di-Higgs scenario~\cite{Huang:2017jws}.


\subsubsection{The $1\ell 2j 2b + \met$ channel}
\label{sec2.3:1l2jMET}

\begin{figure}
\includegraphics[scale=0.2]{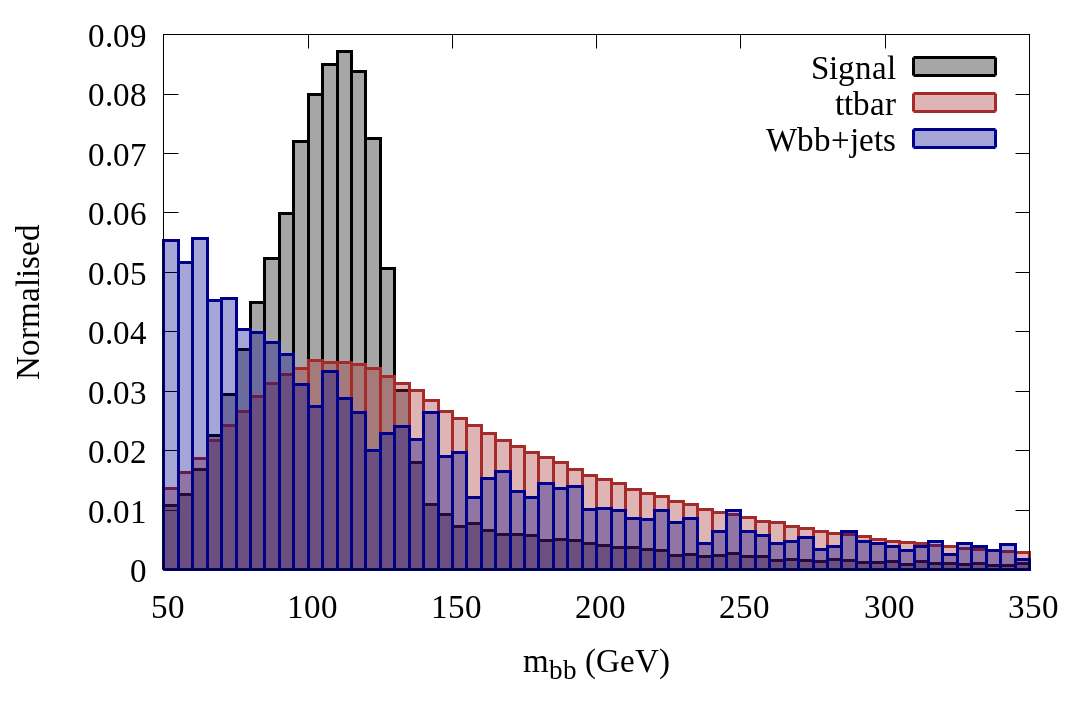}\includegraphics[scale=0.2]{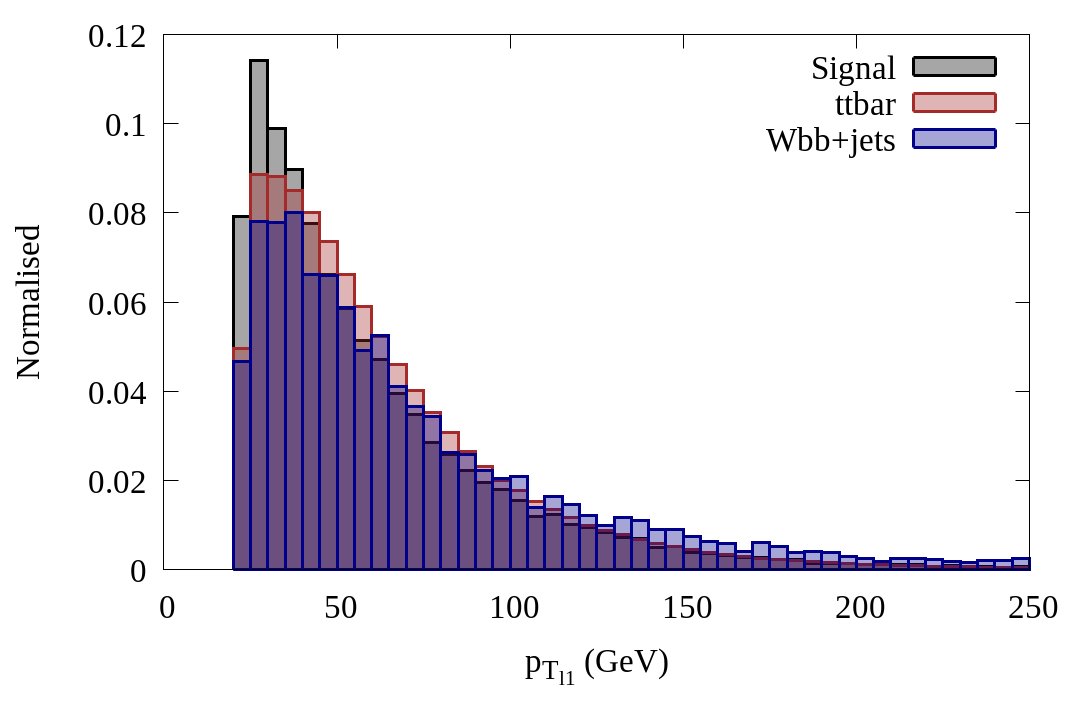}\\
\includegraphics[scale=0.2]{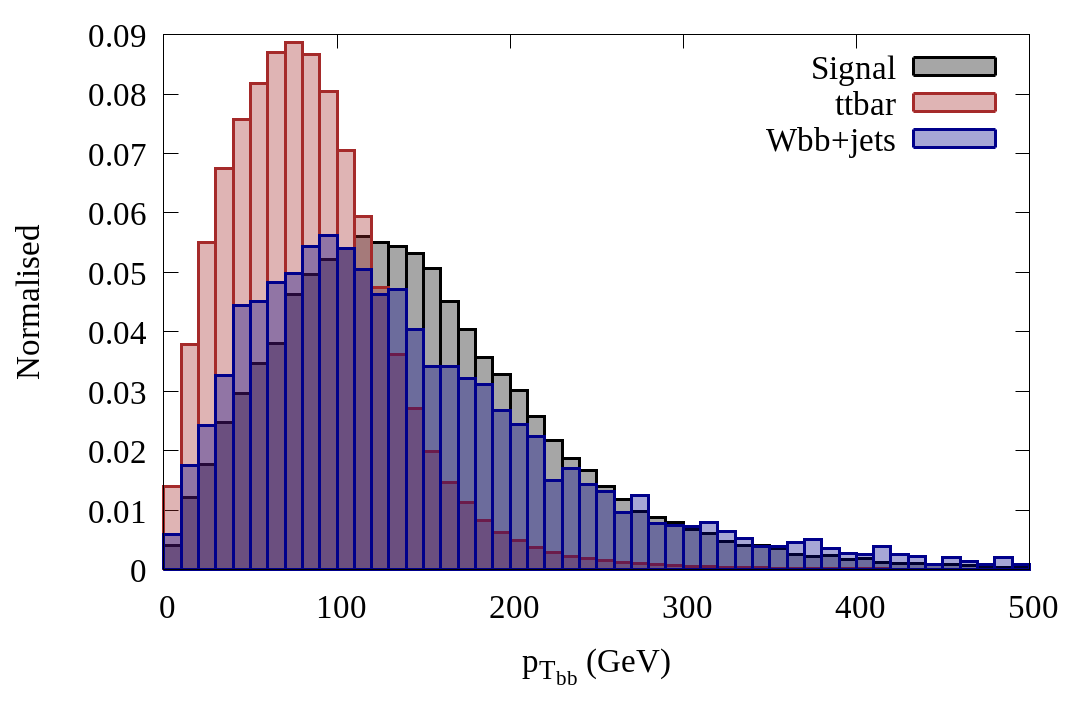}\includegraphics[scale=0.2]{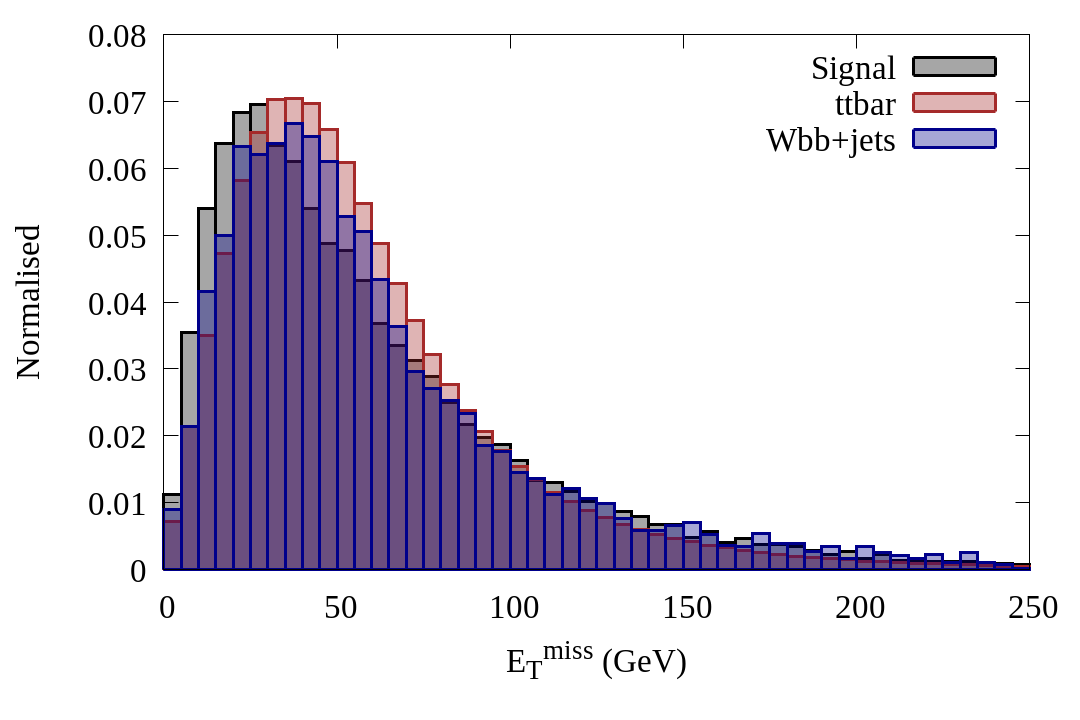}
\caption{Normalised distributions of $m_{bb}, \; p_{T,\ell_1}, \; p_{T, bb}$ and $\met$ for the signal and dominant backgrounds in 
 the $1\ell 2j 2b + \met$ channel after the basic selection cuts.}
\label{fig1:1l2jMET}
\end{figure}

Before concluding this subsection, we make an attempt to decipher the potential of the semi-leptonic final state for the $b\bar{b}WW^*$ 
channel. On the analysis front, we choose events with exactly two $b$-tagged jets, one isolated lepton and at least two light jets 
meeting the trigger criteria as discussed above. We consider the same set of cuts as for the dileptonic channel before performing the 
multivariate analysis. For this case, we find the following variables to have the best discriminatory properties.
\begin{equation}
p_{T, \ell},~\met,~m_{jj},~m_{bb},~\Delta R_{jj},~\Delta R_{bb},~p_{T,bb},~p_{T,\ell jj},~\Delta \phi_{bb \; \ell jj},
 ~\Delta R_{\ell \; jj} \nonumber,
\end{equation}
where $p_{T,\ell jj}, \Delta \phi_{bb \; \ell jj}$ and $\Delta R_{\ell \; jj}$ refer to the visible $p_T$ of the $\ell j j$ system (for 
the signal, ensuing from the $h \to WW^* \to \ell \nu jj$ decay), the azimuthal angle separation between the di-$b$-tagged jet system 
and the $\ell j j$ system and the $\Delta R$ separation between the lepton and the di-jet system respectively. Here the dominant
backgrounds are the semi-leptonic and the leptonic decays of $t\bar{t}$. Hence, in an analogous way to the dileptonic case, we train the 
BDTD with the signal and the $t\bar{t}$ samples, albeit with proper weight factors for the leptonic and semi-leptonic backgrounds. We then
utilise this training for the rest of the backgrounds as well, which are clearly subdominant with respect to the $t\bar{t}$ backgrounds. 
We find a significance of 0.13, however, with a much smaller $S/B$ ratio. The results are summarised in Table~\ref{tab1:1l2jMET}. The 
distributions of the four best observables, \textit{viz.}, $m_{bb}, \; p_{T,\ell_1}, \; p_{T, bb}$ and $\met$ are shown in 
Fig.~\ref{fig1:1l2jMET}. We do not find a promising significance for this scenario. We obtain a negligible $S/B$ and a significance of
0.13 assuming zero systematic uncertainties. A somewhat promising result has been obtained in Ref.~\cite{Papaefstathiou:2012qe} using
jet substructure techniques.

\begin{center}
\begin{table}[htb!]
\centering
\scalebox{0.7}{%
\begin{tabular}{|c|c|c|}\hline
Sl. No. & Process & Events \\ \hline \hline

\multirow{4}{*}{Background}
 & $t\bar{t}$ semi-lep            & $866990.56$ \\ 
 & $t\bar{t}$ lep                 & $96147.82$ \\  
 & $t\bar{t}h$                    & $4508.25$ \\  
 & $t\bar{t}Z$                    & $5192.52$\\  
 & $t\bar{t}W$                    & $2949.65$ \\ 
 & $Wb\bar{b}+\textrm{jets}$ [LO] & $121313.52$ \\  
 & $\ell\ell b\bar{b}$            & $5780.47$ \\ \cline{2-3}  
 & \multicolumn{1}{|c|}{Total}    & $1102882.79$ \\ \hline
\multicolumn{2}{|c|}{Signal ($hh \to 2b2W$)} & $134.34$ \\\hline 
\multicolumn{2}{|c|}{Significance ($S/\sqrt{B}$)} & $0.13$ 
\\ \hline \hline  
 
\end{tabular}}
\caption{Signal, background yields and final significance for the $1\ell 2j 2b + \met$ channel after the BDT analysis.The 
various orders of the signal and backgrounds are same as in Table~\ref{tab1:2b2l2nu}.}
\label{tab1:1l2jMET}
\end{table}
\end{center}

\subsection{The $\gamma\gamma WW^*$ channel}
\label{sec2.3:2W2ga}

In this subsection, we analyse the process $p p \to h h \to WW^{*}\gamma\gamma$ and consider both the pure leptonic ($\ell^+ 
\ell^- \gamma\gamma + \met$) and semi-leptonic ($\ell j j \gamma\gamma + \met$) final states. We abstain from analysing the pure hadronic 
decay mode as it entails an enormous irreducible background, rendering the search hopeless even at the HL-LHC. For the leptons, photons 
and jets, we employ the following trigger level cuts:
\begin{itemize}
 \item For electrons, $p_{T} > 25~{\rm GeV},~|\eta| < 2.5$,
 \item For muons, $p_{T} > 20~{\rm GeV},~|\eta| < 2.5$,
 \item For photons, $p_{T} > 20~{\rm GeV},~|\eta| < 2.5$ and
 \item For jets, $p_{T} > 30~{\rm GeV},~|\eta| < 4.7$.
\end{itemize}
Following this, we then discuss some of the most significant kinematic variables which distinguish the signal and backgrounds most 
efficiently. Finally, we present the results from multivariate analysis. 

\subsubsection{Pure leptonic decay}

The signal yield in this current scenario is much smaller in comparison to the most-studied di-Higgs search channels like 
$b\bar{b}\gamma\gamma$ and $b\bar{b}\tau^{+}\tau^{-}$. However, as we will see below, this channel has a significantly lower 
background yield.

We require each event to have exactly two isolated photons and two isolated leptons having opposite electric charge. Sizeable 
backgrounds to this final state arise from the $t\bar{t}h$ associated production, the Higgs-strahlung $Zh$ process (merged up to three
jets), and from the $\ell \ell\gamma\gamma$ (where $\ell = e, \mu, \tau$ for this case) final state. The irreducible background to this 
search channel comes from $\ell \nu \ell\nu \gamma \gamma$ (mostly from $VV\gamma \gamma$), which has a relatively smaller cross-section 
as compared to the aforementioned backgrounds, and hence has not been considered in the current analysis. While generating the $\ell 
\ell \gamma \gamma$ background, we merge the samples up to one extra jet and we also impose a generation-level cut on the invariant mass 
of the $\gamma\gamma$ pair, \textit{viz.}, $120~{\rm GeV} < m_{\gamma\gamma} < 130~{\rm GeV}$.

\begin{figure}
\includegraphics[scale=0.2]{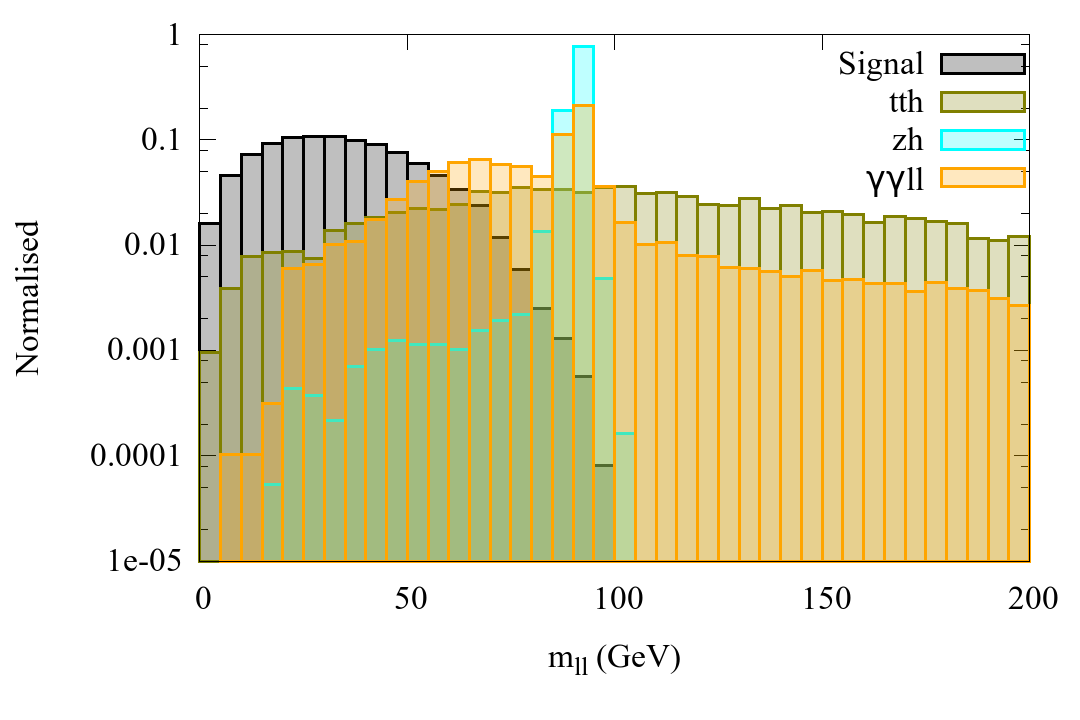}\includegraphics[scale=0.2]{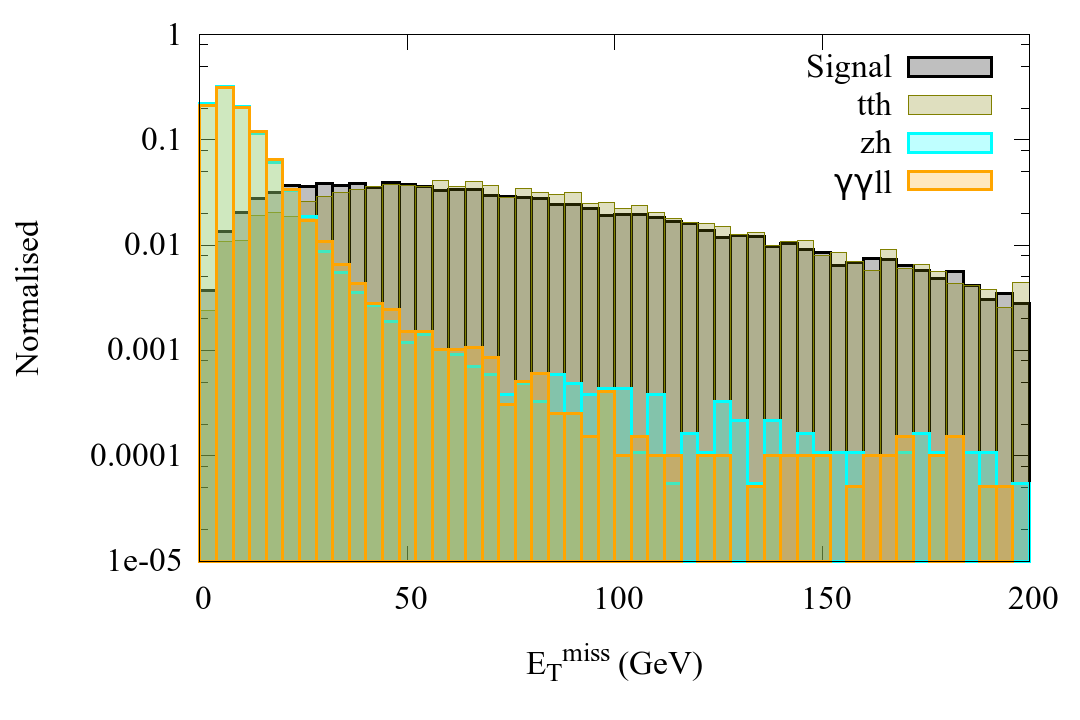}\\
\includegraphics[scale=0.2]{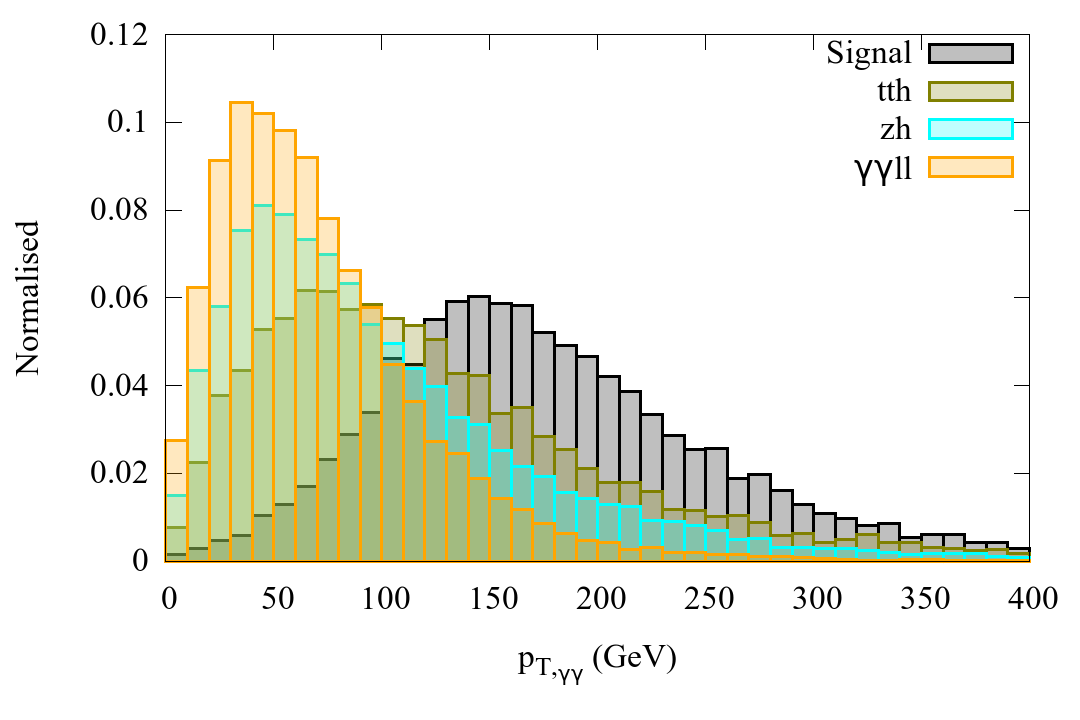}\includegraphics[scale=0.2]{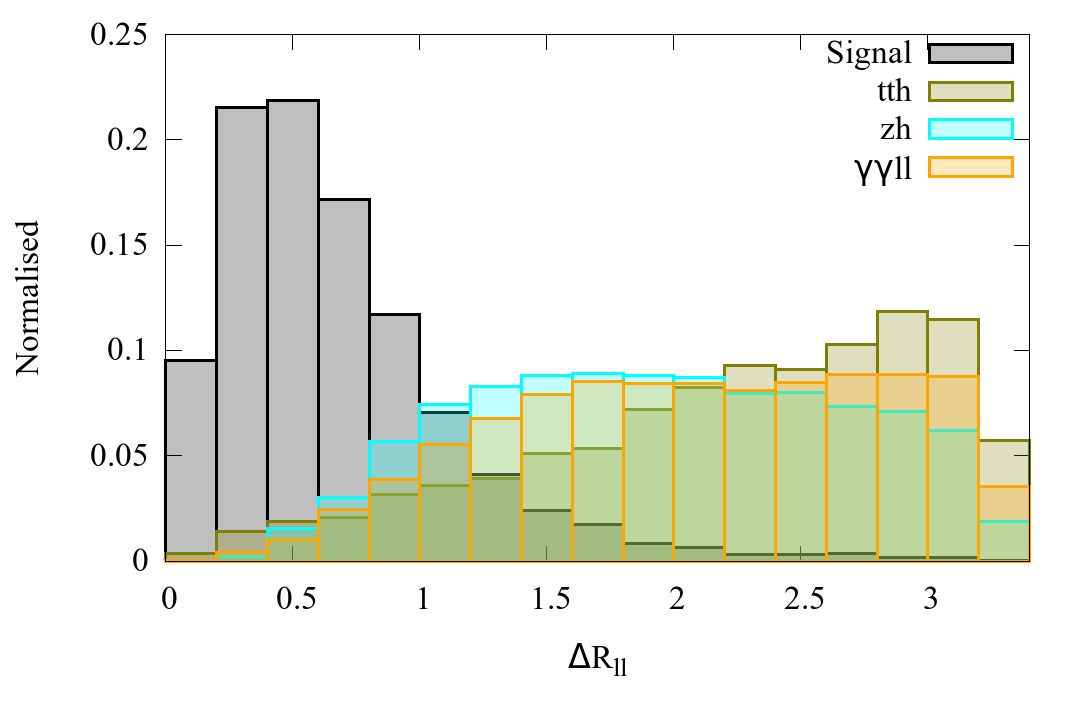}
\caption{Normalised distributions of $m_{\ell \ell}$, $\met$, $p_{T, \gamma \gamma}$ and $m_{\gamma \gamma}$ for the signal 
and all relevant backgrounds in the $2\ell 2\gamma + \met$ channel after the basic selection cuts.}
\label{fig:wwgamgam_lnulnu_1}
\end{figure}

Before listing down the variables we use for the multivariate analysis, we also impose a $b$-jet veto to the events. This reduces
the $t\bar{t}h$ background substantially. For this analysis as well for the semi-leptonic analysis that follows, we require the invariant 
mass of the di-photon system to be $122~{\rm GeV} < m_{\gamma\gamma} < 128~{\rm GeV}$. As an optimised cut-based analysis for this 
channel is not available in the literature, we implement a BDT optimisation approach. The following are the variables used to train the 
signal and background samples. 

\begin{equation}
 p_{T,\ell_{(1,2)}},~\met,~m_{\ell \ell},~m_{\gamma \gamma},~\Delta R_{\gamma \gamma (\ell \ell)},~p_{T,\ell \ell},~p_{T,\gamma \gamma},
 ~\Delta \phi_{\ell \ell \; \gamma \gamma} \nonumber,
\end{equation}
where the last term denotes the azimuthal angle separation between the di-lepton and the di-photon systems. In Fig.~\ref{fig:wwgamgam_lnulnu_1},
we show the kinematic distributions of the four variables, \textit{viz.}, $m_{\ell \ell}$, $\met$, $p_{T, \gamma \gamma}$ and 
$m_{\gamma \gamma}$. These variables help distinguish the signal from the weighted background samples, most efficiently.

We find that upon imposing a cut on the BDT variable, the $S/B$ improves from 4.4$\times 10^{-3}$ (after the basic selection) to 0.40. 
This is a significant improvement and perhaps has of the best signal over background ratios amongst all the channels studied so 
far. Unfortunately for us, this channel is plagued by very small branching ratios rendering a signal yield of less than unity. 
Given the dearth of signal events, we can not define a statistical significance. We must, however, note that this channel can be 
one of the most important channels for a 28 TeV/ 33 TeV collider. The signal and background yields are listed in 
Table~\ref{tab:wwgamgam3}. Hence we conclude that in order for this channel to have a significant contribution in the combination of 
the various final states, one requires either a large luminosity or higher energies. 

\begin{center}
\begin{table}[htb!]
\centering
\scalebox{0.7}{%
\begin{tabular}{|c|c|c|c|}\hline
Sl. No. & Process & Order & Events \\ \hline \hline
\multirow{5}{*}{Background} 
        & $t\bar{t}h$                          & NLO~\cite{bkg_twiki_cs}                   & $0.89$ \\
        & $Zh \; + $ jets                      & NNLO (QCD) + NLO (EW)~\cite{bkg_twiki_cs} & $0.20$ \\
        & $\ell \ell \gamma \gamma \; +$ jets  & LO                                        & $0.33$ \\ \cline{2-4} 
 Total  &                                      &                                           & $1.42$ \\ \hline
 Signal &                                      & NNLO~\cite{hhtwiki}                       & $0.57$ \\\hline 
\end{tabular}}
\caption{Signal and background yields for the $2\ell 2\gamma + \met$ channel after the BDT analysis.}  
\label{tab:wwgamgam3}
\end{table}
\end{center}

\subsubsection{Semi leptonic decay}

This channel has been studied by ATLAS~\cite{ATLAS-CONF-2016-071} with an integrated luminosity of 13.3 fb$^{-1}$. However, given the 
extremely small branching ratio of $h \to \gamma \gamma$, this channel is yet not sensitive and imposes a very weak observed upper
limit on the non-resonant di-Higgs cross-section at 25.0 pb (95\% confidence-level). Here, we concern ourselves with the $\ell \gamma 
\gamma + \; \textrm{jets} \; + \met$ final state. This process, however, has an additional complexity since the kinematics of the final 
state depends on whether the $\ell \nu$ ($jj$) comes from the on-shell or the off-shell $W$-boson decay. Even though the event rate of 
the semi-leptonic scenario is larger than its purely leptonic counterpart, the presence of additional jets lead to considerably larger 
backgrounds. 

For the event selection, we do not follow the analysis sketched in Ref.~\cite{ATLAS-CONF-2016-071} as it is designed to maximise the 
signal events given the dearth in the integrated luminosity for such a process. We perform a multivariate analysis with looser basic 
selection cuts. We demand exactly one isolated lepton, two isolated photons and at least one light jet, with the $p_T$ and $|\eta|$ 
ranges mentioned above. The irreducible background to this process comes from $\ell \nu \gamma\gamma$, merged up to one hard jet 
and has a tree level cross-section of $\sim 3.28~{\rm fb}$. In addition, $\ell \ell \gamma \gamma$ ($\ell = e, \mu, \tau$ for both cases), 
merged up to one hard jet and having a generation level cross-section of 1.05 fb, also contributes to the background when one of the 
leptons goes missing. These two backgrounds have been generated with a hard cut at the generation level as has been discussed for the 
di-leptonic scenario. Similar to the previous analysis for the full leptonic case, $t\bar{t}h$ and $Zh+$jets also contribute significantly 
to the background. In addition, we consider the $Wh$ process, merged up to 3 jets, as an important background.

\begin{figure}
\includegraphics[scale=0.2]{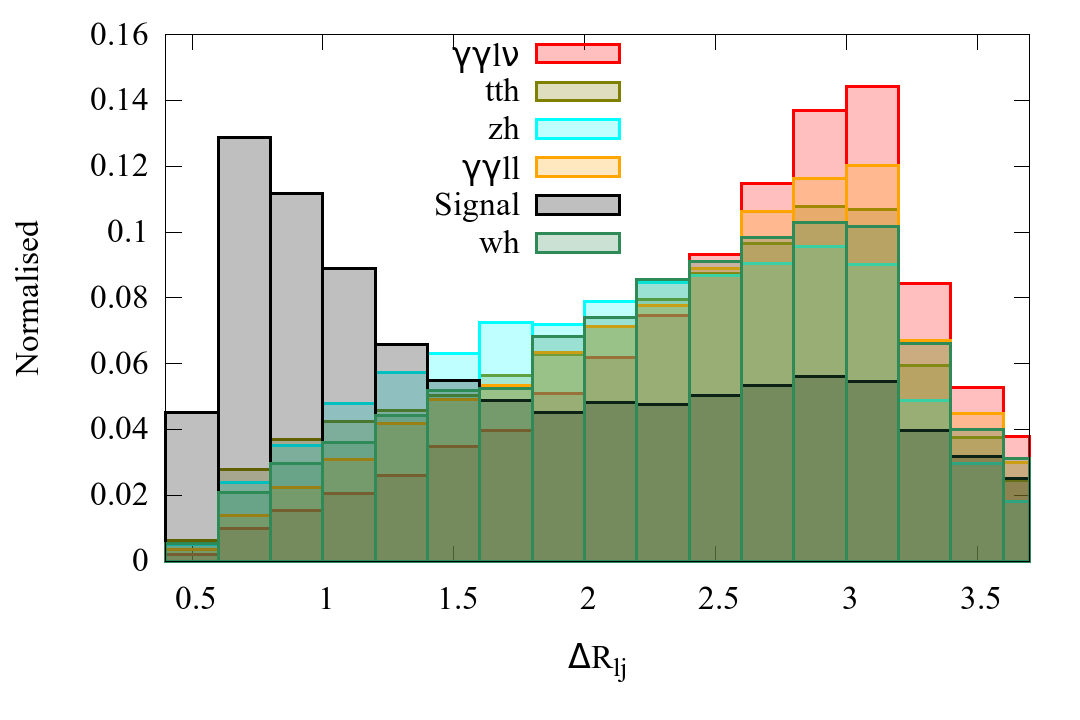}\includegraphics[scale=0.2]{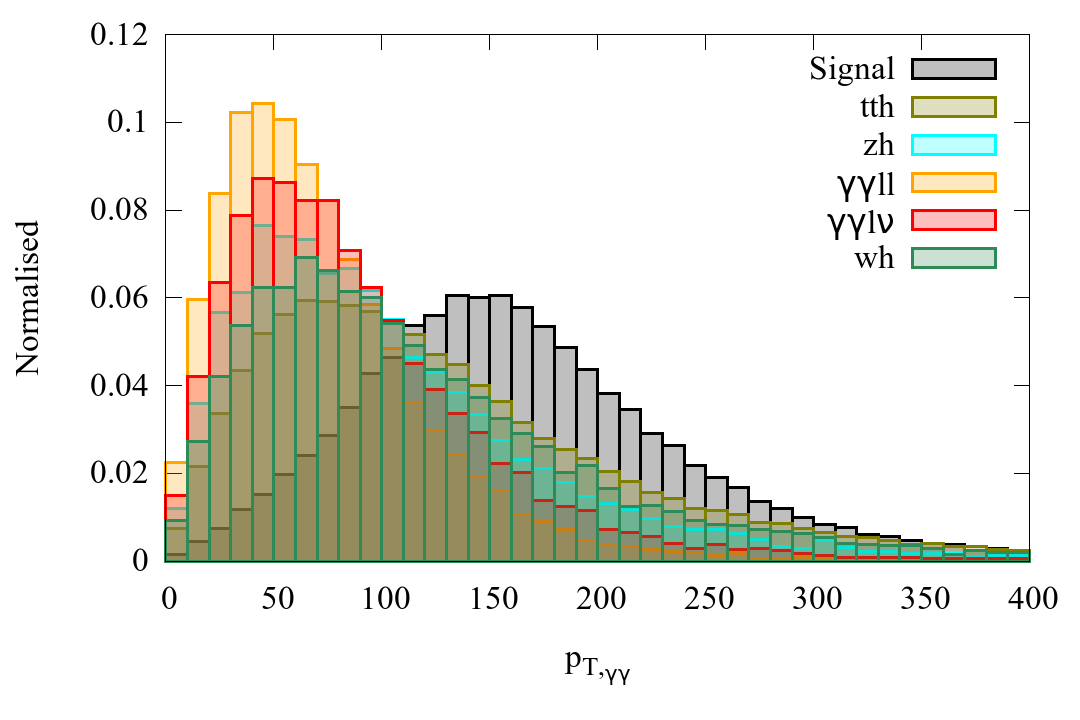}\\
\includegraphics[scale=0.2]{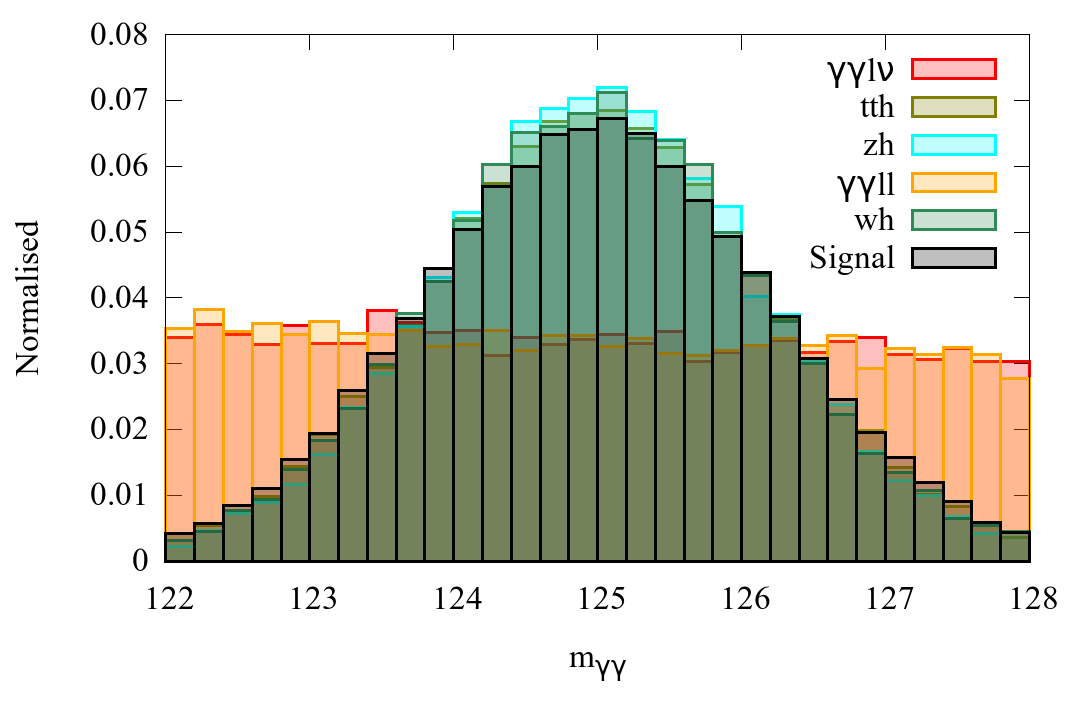}\includegraphics[scale=0.2]{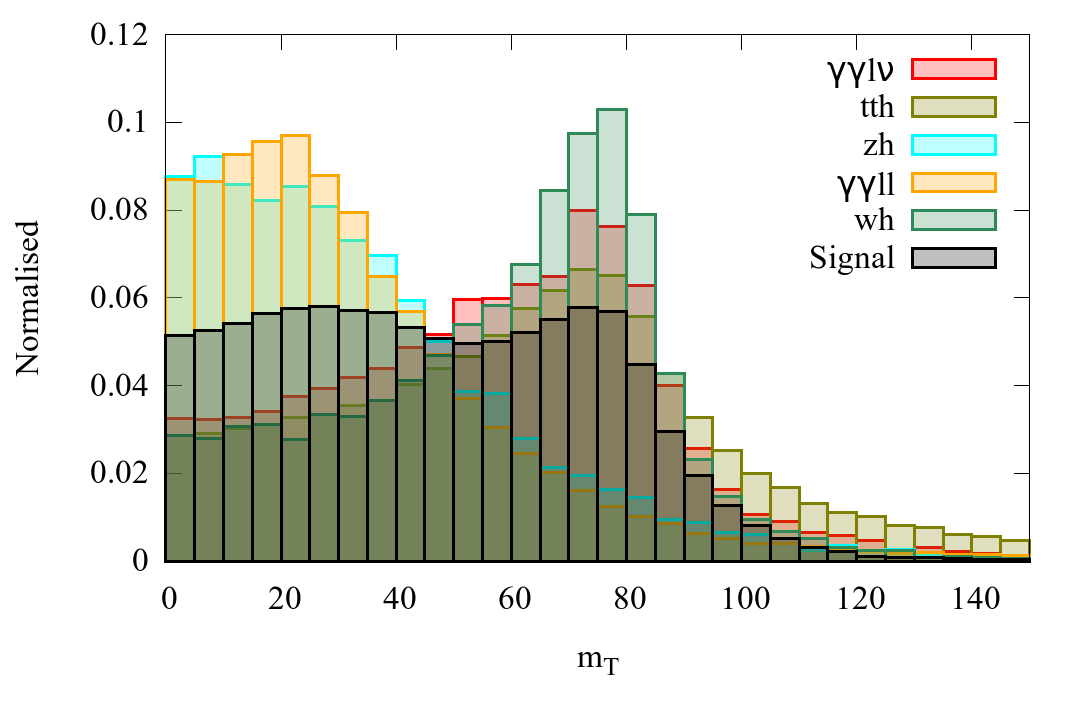}
\caption{Normalised distributions of $\Delta R_{\ell j}$, $p_{T,\gamma\gamma}$, $m_{\gamma\gamma}$ and $m_{T}$  for 
the signal and all relevant backgrounds in the $\ell \gamma \gamma + \; \textrm{jets} \; + \met$ channel after the basic selection 
cuts.}
\label{fig:wwgamgam_lnulnu_2}
\end{figure}

We perform our standard multivariate analysis upon employing these nine kinematic variables.
\begin{equation}
p_{T, \ell_1},~\met,~m_{\gamma\gamma},~\Delta R_{\gamma \gamma},~ \\
~p_{T,\gamma\gamma},~p_{T,\ell j},~\Delta \phi_{\ell j \;\gamma\gamma},~\Delta R_{\ell j},~ m_{T} \nonumber,
\end{equation}
where $\Delta \phi_{\ell j \;\gamma\gamma}$ is the azimuthal angle separation between the $\ell j$ and the reconstructed di-photon 
systems with $j$ being the hardest jet and $m_{T}$ is the transverse mass variable. It is found that $\Delta R_{\ell j}$, 
$p_{T,\gamma\gamma}$, $m_{\gamma\gamma}$ and $m_{T}$ are the most effective variables in distinguishing the signal 
from the backgrounds as can be seen in Fig.~\ref{fig:wwgamgam_lnulnu_2}. We find that after a proper BDT implementation, the signal 
over background ratio improves from 4.8$\times 10^{-3}$ (after basic selection) to 0.11. The signal and background yields after imposing 
an appropriate cut on the BDTD variable are summarised in Table~\ref{tab:wwgamgam2}. Here also we find that similar to its precursor, 
\textit{i.e.}, the purely leptonic scenario, the $S/B$ is much better than most of the channels considered thus far. However, the low 
rate due to the small branching ratio of $h \to \gamma \gamma$ acts as a hindrance to render this final state useful at present. Going 
to high energy machines, higher integrated luminosities of around 5000 fb$^{-1}$ with the 14 TeV collider, performing a combination of 
integrated luminosities from CMS and ATLAS at the HL-LHC, and lastly a modification to the SM cross-section, will enhance this channel's 
potential. In summary, the $\gamma \gamma W W^*$ final states yield extremely good $S/B$ ratios.

\begin{center}
\begin{table}[htb!]
\centering
\scalebox{0.7}{%
\begin{tabular}{|c|c|c|}\hline
Sl. No. & Process & Events \\ \hline \hline
\multirow{4}{*}{Background} 
        & $t\bar{t}h$                                         & $6.49$ \\
        & $Zh \; +$ jets                                      & $1.71$ \\ 
        & $Wh \; +$ jets                                      & $5.13$       \\
        & $\ell \nu \gamma \gamma \; +$ jets                  & $2.57$ \\ 
        & $\ell \ell \gamma \gamma \; +$ jets                 & $1.07$ \\ \cline{2-3} 
        & Total                                               & $16.97$ \\ \hline
 Signal &                                                     & $1.85$ \\\hline 
 
\end{tabular}}
\caption{Signal and background yields for the $\ell \gamma \gamma + \; \textrm{jets} \; + \met$ channel after the 
BDT analysis. The various orders for the signal and backgrounds are same as in Table~\ref{tab:wwgamgam3}. The order for $Wh \; +$ jets 
($\ell \nu \gamma \gamma \; +$ jets) is the same as for $Zh \; +$ jets ($\ell \ell \gamma \gamma \; +$ jets).} 
\label{tab:wwgamgam2}
\end{table}
\end{center}

\subsection{The $4W$ channel}
\label{sec2.4:4w}

In this subsection, we focus on the yet-untouched final states ensuing from the di-Higgs production mode, \textit{viz.}, the $4W$ 
channel~\footnote{The resonant scenario has, however, recently been studied in Ref.~\cite{Ren:2017jbg}.}. For completeness, we consider 
both semi-leptonic and fully leptonic decay modes. We lose cleanliness upon including more and more jets in the final state, 
\textit{i.e.}, upon considering the semi-leptonic decays. On the other hand, for a fully leptonic final state, the cross-section yield 
is extremely small. Considering two, three and four leptons, we choose following final states:

\begin{itemize}
\item Same-sign di-leptons ($SS2 \ell$): $\ell^{\pm} \ell^{\pm} + 4j + \met$,
\item Tri-leptons          ($3 \ell$)  : $3 \ell + 2j + \met$ and 
\item Four leptons         ($4 \ell$)  : $4 \ell + \met$.
\end{itemize}

Before moving on to the multivariate analyses, we impose the following basic cuts:
\begin{itemize}
 \item For jets, $p_T > 30$ GeV and $|\eta| < 4.7$ and
 \item For leptons, $p_T > 10$ GeV and $|\eta| < 2.5$.
\end{itemize}
In the following, we discuss the three cases as listed above.
\subsubsection{The $SS2\ell$ final state}
\label{sec2.4:SS2L}

Before implementing the multivariate analysis, we require each event to have exactly two leptons carrying the same electric charge and
having $p_T > 25$ GeV. Furthermore, we require events with at least two jets with a veto on $b$-tagged and $\tau$-tagged jets. The $WZ$ 
($W \to \ell \nu, Z \to \ell \ell$), $t\bar{t}$ and same-sign $W$-boson pair production constitute the most dominant backgrounds for 
this channel. Besides, we have the $Vh$ production ($V= W^\pm,Z$ decays leptonically and Higgs decays to $WW^*,ZZ^*$), $t\bar{t}X$ 
($X=W^\pm,Z,h$). The $t\bar{t}$ channel is a fake background for this process where either jets fake as leptons or charges are misidentified. 
Save for the same-sign $W$-boson pair, all the other dibosonic backgrounds are merged up to 3 jets. We must also note that by demanding 
a veto on the $b$-tagged jets, we are able to reduce a significant portion of the $t\bar{t}$ and $t\bar{t}X$ backgrounds.

\begin{figure}
 \includegraphics[scale=0.18]{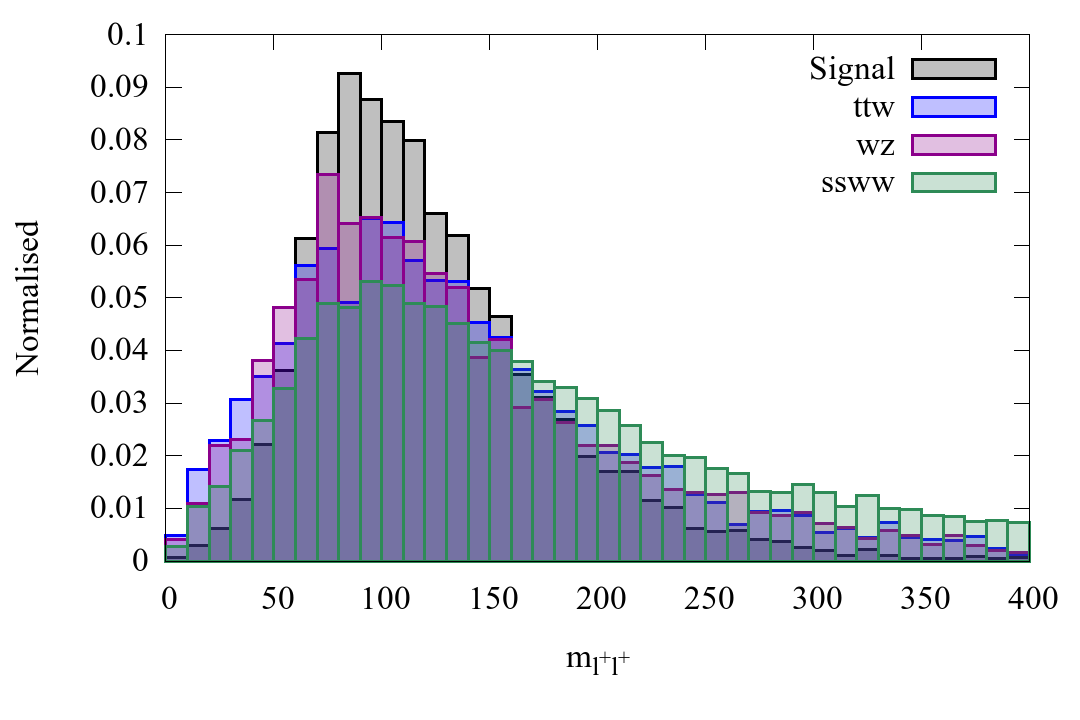}\includegraphics[scale=0.18]{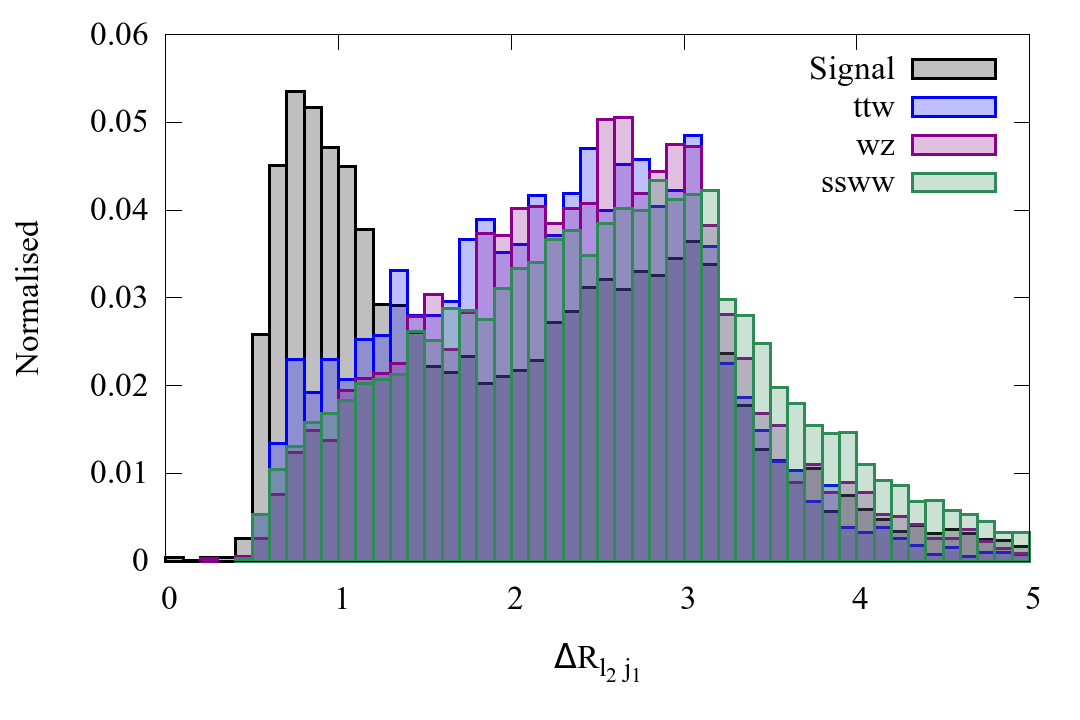}\\
 \includegraphics[scale=0.18]{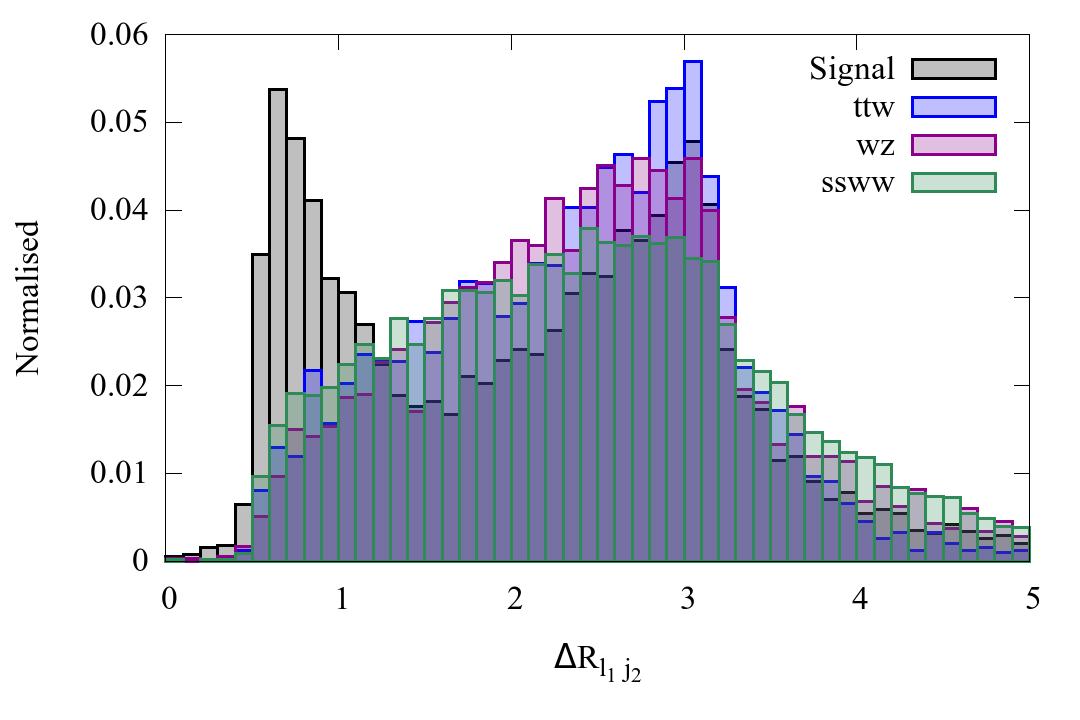}\includegraphics[scale=0.18]{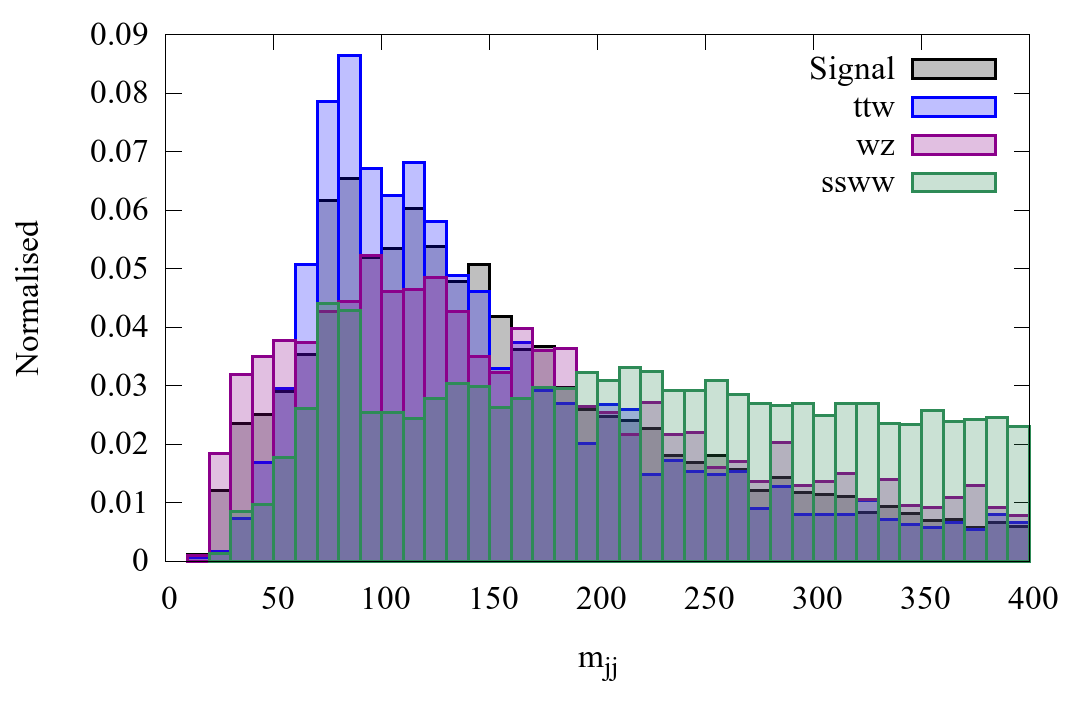}
 \caption{Normalised distributions of $m_{\ell^\pm \ell^\pm}$, $\Delta R_{\ell_2 j_1}$, $\Delta R_{\ell_1 j_2}$ and $m_{jj}$ for the 
 signal and the most relevant backgrounds for the $SS2\ell$ final state.}
 \label{fig:wwww_2l_1}
 \end{figure}

In a similar spirit as in all the previous subsections, we embark upon our multivariate analysis by choosing the six following 
kinematic variables.
\begin{equation}
 m_{\ell^\pm \ell^\pm},~\Delta R_{\ell_i j_k},~m_{jj} \nonumber,
\end{equation}
where $i,k = 1,2$ gives four combinations and $m_{jj}$ signifies the invariant mass constructed out of the hardest two jets. We show
the four most discriminatory variables in Fig.~\ref{fig:wwww_2l_1} and list down the final signal, background yields along with the
zero-systematics significance in Table~\ref{tab:wwww2}. We find that upon performing a BDT optimisation, the $S/B$ ratio improves
from $ 2.2 \times 10^{-4}$ (after basic selection cuts) to $ 9.7 \times 10^{-4}$. Unless the production cross-section is increased significantly or we find 
better techniques to control the $S/B$, this channel does not have much hope for a standard di-Higgs search. A drastic change in 
kinematics might change the picture altogether.

\begin{center}
 \begin{table}[htb!]
 \centering
\scalebox{0.7}{%
 \begin{tabular}{|c|c|c|c|}\hline
 Sl. No. & Process & Order & Events \\ \hline \hline
 \multirow{5}{*}{Background} 
        & $4\ell$               & LO                                        & $ 234.92 $ \\
        & $VVV~,(V=W,Z)$        & LO                                        & $ 291.80 $ \\
        & $Zh \; +$ jets        & NLO~\cite{bkg_twiki_cs}                   & $ 141.13 $ \\
        & $W^{\pm}W^{\pm}$      & LO                                        & $ 1896.03 $ \\
        & $Wh \; +$ jets        & NNLO (QCD) + NLO (EW)~\cite{bkg_twiki_cs} & $ 682.53 $ \\ 
        & $WZ \; +$ jets        & LO                                        & $ 6012.37 $ \\
        & $t\bar{t}W$           & NLO~\cite{Campbell:2012dh}                & $ 652.95 $ \\
        & $t\bar{t}h$           & NLO~\cite{bkg_twiki_cs}                   & $ 273.68 $ \\
        & $t\bar{t}Z$           & NLO~\cite{Lazopoulos:2008de}              & $ 293.31 $ \\
        & $t\bar{t}$ lep        & NNLO~\cite{ttbarNNLO}                     & $ 366.49 $ \\
        & $t\bar{t}$ semi-lep   & NNLO~\cite{ttbarNNLO}                     & $ 1521.32 $ \\ \cline{2-4} 
        & Total                 &                                           & $ 12366.53 $ \\ \hline
 Signal &                       & NNLO~\cite{hhtwiki}                       & $ 11.96 $ \\\hline 
\multicolumn{3}{|c|}{Significance ($S/\sqrt{B}$)} & $0.11$ \\\hline \hline
 \end{tabular}}
 \caption{Signal and background yields for the $SS2\ell$ channel after the BDT optimisation.}  
 \label{tab:wwww2}
 \end{table}
 \end{center}


\subsubsection{The $3 \ell$ final state}
\label{sec2.4:3L}

The trilepton analysis is somewhat similar in spirit to its $SS2\ell$ counterpart. For the $p_T$ cuts on the lepton, we relax them 
somewhat in this analysis. We require $p_{T, \ell_1} > 25$ GeV,  $p_{T, \ell_2} > 20$ GeV and $p_{T, \ell_3} > 15$ GeV, in order not 
to make the basic selection cuts too stringent. The pseudorapidity requirements for the leptons and the various requirements for the 
jets are as before. Furthermore, in order to remove events with leptons ensuing from the $Z$-boson, we require $|m_Z - m_{\ell \ell}| > 
20$ GeV for leptons having opposite sign and same flavour. The main backgrounds for this channel come from $Wh$, diboson production 
(mainly $WZ$) and the fake backgrounds coming from $t\bar{t}$. Apart from these, the $Zh$ ($Z \to \ell \ell, h \to W^+ W^-$), 
$t\bar{t}X$ ($X=W^\pm,Z,h$) and $ZZ$ backgrounds also contribute significantly. All the dibosonic processes are merged up to three jets.

\begin{figure}
\includegraphics[scale=0.2]{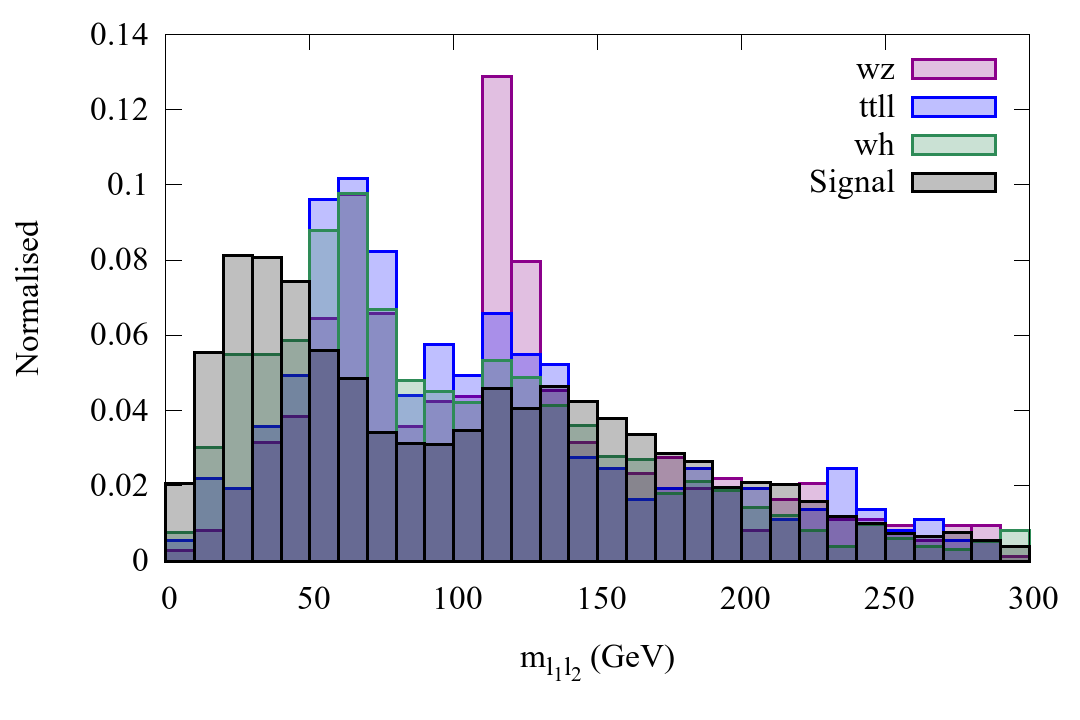}\includegraphics[scale=0.2]{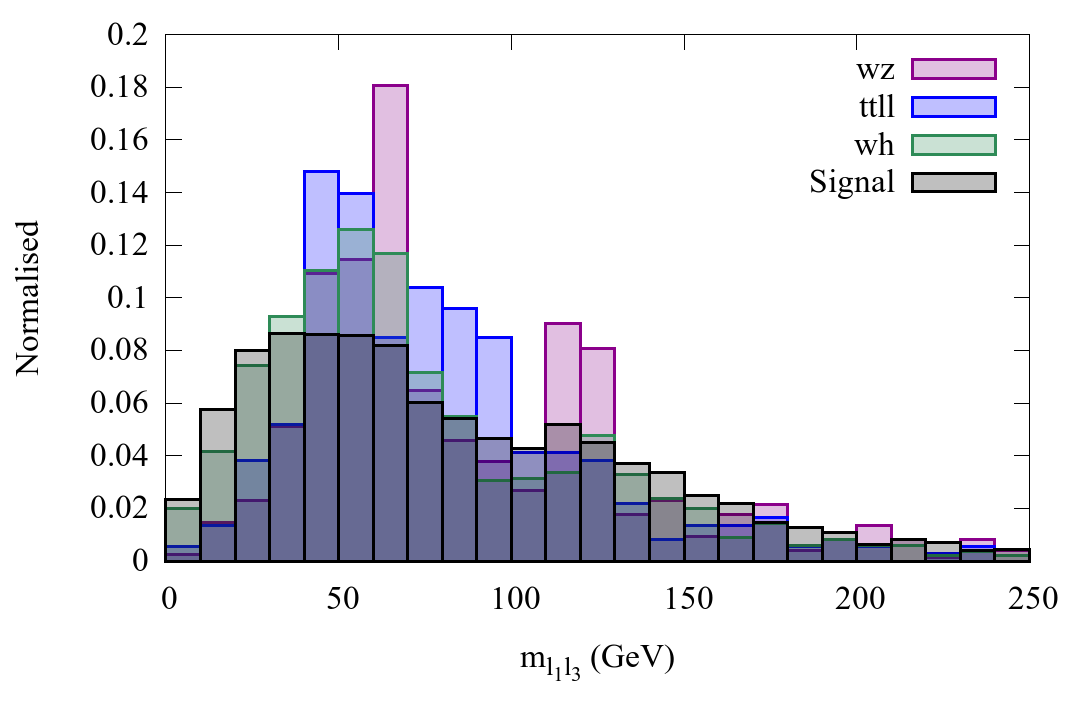}\\
\includegraphics[scale=0.2]{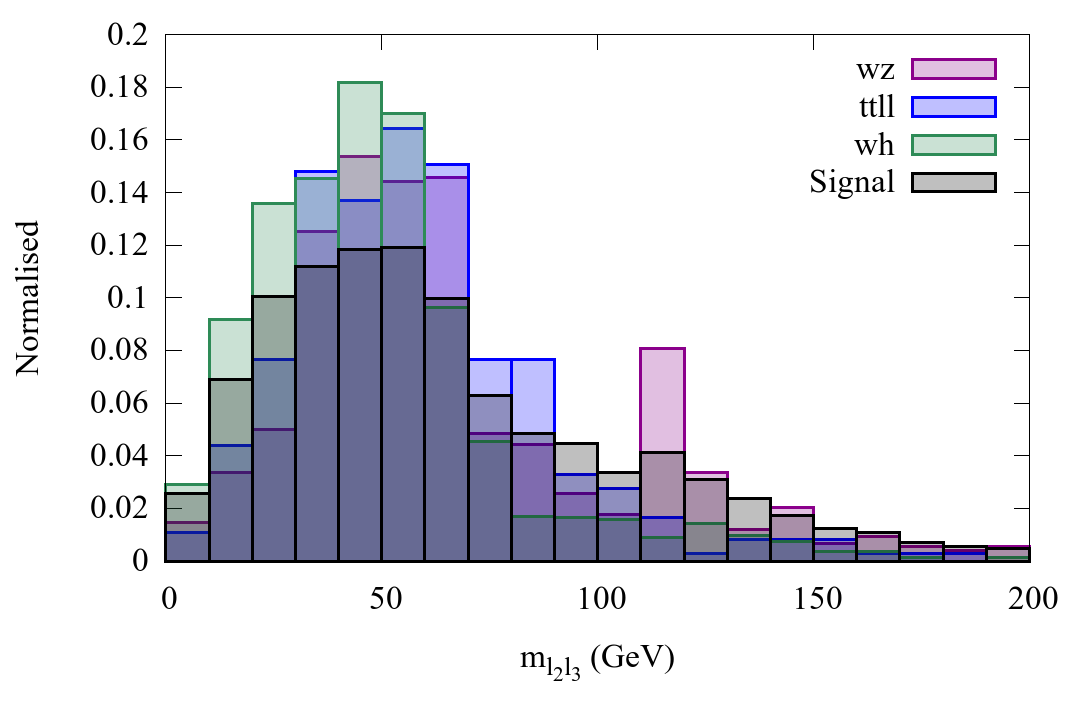}\includegraphics[scale=0.2]{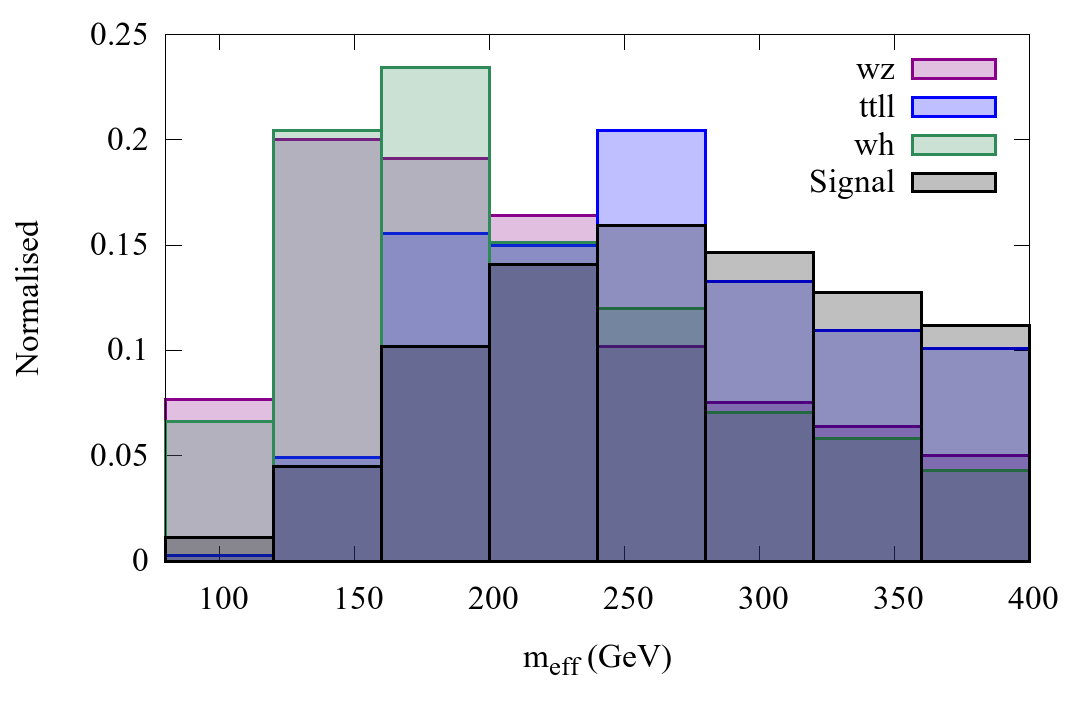}\\
\caption{Normalised distributions of $m_{\ell_1 \ell_2}$, $m_{\ell_1 \ell_3}$, $m_{\ell_2 \ell_3}$  and $m_{\textrm{eff}}$ for the signal and the 
most relevant backgrounds for the trilepton analysis.}
 \label{fig:wwww_3l_1}
\end{figure}

For this installment, we choose the following kinematic variables to train our BDTD algorithm.
\begin{equation}
 m_{\ell_i \ell_j},~\Delta R_{\ell_i \ell_j},~m_{\ell \ell \ell},~m_{\textrm{eff}},~\met,~p_{T,\ell_i},~n_{\textrm{jet}},\nonumber
\end{equation}
where $i,j$ runs from 1 to 3, $m_{\textrm{eff}}$ is the effective mass summing the $\met$, the scalar $p_T$ of the three leptons and all
the jets in the event. Lastly, $n_{\textrm{jet}}$ is the count of the number of jets per event. The four best variables are shown in
Fig.~\ref{fig:wwww_3l_1}. The event yields and final significance are shown in Table~\ref{tab:wwww4}. In this case, the $S/B$ changes
from $ 7.3 \times 10^{-4}$ (after basic selection cuts) to $2.8 \times 10^{-3}$. We find that there is a slight improvement 
compared to the $SS2\ell$ scenario. Finally, we end up with a statistical significance of 0.20.

\begin{center}
\begin{table}[htb!]
\centering
\scalebox{0.7}{%
\begin{tabular}{|c|c|c|}\hline
Sl. No. & Process & Events \\ \hline \hline
\multirow{4}{*}{Background} 
       & $4\ell$              & $451.14$ \\
       & $VVV~(V=W,Z)$        & $158.53$ \\
       & $Wh \; +$ jets       & $668.49$ \\ 
       & $WZ \; +$ jets       & $1384.51$ \\
       & $t\bar{t}W$          & $244.76$ \\
       & $t\bar{t}h$          & $301.01$ \\
       & $t\bar{t}Z$          & $157.54$ \\
       & $t\bar{t}$ lep       & $1635.09$ \\
       & $t\bar{t}$ semi-lep  & $240.21$ \\
       & $Zh \; +$ jets       & $133.17$ \\\cline{2-3} 
       & Total                & $5374.45$ \\ \hline
Signal &                      & $ 15.01 $ \\\hline 
\multicolumn{2}{|c|}{Significance ($S/\sqrt{B}$)} & $ 0.20 $ \\ \hline \hline   
\end{tabular}}
\caption{Signal, background yields and final significance for the trilepton channel after applying the most optimised BDT cut. The 
various orders for the signal and the backgrounds are same as those in Table~\ref{tab:wwww2}. The order for $Zh \; +$ jets ($ZZ \; +$ 
jets) is the same as that for $Wh \; +$ jets ($WZ \; +$ jets).}
\label{tab:wwww4}
\end{table}
\end{center}

\subsubsection{The $4 \ell$ final state}
\label{sec2.4:4L}

This brings us to our final non-resonant analysis. For this analysis, we perform a simple cut-based analysis. We require each event to 
have four isolated leptons. The dominant backgrounds are $Wh$, $t\bar{t}h$, $t\bar{t}$, $ZZ$ and $Zh$. Besides, we have non-negligible
contributions from $t\bar{t}V$ ($V = W^\pm, Z$). All the dibosonic backgrounds are merged up to three jets save for the $ZZ$ sample which 
is merged up to one extra jet. The leading and sub-leading leptons are required to have $p_T > 20$ GeV. For the remaining two softer 
leptons, we demand $p_T > 10$ GeV. Besides, we also employ the $|m_Z - m_{\ell_i \ell_j}| > 20$ GeV cut in order to reduce backgrounds 
having a pair of opposite sign same flavour leptons coming from $Z$-bosons. Furthermore, we apply a cut on the missing transverse energy, 
\textit{viz.}, $\met > 50$ GeV to greatly reduce the $4 \ell$ background. These cuts are extremely helpful in reducing the backgrounds 
by a great deal. However, the extremely small signal yield reduces to an even smaller number which is not statistically significant for 
all practical purposes. In Table~\ref{tab:wwww5}, we find an $S/B$ of $\sim 2.5 \times 10^{-4}$ after imposing the aforementioned cuts. 
On adding the $\met$ cut the $S/B$ increases to 7.8$\times10^{-3}$. However, upon having such small cross-sections, we do not perform 
a BDT analysis for this scenario.
\begin{center}
\begin{table}[htb!]
\centering
\small
\begin{tabular}{|c|c|c|c|}\hline
Sl. No. & Process & Events & Events \\ 
 & & & $\met~>~50~{\rm GeV}$ \\\hline \hline
\multirow{4}{*}{Background} 
       & $4\ell$                & $5736.77$   & $34.18$ \\
       & $Wh$                   & $12.28$     & $1.75$ \\ 
       & $VVV$                  & $4.59$      & $3.60$ \\   
       & $t\bar{t}W$            & $0.78$      & $0.78$ \\
       & $t\bar{t}h$            & $36.44$     & $23.74$ \\
       & $t\bar{t}Z$            & $5.12$      & $5.12$ \\
       & $t\bar{t}$ lep         & $56.38$     & $56.38$ \\
       & $t\bar{t}$ semi-lep    & $~0.00$     & $0.00 $\\
       & $Zh$                   & $23.85$     & $5.96$ \\ \cline{2-4} 
       & Total                  & $5876.22$   & $131.51$ \\ \hline
Signal &                        & $2.02$      & $1.42$ \\\hline \hline 
\end{tabular}
\caption{Signal and background yields after applying the selection cuts for the $4\ell$ final state.}
\label{tab:wwww5}
\end{table}
\end{center}

\subsection{Summarising the non-resonant search results}
\label{sec2.5:summary}

To summarise this long section, we find that the prospects of discovering the SM non-resonant di-Higgs channel at the HL-LHC (14 TeV
with 3 ab$^{-1}$ of integrated luminosity) are bleak. The most promising channel comes in the form of $b\bar{b} \gamma \gamma$ yielding
an $S/B$ ratio of $\sim 0.19$ and a statistical significance of 1.76. The situation for the $b\bar{b} \tau^+ \tau^-$ channels is more
challenging unless we find an excellent algorithm to reconstruct the di-tau system. The purely leptonic final state of the 
$b\bar{b} WW^*$ mode shows promise but one will either require data-driven techniques to reduce systematic uncertainties on the
backgrounds or even better ways to curb the backgrounds. Both the leptonic and semi-leptonic decay modes for the $\gamma \gamma WW^*$
channel yield excellent signal to background ratios. However, the extremely small event yields render these channels unimportant with
the planned luminosity upgrade. The $4W$ channel has three distinct final states with leptons. Upon doing detailed analyses, we find 
that the signal yields are very small. The $S/B$ improves upon increasing the number of leptons but the signal yields fall rapidly. 
Upon combining all the statistically significant searches with at least 5 signal events after all the cuts, we end up
with a combined significance of 2.08$\sigma$ at the HL-LHC. We expect that in the event of running the LHC till higher luminosity or upon considering the CMS and ATLAS results to be statistically 
independent (giving us 6 ab$^{-1}$ data), one can reach close to 2.95$\sigma$ (with 6 ab$^{-1}$ luminosity, we gain by a factor
of $\sqrt{2}$) upon combining all the statistically significant channels. We must note that if we consider 
a flat systematic uncertainty on the background estimation, then upon using the formula $\mathcal{S} = N_S/\sqrt{N_S + N_B + 
\kappa^2 N_B^2}$, with $\mathcal{S}, N_S, N_B$ and $\kappa$ being respectively the significance, number of signal and background events
after all possible cuts and the systematic uncertainty, we will face a reduction in the quoted statistical significance depending on the 
value of $\kappa$. Even $\kappa = 0.1, 0.2$, \textit{i.e.}, a 10\%-20\% systematic uncertainty, may completely dilute our 
significance. Hence, we need excellent control over systematics in order for us to observe any hints coming from the di-Higgs channels. 
A 100 TeV collider has the potential of measuring the di-Higgs channel to a greater degree of accuracy. We also note that, in some channels, 
an enhancement in the production cross-section by a factor of 3 may help the discovery with the HL-LHC. Lastly, modified kinematics
will alter this picture completely and we may see encouraging results with lesser integrated luminosities. In the following section, 
we discuss various BSM scenarios yielding the same final states as have been discussed in the present section.

\section{Ramifications of varying the Higgs self-coupling}
\label{sec2-2}

\begin{figure}
\centering
\includegraphics[scale=0.8]{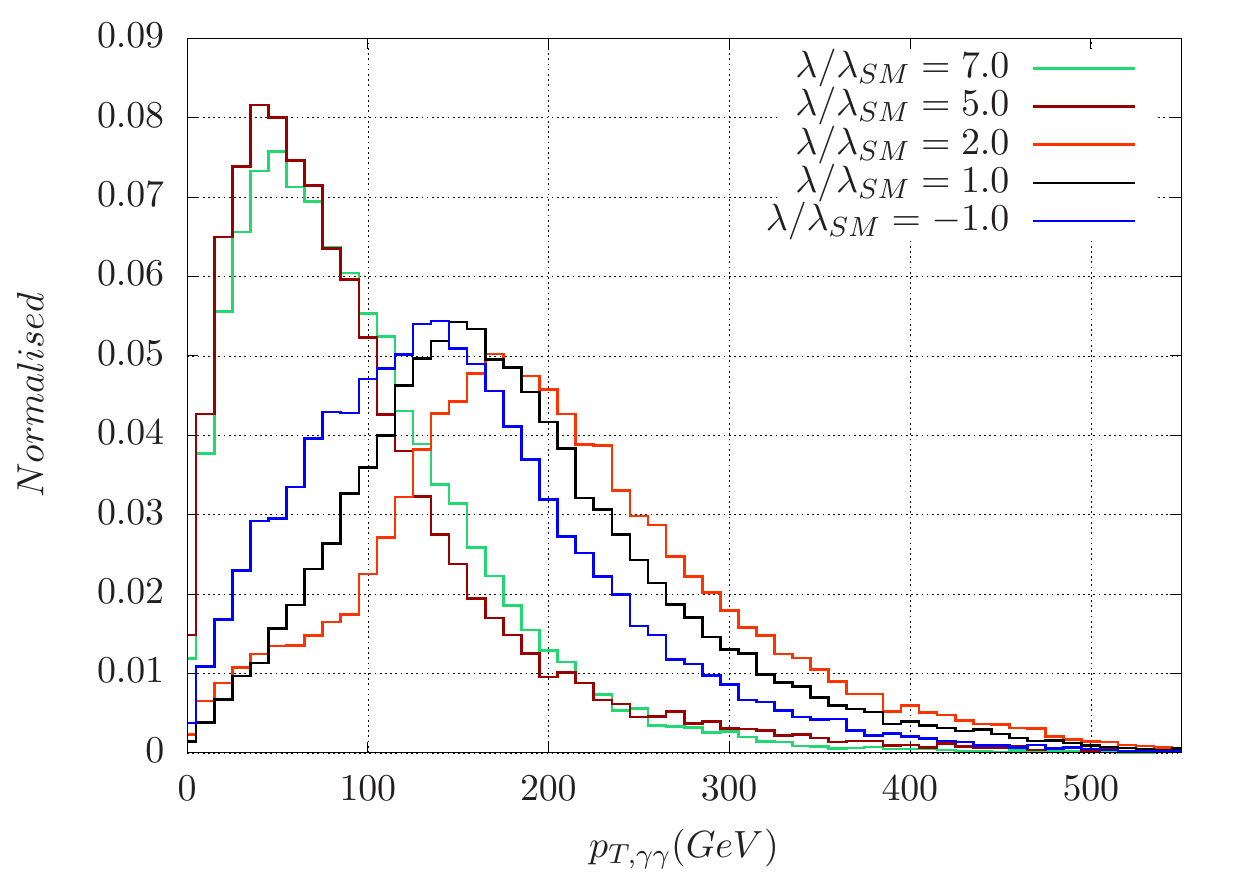}
\caption{Normalised distributions of $p_{T,\gamma\gamma}$ for the signal with different $\lambda_{hhh}/\lambda_{SM}$ values after the basic selection cuts.}
\label{fig:lambda}
\end{figure}

Before discussing the contaminations from various BSM scenarios to the standard double Higgs channels,
we address the issue of the variation of the Higgs self-coupling from its SM expectation. The 
Higgs self-coupling in the SM is an extremely small number and the HL-LHC study by 
ATLAS~\cite{ATL-PHYS-PUB-2017-001} predicts a sensitivity of $-0.8 < \lambda_{hhh}/\lambda_{SM} < 7.7$ upon 
assuming SM-like couplings for the remaining. In this regard, we must be wary of the differences in
the kinematic distributions upon changing $\lambda_{hhh}$ because it changes the magnitude of the
destructive interference with the SM box-diagram as we shall see below. This not only modifies the 
rate of the double Higgs production, but also alters the kinematics significantly. For the present 
study, we will consider the following six values of $\lambda_{hhh}/\lambda_{SM}$, \textit{viz.}, 
-1, 1, 2, 5 and 7. Because we have seen that the $b\bar{b} \gamma \gamma$ channel is the most sensitive channel for di-Higgs studies at the 
HL-LHC, we will restrict the anomalous self-coupling study to only this channel. Hence, referring to
section~\ref{sec2.1:2b2gamma}, we tread the following three steps. First, we consider double
Higgs production with each of the aforementioned $\lambda_{hhh}$ values (one at a time) as our signal and 
pass them through the cut-based analysis which has been optimised (with the 
cuts listed in Table~\ref{tab:bbgamgam_sel_cut}) to maximise the SM ($\lambda_{hhh}/\lambda_{SM}=1$) 
signal. Following this, we pass each of the $\lambda_{hhh}$ samples through the BDT framework optimised for
the SM double Higgs production (see Table~\ref{tab:bbgamgam_tmva}). Thereafter, we train all the 
samples with an alternative $\lambda$, \textit{viz.} $\lambda_{hhh}/\lambda_{SM}=5$. Finally, we train the BDT for each $\lambda_{hhh}$ point and compute
the significance. We list the results in Table~\ref{tab:lambda}. The cross-sections are for the 
process $p p \to h h \to b \bar{b} \gamma \gamma$ as a function of $\lambda_{hhh}/\lambda_{SM}$. The 
efficiencies are computed as the ratios of the final number of events (after the cut and count or the 
multivariate analysis) to the number of generated events. Finally, the yields are given for the 
signal and background samples for an integrated luminosity of 3 ab$^{-1}$. The cut-efficiency is
shown to be the maximum for the value of $\lambda_{hhh}/\lambda_{SM}=2$ where incidentally the 
cross-section is the smallest. We had already seen that going from a simple cut and count analysis
to a BDT analysis, rigorously trained to segregate the signal from background, we gain in significance.
This already holds true for the first two sub-tables, with an improvement varying between 13\%-23\%.
However, when we train the BDT with the corresponding $\lambda_{hhh}$ samples, the BDT becomes more
tuned to the modified kinematic distributions and in almost all cases, we find an improvement in
significance compared to its counterpart where the training was performed with the SM signal sample.
We can see the results in the fourth sub-table in Table~\ref{tab:lambda}. Also, in order to quantify
the difference in distributions for the variation of the Higgs trilinear coupling, we show the 
normalised distributions of the reconstructed Higgs $p_T$ in the di-photon channel ($p_{T,\gamma\gamma}$)
upon varying $\lambda_{hhh}/\lambda_{SM}$ (see Fig.~\ref{fig:lambda}).
Finally, we employ the log-likelihood CLs hypothesis test~\cite{Junk:1999kv,Read:2000ru,Read:2002hq} upon assuming the SM (and also $\lambda_{hhh}/\lambda_{SM}=5$) 
to be the null hypothesis. We obtain the following ranges of $\kappa = \lambda_{hhh}/\lambda_{SM}$:
\begin{align*}
\label{eq:resolvedresult}
-0.86 < \kappa < 7.96 \quad & \text{CBA for~} \kappa = 1 \text{~optimisation; SM null hypothesis} \\ \nonumber
-0.63 < \kappa < 8.07 \quad & \text{BDT analysis for~} \kappa = 1 \text{~optimisation;~SM null hypothesis} \\ \nonumber
-0.81 < \kappa < 6.06 \quad & \text{BDT analysis for~} \kappa = 5 \text{~optimisation;~SM null hypothesis} \\ \nonumber
-1.24 < \kappa < 6.49 \quad & \text{BDT analysis for~} \kappa = 5 \text{~optimisation;~} \kappa = 5 \text{~null hypothesis.} \nonumber
\end{align*}
Note that for $\kappa = 1$,  we are quite close in reproducing the HL-LHC prediction by ATLAS (i.e., $-0.8 < \lambda_{hhh}/\lambda_{SM} < 7.7$) in both the cut-based
(CBA) and BDT optimisation procedures. However, $\kappa$ is an unknown parameter (as the Higgs trilinear coupling has still not been measured) and hence, 
in principle, should be varied as well. Upon training with a different value of $\kappa$ other than $1$, viz.,  $\kappa = \lambda_{hhh}/\lambda_{SM} = 5$, a shift in the allowed
ranges for $\kappa$ has been obtained, which further depends on the hypothesis chosen.  We find a rather stronger upper-limit on the allowed range of the
trilinear coupling upon training with  the $\lambda_{hhh}/\lambda_{SM} = 5$ sample.
To conclude this section, we emphasise the fact that
we must be geared to tackle variations of the trilinear couplings from the SM expectations and
must be able to segregate them with the help of various kinematic distributions up to a certain
uncertainty.

\begin{center}
\begin{table}[htb!]
\centering
\scalebox{0.7}{
\begin{tabular}{|c|c|c|c|c|c|}\hline
\multicolumn{6}{|c|}{Cut Based (optimised for $\lambda_{hhh}/\lambda_{SM}=1$)}\\\hline
$\lambda/\lambda_{SM}$ & \makecell{Signal cross-\\ section (fb)} & Efficiency & Signal yield & Background yield & $S/\sqrt{B}$ \\\hline  
                                        
$-1$      &  $0.40$    &   $0.027$    &  $32.40$    & \multirow{7}{*}{$70.81$}  &  $3.85$ \\ \cline{1-4} \cline{6-6}                                   
$1$       &  $0.105$   &   $0.039$    &  $12.28$    &                           &  $1.46$ \\   \cline{1-4} \cline{6-6} 
$2$       &  $0.05$    &   $0.046$    &  $6.90$     &                           &  $0.82$ \\  \cline{1-4} \cline{6-6}
$5$       &  $0.26$    &   $0.008$    &  $6.24$     &                           &  $0.74$ \\   \cline{1-4} \cline{6-6}
$7$       &  $0.70$    &   $0.010$    &  $21.00$    &                           &  $2.49$ \\   \cline{1-4} \cline{6-6} \hline
\end{tabular}
}

\quad
\bigskip

\scalebox{0.7}{
\begin{tabular}{|c|c|c|c|c|c|}\hline
\multicolumn{6}{|c|}{BDT (optimised for $\lambda_{hhh}/\lambda_{SM}=1$)}\\\hline
$\lambda/\lambda_{SM}$ & \makecell{Signal cross-\\ section (fb)} & Efficiency & Signal yield & Background yield & $S/\sqrt{B}$ \\\hline 
                                      
$-1$      &  $0.40$    &   $0.035$    &  $41.76$    &  \multirow{7}{*}{$87.05$}  &  $4.48$ \\ \cline{1-4} \cline{6-6}                                      
$1$       &  $0.105$   &   $0.052$    &  $16.46$    &                           &  $1.76$ \\   \cline{1-4} \cline{6-6} 
$2$       &  $0.05$    &   $0.063$    &  $9.42$     &                           &  $1.01$ \\  \cline{1-4} \cline{6-6}
$5$       &  $0.26$    &   $0.010$    &  $7.84$     &                           &  $0.84$ \\   \cline{1-4} \cline{6-6}
$7$       &  $0.70$    &   $0.011$    &  $23.10$    &                           &  $2.48$ \\   \cline{1-4} \cline{6-6} \hline
\end{tabular}
}

\quad
\bigskip

\scalebox{0.7}{
\begin{tabular}{|c|c|c|c|c|c|}\hline
\multicolumn{6}{|c|}{BDT (optimised for $\lambda_{hhh}/\lambda_{SM}=5$)}\\\hline
$\lambda/\lambda_{SM}$ & \makecell{Signal cross-\\ section (fb)} & Efficiency & Signal yield & Background yield & $S/\sqrt{B}$ \\\hline 
 
$-1$      &  $0.40$    &   $0.060$    &  $72.00$    & \multirow{7}{*}{$455.51$} &  $3.37$ \\  \cline{1-4} \cline{6-6} 
$1$       &  $0.11$    &   $0.068$    &  $21.42$    &                           &  $1.00$ \\  \cline{1-4} \cline{6-6} 
$2$       &  $0.05$    &   $0.073$    &  $10.95$    &                           &  $0.51$ \\  \cline{1-4} \cline{6-6}
$5$       &  $0.26$    &   $0.046$    &  $35.88$    &                           &  $1.69$ \\  \cline{1-4} \cline{6-6}
$7$       &  $0.70$    &   $0.047$    &  $98.70$    &                           &  $4.62$ \\  \cline{1-4} \cline{6-6} \hline
\end{tabular}
}

\quad
\bigskip

\scalebox{0.7}{
\begin{tabular}{|c|c|c|c|c|c|}\hline
\multicolumn{6}{|c|}{BDT (optimised for each $\lambda_{hhh}$)}\\\hline
$\lambda/\lambda_{SM}$ & \makecell{Signal cross-\\ section (fb)} & Efficiency & Signal yield & Background yield & $S/\sqrt{B}$ \\\hline  

$-1$      &  $0.40$    &   $0.049$    &  $58.80$   & $166.13$    & $4.55$  \\\hline                                       
$1$       &  $0.105$   &   $0.052$    &  $16.46$   & $87.05$     & $1.76$  \\\hline    
$2$       &  $0.05$    &   $0.068$    &  $10.20$   & $85.54$     & $1.10$  \\\hline  
$5$       &  $0.26$    &   $0.046$    &  $35.88$   & $455.51$    & $1.69$  \\\hline
$7$       &  $0.70$    &   $0.049$    &  $102.90$  & $466.97$    & $4.76$  \\\hline

\end{tabular}
}
\caption{Table showing the cross-sections, signal efficiencies, signal and background yields and
significances as a function of $\lambda_{hhh}/\lambda_{SM}$ for (a) cut and count analysis optimised 
for $\lambda_{SM}$, (b) BDT analysis optimised for $\lambda_{SM}$ and (c) BDT analyses optimised for
each $\lambda_{hhh}$.}
\label{tab:lambda}
\end{table}
\end{center}

\section{Contaminations to non-resonant di-Higgs processes}
\label{sec3}

Measuring the trilinear Higgs coupling has been the primary focus for all di-Higgs searches. However, as we have seen in details in
the previous section, the SM Higgs pair production cross-section being extremely small, makes it a challenging job to look for its 
signatures even at the HL-LHC. In the previous section, we found that the combined significance upon assuming zero systematic 
uncertainties is $\sim 2.1\sigma$. However, up until now, we reserved ourselves from introducing any BSM effects. We saw that the 
number of signal events (or rather the $S/B$) is small for most of the final states and hence small contributions from any BSM 
physics can potentially distort or contaminate the signal. Statistically significant deviations from the expected SM di-Higgs yields may be 
considered as signatures of new physics. On the one hand, such deviations can be attributed solely to modifications in $\lambda_{hhh}$ 
or $y_t$ with respect to their SM values. On the other hand, markedly different new physics processes can also be responsible for the 
modification in the event rate in a particular production mode. Having performed boosted decision tree analyses designed solely to 
maximise the SM di-Higgs yield, a fair question to ask at this stage is whether any new physics can at all mimic the SM signatures. The 
answer is twofold. If perchance the primarily discriminatory kinematic variables of the new physics scenario in question, overlap with 
their SM counterparts to a good degree, then there is a good chance of the new physics mimicking this SM signal. Secondly, even if the 
overlap is not significant then the largeness of the new physics cross-section may determine the degree of contamination. The purpose 
of this section is to study some such imposters ensuing from various well-motivated new physics scenarios which may potentially 
contaminate the non-resonant SM Higgs pair event yields in various final states. We will study the extent of these contaminations upon 
considering various benchmark scenarios. We will also find correlated channels during our quest of extracting the effects of 
contamination. The effect of correlation simply means that some search channels for the non-resonant di-Higgs searches will allow for 
more contaminating new physics scenarios compared to some other search channels. Broadly, the following are the three scenarios which 
can contaminate the non-resonant Higgs pair production in certain final states:

\begin{itemize}
\item Double Higgs production, $p p \to hh (+ X)$ through resonant or non-resonant production modes,
\item Single Higgs production in association with some other particles, $p p \to h+X$ and
\item Null Higgs scenario, $p p \to X$, yielding some of the final states as has been discussed in section~\ref{sec2},
\end{itemize}
where $X$ is an object or a group of objects not coming from an SM Higgs boson decay. In the following subsections we detail these three
broad scenarios citing examples from specific new physics models.

\subsection{The $hh(+X)$ channels}
\label{sec3:1}
Several extensions of the SM, primarily with an extended Higgs sector, may significantly enhance the Higgs pair production cross section
and may also alter the kinematics of certain observables. More specifically, two Higgs doublet models (2HDM)~\cite{Hespel:2014sla,Bian:2016awe}
and complex scalar extensions~\cite{No:2013wsa,Kotwal:2016tex,Huang:2017jws} are some prime examples. In the type-II 2HDM 
scenarios, which can be embedded in an MSSM, there is a $CP$-even Higgs, a $CP$-odd Higgs and two charged Higgs bosons on top of the 
SM-like Higgs with $m_h=125$ GeV. The SM-like Higgs pair can be produced from the decay of a heavy $CP$-even Higgs boson, $H$. The couplings of the various Higgses in 2HDM scenarios 
depend mainly on the Higgs mixing parameter, $\alpha$ and the ratio of the two vacuum expectation values (vevs), $\tan{\beta}$ of the 
two Higgs doublets. In order to abide by the LHC results and constraints pertaining to the discovered scalar at $\sim 125$ GeV, one has 
to invoke the so-called alignment limit, where the lightest $CP$-even Higgs automatically aligns itself with the SM-like Higgs, 
having couplings close to the SM predictions. The allowed masses of the pseudo-scalar ($A$) and the $CP$-even heavy Higgs lie in the 
range of a few hundred GeVs. In the low $\tan{\beta}$ regime, the rate for the $CP$-even heavier Higgs decaying to a pair of SM-like 
Higgs bosons can become significant and may even surpass the SM di-Higgs cross-section~\cite{Hespel:2014sla,Bian:2016awe}. The resonant production 
of a heavy $CP$-even Higgs can, in principle, contaminate the SM di-Higgs signal thus affecting the measurement of the Higgs 
self-coupling. In particular, the low $\tan{\beta}$ region can affect the Higgs trilinear coupling measurement. For large $\tan{\beta}$, 
the $H \to bb$ and $H \to \tau \tau$ modes become dominant as the coupling scales as $m_b (m_\tau) \times \tan \beta$. Hence, we do not 
concern ourselves with the large $\tan{\beta}$ regime. We must also note that high $\tan{\beta}$-low $m_A$ regions are 
excluded~\cite{Aaboud:2017sjh}.

\begin{center}
\begin{figure}
\centering
\includegraphics[scale=0.33]{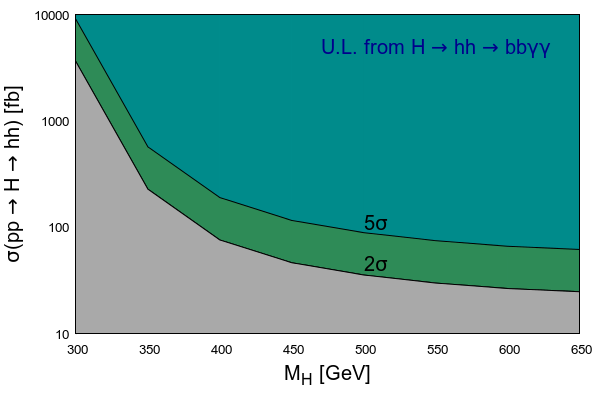}\includegraphics[scale=0.33]{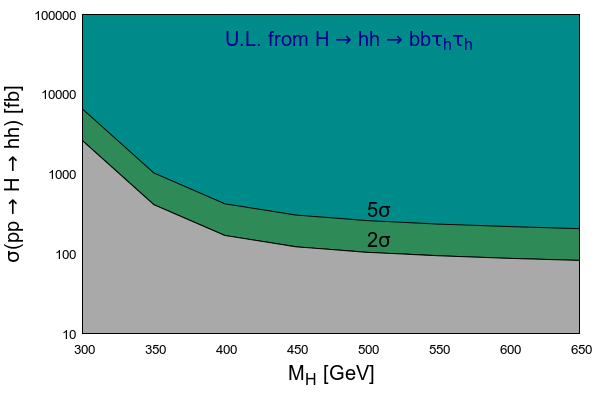}\\
\includegraphics[scale=0.33]{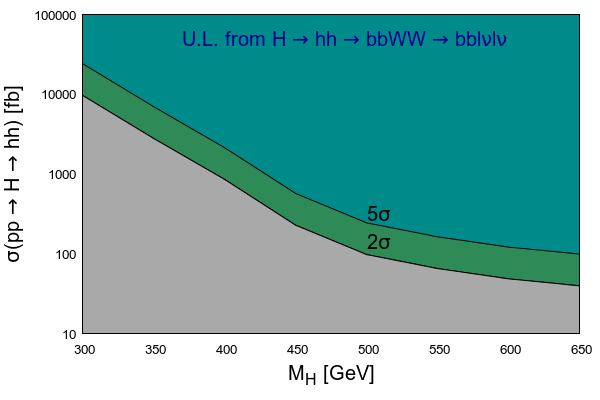}\includegraphics[scale=0.33]{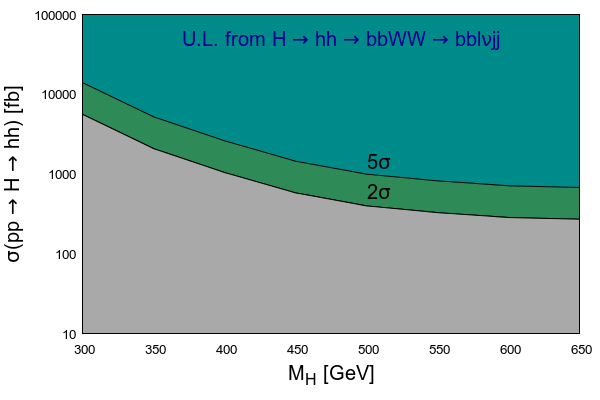}\\
\includegraphics[scale=0.33]{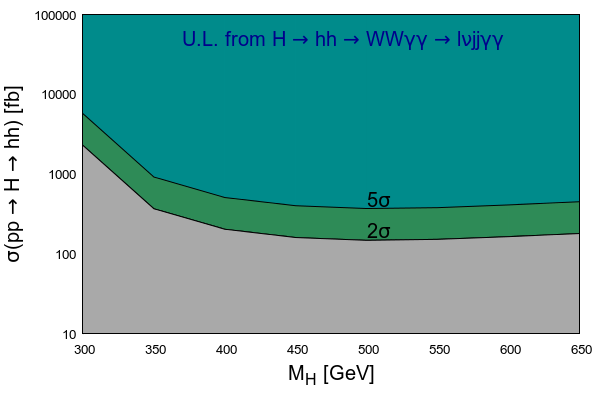}\includegraphics[scale=0.33]{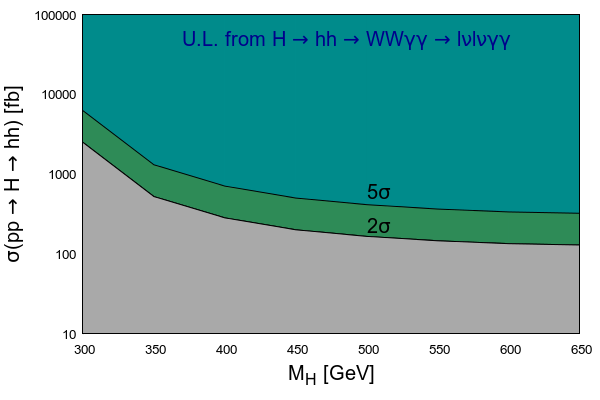}\\
\includegraphics[scale=0.33]{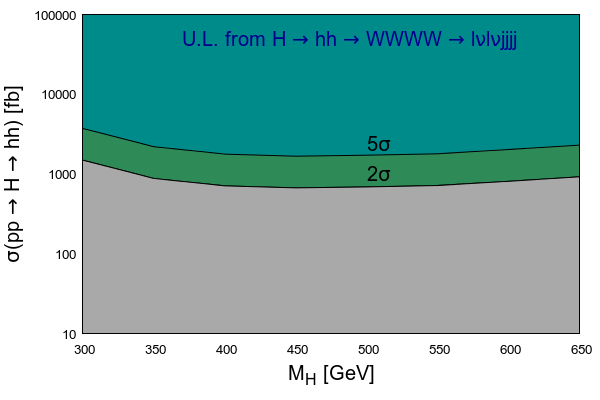}\includegraphics[scale=0.33]{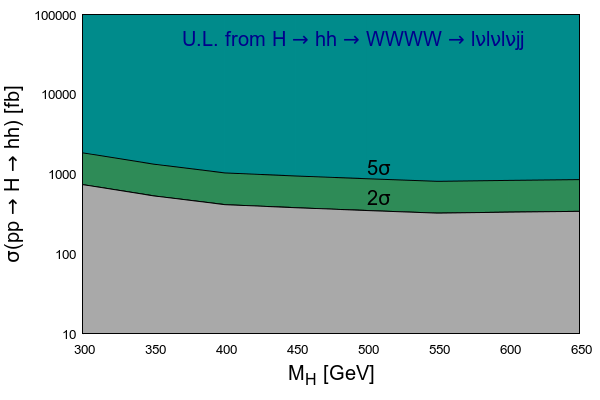}
\caption{Upper limits on $\sigma(pp\to H \to hh)$ [fb] from searches corresponding to various final states, as functions of $m_{H}$ 
[GeV].}
\label{fig:Htohh}
\end{figure}
\end{center}

In order to study the contamination from the process $p p \to H \to h h$, we generate the 
signal samples in \texttt{Pythia-6} and demand a narrow-width for $H$, \textit{i.e.}, in the GeV range, less than the detector resolution. 
The results are shown in Fig.~\ref{fig:Htohh} as upper limits on the cross-section $p p \to H$ times the branching ratio of $H \to h h$,
\textit{viz.}, $\sigma(p p \to H \to h h)$, as functions of the heavy Higgs mass, $m_H$. We try to present the results in a somewhat
model independent fashion. One can imagine the effects of $\tan{\beta}$ or any other theory parameter to have been absorbed in the 
upper limit of the cross-section. The green (blue) region signifies the upper limit on the cross-section required to contaminate the SM 
yield at 2$\sigma$ (5$\sigma$), where the cross-section upper limits are derived using the inequality
\begin{equation}
 S_{\textrm{NP}}^{\textrm{UL}}/\sqrt{B_{\textrm{SM}}} \geq N\sigma,
\end{equation}
where $S_{\textrm{NP}}^{\textrm{UL}}$ is the computed upper limit at $N\sigma$ on the new physics (NP) scenario upon considering a 
background which includes the SM di-Higgs contribution as well.
The grey region is part of the new physics parameter space which does not contaminate the SM 
expectations. As we know, the invariant mass of the SM di-Higgs system peaks around 400 GeV and hence because
of our robust BDT optimisation, which captures to a very precise degree the shape of the non-resonant SM observables, a heavy Higgs boson of 
mass $m_H \lesssim 400$ GeV gets literally treated as a background. Hence, as seen in Fig.~\ref{fig:Htohh}, one requires larger 
cross-sections for $m_H \lesssim 400$ GeV in order to contaminate the SM signal even at the 95\% confidence level. We see that the 
strongest bound on the upper limit on $\sigma(p p \to H \to h h)$ comes about from the $b \bar{b} \gamma \gamma$ channel. The upper 
limit varies between 76 fb and 25 fb between $m_H = 400$ GeV and 650 GeV. This is followed by $b\bar{b}\tau^+\tau^-$. We find the 2$\sigma$
upper limit on the cross-section varying between 170 fb and 83 fb for the aforementioned mass range. The limit is also considerably strong in the fully 
leptonic decay of $b\bar{b}WW^*$, varying between 228 fb and 40 fb for $m_H$ varying between 450 GeV and 650 GeV. The upper limits from the 
$WW^*\gamma \gamma$ channels are fairly strong as well. The 2$\sigma$ upper limit plateaus between 129 fb and 282 fb for the fully leptonic case. Bounds from the other 
modes, especially from the $4W$ modes are much weaker. Hence, we see that the channels where we obtained the best $S/\sqrt{B}$ 
values have the strongest bounds on the upper limits of the cross-section. Thus, for the best optimised modes, one requires lesser cross-sections from 
the heavy Higgs production in order to contaminate the non-resonant Higgs pair production. We must emphasise once again that our BDT 
optimisation was done solely for the SM non-resonant Higgs pair production modes and this subsection is only showing the effects of the 
new physics contamination to the SM signal. In order to search for such a resonance, one needs to redo the optimisation upon treating 
it as a signal. This will be the subject matter of our forthcoming work. To summarise this part, we find that an order 100 fb of 
cross-section for a resonant Higgs mass $\gtrsim 400$ GeV will contaminate the SM di-Higgs expectation to at least 2$\sigma$.
Similarly, Higgs pair production in supersymmetric models~\cite{Ellwanger:2013ova,Cao:2013si,Bhattacherjee:2014bca,Costa:2015llh} are 
also very well motivated. To put things into perspective, in this work we restrict ourselves to MSSM which predicts supersymmetric 
partner(s) for each SM particle. The theory also requires two Higgs doublets. The decays of some of the supersymmetric scalar particles 
result in the SM-like Higgs along with their fermionic counterparts. The processes which can contaminate the di-Higgs search channels, 
other than the heavy Higgs resonance mentioned above, come from the squark (anti-squark) pair production. Although LHC has already 
imposed stringent bounds on the first and second generation squark masses, \textit{viz.}, $\geq \mathcal{O}(\textrm{TeV})$, still this particular channel can attain sizeable cross-sections owing to 
the strong couplings and contribution from each light flavour. We choose a benchmark point (BP1) to study squark pair production 
($\tilde{q}_{L}\tilde{q}_{L}, \tilde{q}_{L}\tilde{q}_{L}^{*}, \tilde{q}_{L}^{*}\tilde{q}_{L}^{*}$) followed by subsequent decay of the 
squark to a light quark and Higgs boson accompanied by $\chi_{1}^{0}$. This yields a final state of $h h + \met + \; \textrm{jets}$. 
In Table~\ref{tab:susy_bpt}, we list three benchmark points which are still allowed by all experimental constraints,
particularly the constraints coming from the Higgs mass and couplings measurements. The first of these 
is relevant for our discussion in this subsection. The common parameters for the three benchmark points are as follows:
\begin{eqnarray}
M_A = 1000 \; \textrm{GeV},~\tan{\beta} = 10,~A_t = 2500 \; \textrm{GeV}, \nonumber \\ 
~m_{\tilde{Q}_{3\ell}} =  m_{\tilde{b}_{R}} = 3000 \; \textrm{GeV}, ~A_{b} = A_{\tau} = 0,~M_3 = 3000 \; \textrm{GeV}. \nonumber
\end{eqnarray}

\begin{center}
\begin{figure}
\centering
\includegraphics[scale=0.33]{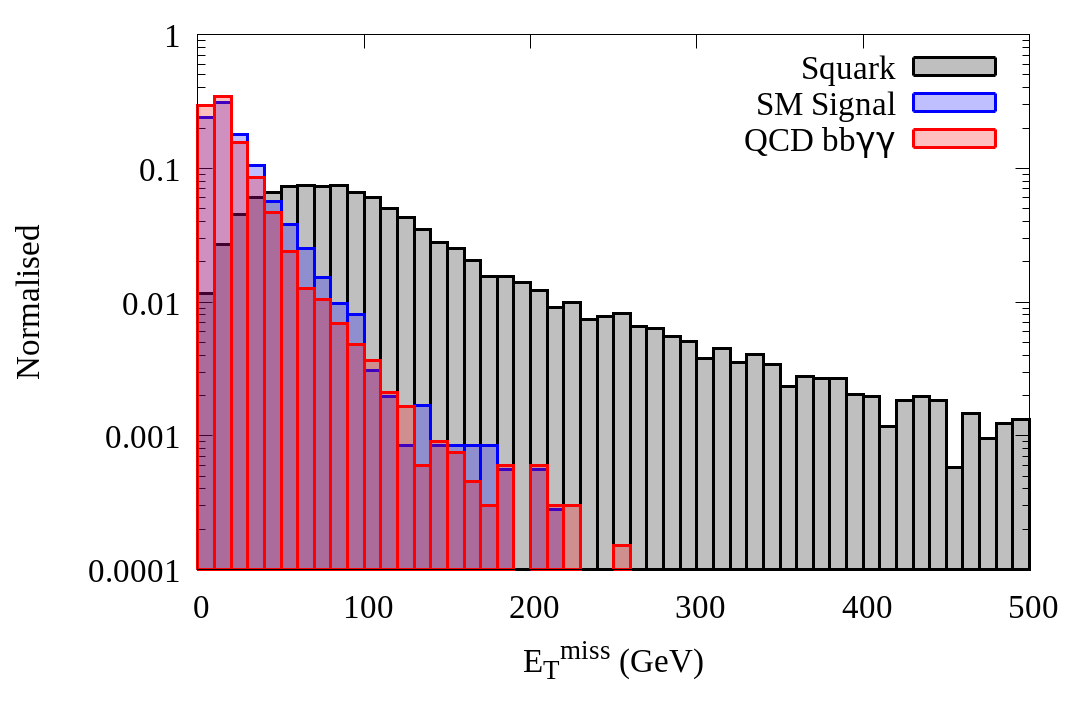}
\caption{Normalised $\met$ distribution for SM di-Higgs pair production, dominant QCD+QED background and for BP1 in the $b\bar{b} \gamma 
\gamma$ final state.}
\label{fig:sqsq}
\end{figure}
\end{center}

\begin{center}
\begin{table}
\begin{tabular}{|c|c||c||c|c|} \hline
Benchmark                                                                    & Parameters (GeV)                              & Mass (GeV)                 & Processes                                  & Branching \\ 
Points                                                                       &                                               &                            &                                            & Fraction\\ \hline\hline
                                                                             & $M_1 = 700, M_2 = 840$                            & $m_{\tilde{u}_{L}}=850.1$  & $\tilde{u}_{L} \to \chi_{2}^{0} u_{L}$     & 13.8\% \\ 
BP1                                                                          & $\mu = 3000, m_{\tilde{t}_{R}} = 3000$           & $m_{\tilde{d}_{L}}=850.1$  & $\tilde{d}_{L} \to \chi_{2}^{0} d_{L}$     & 15.4\% \\
$pp \to \tilde{q}_{L}^{(*)}\tilde{q}_{L}^{(*)}$                              &                                               & $m_{\tilde{c}_{L}}=850.1$  & $\tilde{c}_{L} \to \chi_{2}^{0} c_{L}$     & 13.8\% \\
(Cross-section:                                                              &                                               & $m_{\tilde{s}_{L}}=850.1$  & $\tilde{s}_{L} \to \chi_{2}^{0} s_{L}$     & 15.4\% \\
128.5 fb)                                                                      &                                               & $m_{H}=1000.0$             & $\chi_{2}^{0}  \to \chi_{1}^{0} h$         & 98.7\% \\
$\tilde{q}_{L} = \tilde{u}_{L}, \tilde{d}_{L}, \tilde{c}_{L}, \tilde{s}_{L}$ &                                               & $m_{H^{\pm}}=1003.0$       &                                            &        \\ 
                                                                             &                                               & $m_{\chi_{2}^{0}}=836.0$   &                                            &        \\
                                                                             &                                               & $m_{\chi_{1}^{0}}=700.0$   &                                            &        \\ \hline 
                                                                             & $M_1 = 150, M _{2} = 300$                     & $m_{\chi_{2}^{0}}=296.7$   & $\chi_{1}^{\pm} \to \chi_{1}^{0}\,W^{\pm}$ & 100\%  \\ 
BP2                                                                          & $\mu = 1000, m_{\tilde{t}_{R}} = 3000$        & $m_{\chi_{1}^{\pm}}=296.7$ & $\chi_{2}^{0} \to \chi_{1}^{0}\,h$         & 93.5\% \\
$pp \to \chi_{1}^{\pm} \chi_{2}^{0}$                                         &                                               & $m_{\chi_{1}^{0}}=149.3$   &                                            &        \\
(Cross-section:                                                              &                                               & $m_{h} = 125.0$            &                                            &        \\
420 fb)                                                                      &                                               & $m_{H^{\pm}} = 1003.0$     &                                            &        \\
                                                                             &                                               & $m_{H} = 1000.0$           &                                            &        \\ \hline
                                                                             & $M_1 = 500, M _{2} = 1000$                    & $m_{\tilde{t}_{1}}=609.3$  & $\tilde{t}_{1} \to \chi_{1}^{0}\,b\,W^{+}$ & 99.9\% \\ 
BP3                                                                          & $\mu = 1000, m_{\tilde{t}_{R}} = 625$         & $m_{\chi_{1}^{0}}=498.1$   &                                            &        \\
$pp \to \tilde{t}_{1}\tilde{t}^{*}_{1}$                                      &                                               & $m_{h} = 125.0$            &                                            &        \\
(Cross-section:                                                              &                                               & $m_{H^{\pm}} = 1003.0$     &                                            &        \\
200 fb)                                                                      &                                               & $m_{H} = 1000.0$           &                                            &        \\
                                                                             &                                               &                            &                                            &        \\ \hline
\end{tabular}
\caption{SUSY benchmark points for studying effects of contamination on SM di-Higgs yields at the HL-LHC.}
\label{tab:susy_bpt}
\end{table}
\end{center}
From BP1, we see that the cross-section of $hh+X$ is $\sim 10.8$ fb, which is less than a third of the SM expectation. Moreover, we find
that the $\met$ distribution from the squark pair production is significantly different from the signal as well as from the dominant SM 
background, as shown in Fig.~\ref{fig:sqsq}. After applying the BDT cuts for the $b\bar{b} \gamma \gamma$ analysis, we are left with 
$\sim 0.60$ events, which is much smaller compared to the SM expectation and not statistically significant. Hence in order to minimise the contamination to the
$b\bar{b} \gamma \gamma$ final state ensuing from an SM di-Higgs production, one may perhaps impose certain exclusive cuts, especially 
on the $\met$ distribution. This will help reduce new physics contaminations with large $\met$. Moreover, for certain SUSY
scenarios, we may have cascade decays giving rise to multiple jets. Hence, the cut $N_j < 6$ can come in handy to reduce such 
backgrounds and we may also require to optimise this cut further in order to reduce such contamination effects. In other words, 
removing contamination effects can be tricky and can be somewhat model dependent if we are studying inclusive final states.

\subsection{The $h+X$ channels}
\label{sec3:2}
In the previous subsection, the heavy resonance production and the di-Higgs production ensuing from subsequent decays of a pair of 
(anti-)squarks, potentially contaminate all the SM di-Higgs search channels that are studied in section~\ref{sec2}. In this subsection,
we will look into two specific candidates which will contaminate some di-Higgs final states and not the others. After the HL-LHC run
if one finds excesses in certain di-Higgs like final states and not in the others, then it might be possible to narrow down the new
physics possibilities to a greater degree.

\begin{center}
\begin{figure}
\centering
\includegraphics[scale=0.33]{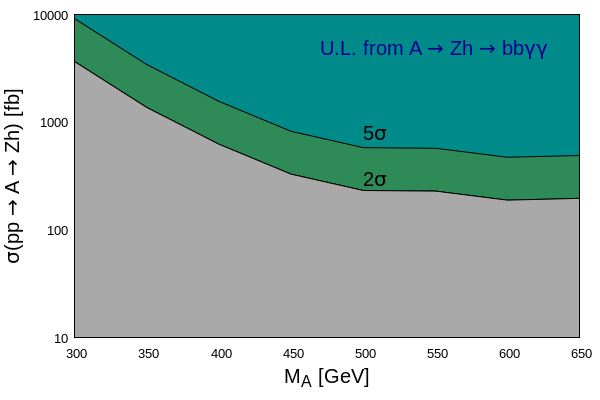}\includegraphics[scale=0.33]{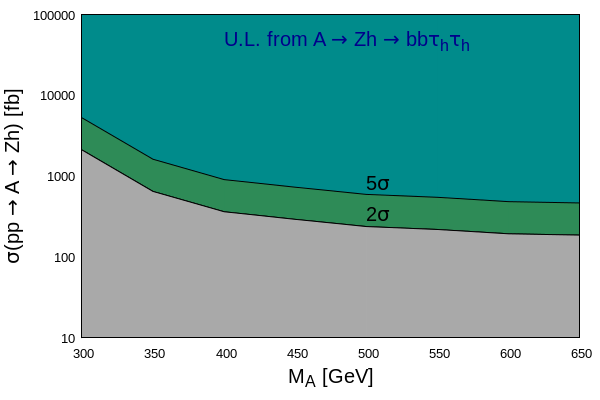}\\
\includegraphics[scale=0.33]{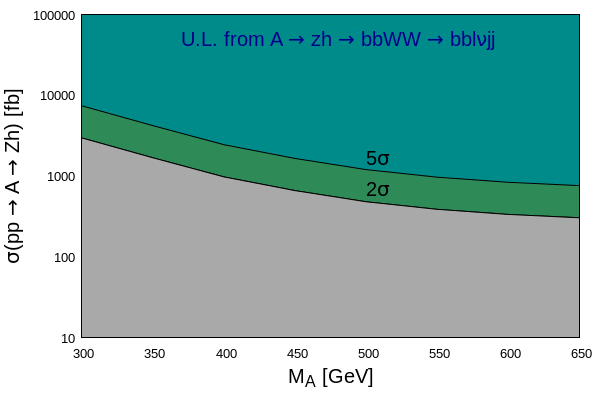}\includegraphics[scale=0.33]{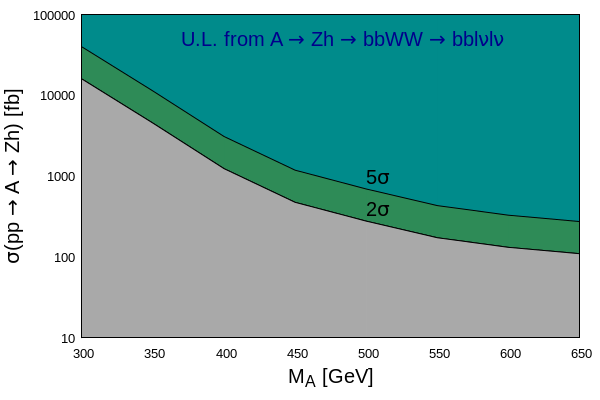}\\
\includegraphics[scale=0.33]{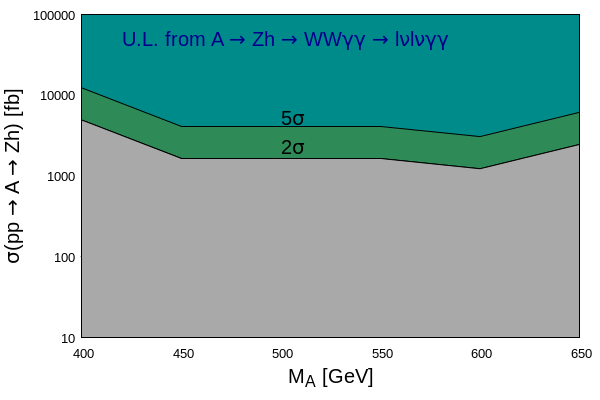}\includegraphics[scale=0.33]{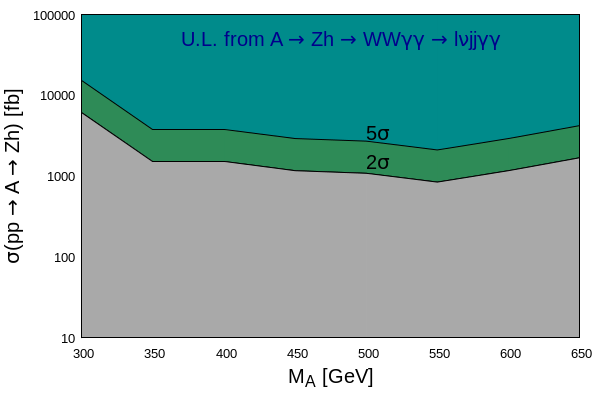}\\
\includegraphics[scale=0.33]{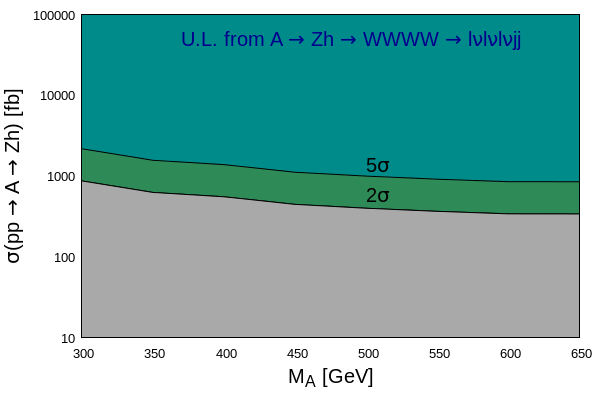}\includegraphics[scale=0.33]{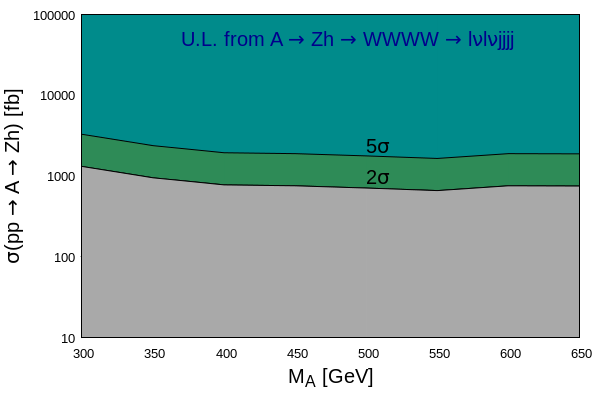}
\caption{Upper limits on $\sigma(pp\to A \to Zh)$ [fb] from searches corresponding to various final states, as functions of $m_{A}$ 
[GeV].}
\label{fig:AtoZh}
\end{figure}
\end{center}

In 2HDMs, a resonant production of the pseudoscalar Higgs production, \textit{viz.}, $pp \to A \to Z h$ followed by $Z$ and $h$ decaying 
to all possible final states, can, in principle, imitate various final states as shown in Fig.~\ref{fig:AtoZh}. The decay rate of the 
pseudoscalar, $A \to Z h$ is appreciable with $M_A$ below the $t \bar{t}$ threshold and for low values of $\tan \beta \; (\lesssim 5)$. 
The upper limits on the cross-sections are weaker than those from the resonant scalar production. One of the strongest bounds arise 
from $b \bar{b} \gamma \gamma$, varying from 330 fb (450 GeV) to around 197 fb (650 GeV). The strongest upper limits, however, comes 
from the $b\bar{b} \tau^+ \tau^-$ search, varying between 292 fb and 186 fb in the aforementioned mass range. For the di-leptonic $b\bar{b}WW^*$ channel, 
the bound strengthens from 1236 fb at $m_A=400$ GeV to $\sim$ 110 fb for $m_A=650$ GeV. From the final state tailored for 
the $3\ell$ mode coming from the $4W$ scenario, the 2$\sigma$ upper limit varies between 555 fb (400 GeV) and around 341 fb (650 
GeV). The upper-limits on the cross-section required for contamination from the remaining final states are rather weak. 
In summary, the $A \to Z h$ channel contaminates in a slightly weaker fashion as compared to the $H \to h h$ channel. One of the possible 
reasons is that the reconstructed $Z$-peak is shifted from the reconstructed Higgs peak as $m_{bb}$ serves as an important discriminatory 
variable in all the searches involving a $b$-jet pair. Hence, more cross-section is required here in order to contaminate the SM di-Higgs 
channels to a similar degree as in the $H \to h h$ channel. As an aside, we would like to mention that the process $p p \to A h$ may 
also potentially contaminate the same final states as for the $A \to Z h$ case. We however, do not consider the details of this channel, for brevity.
As an extended scenario, we now shift our focus to supersymmetry. In MSSM, electroweakino pair production often results in mono-Higgs 
type signals. LHC has come down heavily on such SUSY scenarios constraining much of its parameter space. The bounds on squarks and 
gluino masses have already surpassed a TeV. In this situation, the observation of a SUSY signature will heavily rely on its electroweak 
sector, composed of charginos ($\chi_i^{\pm}$) and neutralinos ($\chi^{0}_j$). In the presence of a decoupled Higgs sector, the 
chargino-neutralino pair production is mediated through the $W$-boson propagator, with the $W^{\pm} \chi^{\mp} \chi^{0}_{1}$ coupling 
containing terms which depend on both the wino and the higgsino components of the electroweakinos involved. However, it is to be noted 
that the contributions from the wino components dominate over the contributions from the higgsino terms. ATLAS and CMS have also 
performed searches for chargino-neutralino pair production in the $3 \ell + \met$ and the same-flavour opposite-sign $2 \ell + \met$ 
final states for a non-generic scenario where both $\chi^{\pm}_{1}$ and $\chi_{2}^{0}$ are dominantly wino-like and mass degenerate. 
They have obtained correlated bounds on the masses of LSP and NLSP~\cite{Aad:2015jqa,Khachatryan:2014mma,Sirunyan:2017eie,
Sirunyan:2017zss}~\footnote{Much stronger limits have been obtained from the 13 TeV results from separate final states involving 
$\tau$-leptons~\cite{Aaboud:2017nhr}. We do not however, consider these limits in the present work.}. We carefully select a benchmark 
point where the wino mass parameter, $M_2$ is much smaller compared to the higgsino mass parameter, $\mu$ making the lightest chargino 
and second lightest neutralino, wino-like. A wino-dominated $\chi_{2}^{0}$ and  $\chi_{1}^{\pm}$ yields much larger cross-section for 
the process $p p \to \chi_{2}^{0} \chi_{1}^{\pm}$ compared to other electroweakino production process, for example, $\chi_2^0$ pair 
production etc. Hence, we will not consider the latter process although it can, in principle, mimic di-Higgs signal through cascade 
decay of $\chi_2^0$. The benchmark point (BP2) is tabulated in Table~\ref{tab:susy_bpt} and is marginally outside the projected exclusion
obtained by ATLAS for the HL-LHC~\cite{ATL-PHYS-PUB-2014-010}. In this parameter space $\chi_{2}^{0}$ dominantly decays to $h \chi_{1}^{0}$, while 
$\chi_{1}^{\pm}$ has a 100\% branching ratio to $W^{\pm} \chi_{1}^{0}$. This essentially produces a $Wh +\met$ final state with a 
cross-section of $\sim 400$ fb, thus generating $h+X$ signatures. Hence, the $Wh+\met$ final state from the chargino-neutralino 
pair production can modestly contaminate some of the di-Higgs search channels, \textit{viz.}, the $b \bar{b} W W^* \to b \bar{b} \ell  
jj+ \met$, $\gamma\gamma W W^* \to \gamma\gamma \ell jj + \met$, $4W \to \ell^{\pm} \ell^{\pm} jjjj + \met, 3\ell jj + \met$. In 
Table~\ref{tab:cpm_c0}, we present the event yields for the benchmark point BP2, in three of the concerned di-Higgs channels, 
corresponding to the most optimised BDT score obtained for the non-resonant SM di-Higgs searches. We find that the contaminations are
large in these channels reminding us that a possible future observation of significant number of events in these channels must be 
treated carefully. We also mention here that the SM di-Higgs expectations from these channels are insignificant leading to negligible
signal over background ratios. Thus, observations of significant numbers of events over and above the SM backgrounds can be potential
signatures for new physics.
\begin{center}
\begin{table}[htb!]
\centering
\small
\begin{tabular}{|c|c|c|c|}\hline
Channel             & SM background & SM $hh$ production & BP2 contamination \\ \hline \hline
$bb \ell jj + \met$ & 1103017.13    & 134.34           & 382.88            \\ \hline
SS$2\ell jj + \met$ & 12378.49      & 11.96            & 270.31             \\ \hline
$3 \ell jj + \met$  & 5389.46       & 15.01            & 291.91            \\ \hline
\end{tabular}
\caption{New physics contaminations from chargino-neutralino pair production in the $bb \ell jj + \met$, SS$2\ell + \met$ and $3 \ell jj + \met$ 
final states. The table shows the number of events at the HL-LHC after the MVA cuts.}
\label{tab:cpm_c0}
\end{table}
\end{center}

\subsection{Null Higgs channels}
\label{sec:3:3}
\begin{center}
\begin{figure}
\centering
\includegraphics[scale=0.33]{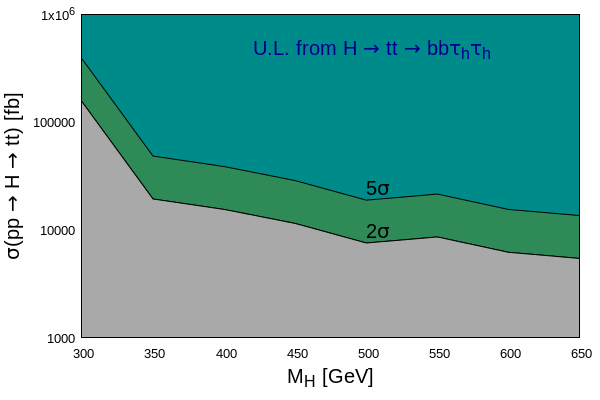}\\
\includegraphics[scale=0.33]{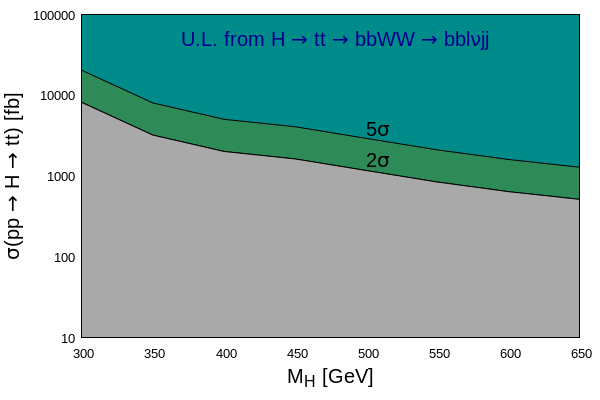}\includegraphics[scale=0.33]{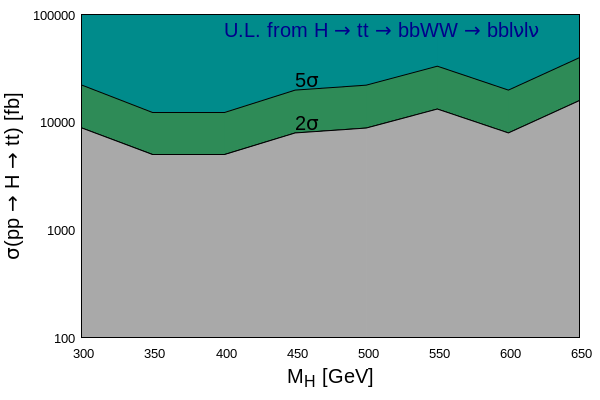}
\caption{Upper limit on $\sigma(pp\to H \to t\bar{t})$ [fb] from searches corresponding to the $b\bar{b} \tau^+ \tau^-$ and 
$b\bar{b} W W^*$ states, as a function of $m_{H}$ 
[GeV].}
\label{fig:Htottbar}
\end{figure}
\end{center}

\begin{center}
\begin{figure}
\centering
\includegraphics[scale=0.23]{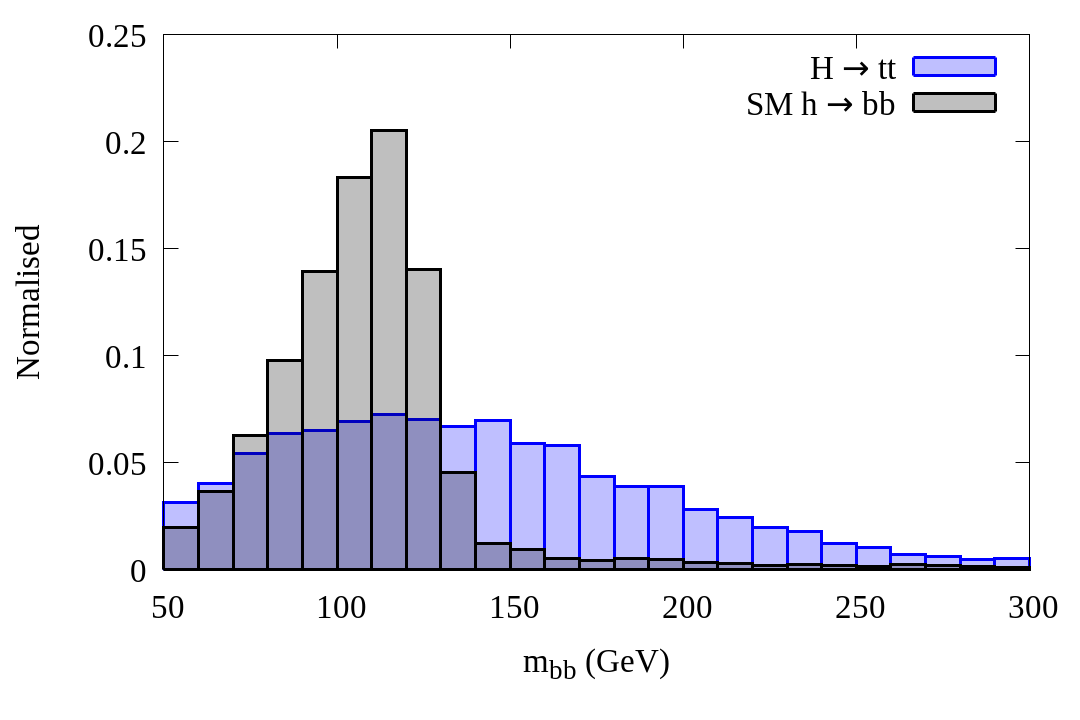}
\caption{Normalised distribution of $m_{bb}$ for $p p \to H \to t \bar{t}$ (shown in blue) and the SM di-higgs signal (shown in black)}.
\label{fig:Htottbar_mbb}
\end{figure}
\end{center}

Before closing this section, we discuss the final category of potential contaminants, \textit{viz.}, the ones with no SM-like Higgs
bosons in the production or decay modes. We start by revisiting the classic heavy resonant (pseudo-)scalar production. This 
(pseudo-)scalar is dominantly produced by the gluon-fusion production mode and in the case where its mass is greater than the $t\bar{t}$ threshold, it can decay to 
a pair of top quarks, the branching ratio depending on the $H(A) t \bar{t}$ Yukawa coupling.
 This channel can potentially contaminate the $b \bar{b} \tau^+ \tau^-$ and $b \bar{b} W W^*$ channels. We find from Fig.~\ref{fig:Htottbar} that the upper limits on the cross-section 
times branching ratio ($\sigma(p p \to H(A) \to t \bar{t})$) from the relatively clean $b \bar{b} W W^* \to b \bar{b} \ell^+ \ell^- +\met$ 
channel, is visibly weak. The upper limits from the semi-leptonic decay mode, \textit{viz.}, $b \bar{b} W W^* \to b \bar{b} \ell +\met + jj$ gives slightly stronger 
2$\sigma$ upper limits on the contamination cross-section, varying between $\sim$ 1.2 pb ($m_H = 500$ GeV)
and $\sim$ 0.5 pb ($m_H = 650$ GeV). The upper limits from $b\bar{b} \tau^+ \tau^-$ also does not fare well. Hence, the $H \to t \bar{t}$ channel 
does not contaminate the SM di-Higgs channels to any considerable degree. One of the prime reasons is the fact that the BDT variable 
$m_{bb}$ is strongly discriminating, peaking at the SM-like Higgs boson mass for the non-resonant Higgs pair production, with the $b$-quark pair from the $t\bar{t}$ mode having a distinct feature as shown in Fig.~\ref{fig:Htottbar_mbb}. 
Hence, one will require a very large production cross-section for the heavy resonant scalar in order to contaminate the SM signature 
significantly.

\begin{center}
\begin{figure}
\centering
\includegraphics[scale=0.33]{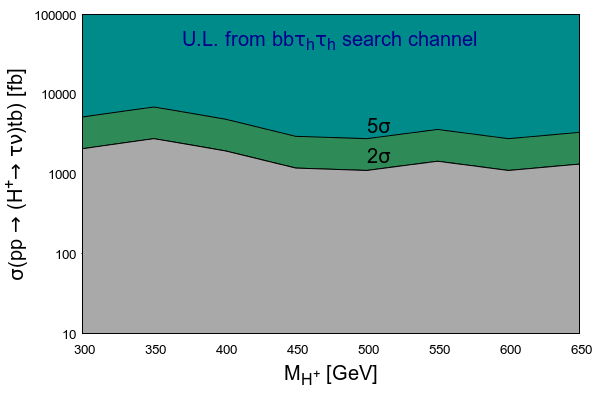}\includegraphics[scale=0.33]{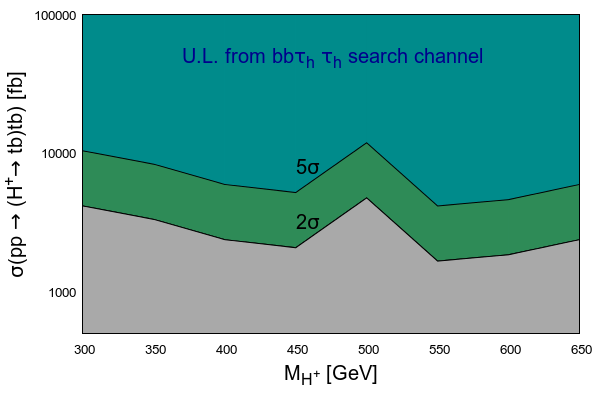}\\
\includegraphics[scale=0.33]{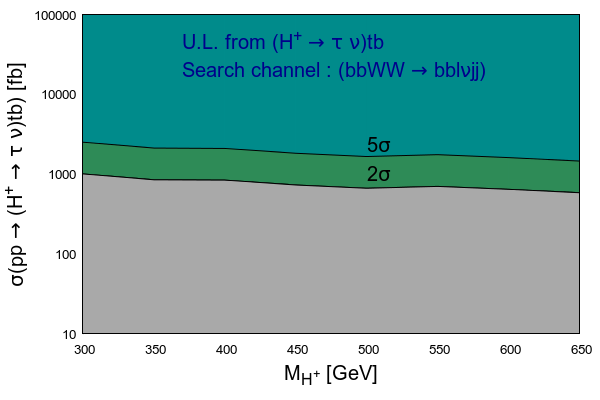}\includegraphics[scale=0.33]{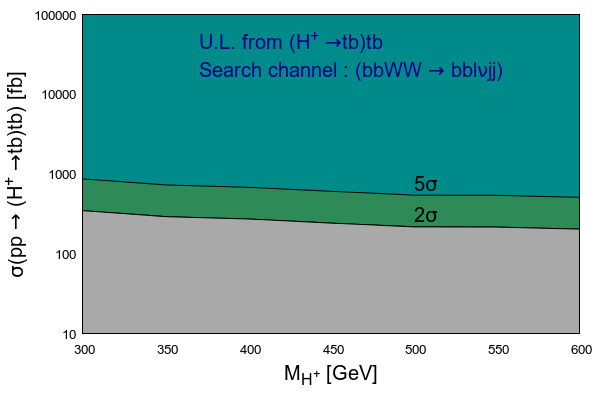}\\
\includegraphics[scale=0.33]{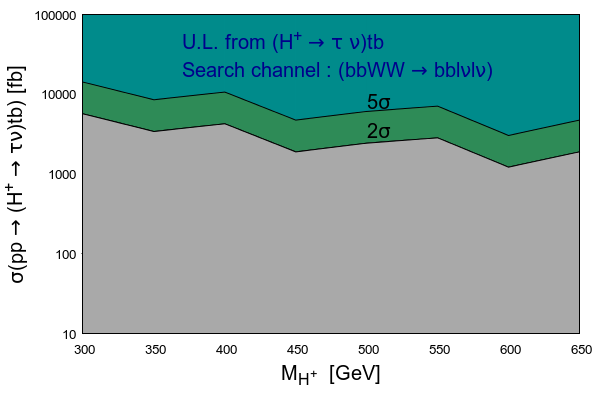}\includegraphics[scale=0.33]{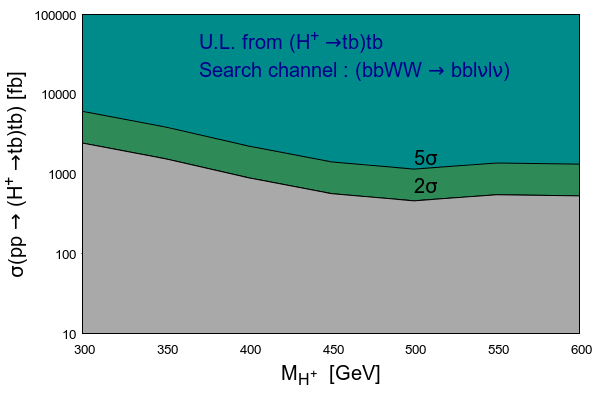}\\
\caption{Upper limits on $\sigma(pp\to \to \bar{t} b H^{+}/t \bar{b} H^-)$ [fb] from searches corresponding to $b \bar{b} W W^*$ and 
$b \bar{b} \tau^+ \tau^-$, as functions of $m_{H^{\pm}}$ [GeV].}
\label{fig:tbHpm}
\end{figure}
\end{center}

Another interesting category can be accommodated in various extensions of the SM involving singly charged Higgs bosons. One can consider 
a scenario where a singly charged Higgs is produced in association with a top quark and a bottom quark, \textit{viz.}, $p p \to 
\bar{t} b H^{+}/t \bar{b} H^-$ and the charged Higgs either decays to $\tau \nu_{\tau}$ or $t \bar{b}$ depending on its mass. These 
channels may adversely contaminate the $b \bar{b} W W^*$ and $b \bar{b} \tau^+ \tau^-$ modes. We find from Fig.~\ref{fig:tbHpm} that 
the $t \bar{b} \bar{t} b$ channel poses the strongest contamination to the $b \bar{b} \ell j j + \met$ final state.
The 2$\sigma$ contamination cross-section for this final state varies between 
393 fb ($m_{H^+} = 250$ GeV) and 204 fb ($m_{H^+} = 650$ GeV). The limits from the 
other channels are weaker. We also note in passing that all the aforementioned processes essentially affect the low $\tan \beta$ region of the parameter space. 
As a final example, we study the stop pair production, $pp \to \tilde{t}_1 \tilde{t}^*_1$ which can potentially mimic some of the 
di-Higgs signatures. The stop pair-production cross-section is fairly large for stop masses of the order of several hundreds of GeVs. 
With an appropriate choice of parameters listed as BP3 in Table~\ref{tab:susy_bpt}, $\tilde{t}_1$ can have a dominant branching ratio 
to $b \chi^{+}_{1}$, with $\chi^{+}_{1}$ eventually decaying to $W^{+} \chi^{0}_{1}$. This gives us a final state of $2b + 2W + \met$
which potentially affects the $h h \to b \bar{b} W W^+$ and $h h \to b \bar{b} \tau^+ \tau^-$ search channels. We choose BP3 such that 
the mass difference between $\tilde{t}_1$ and $ \chi_{1}^{0}$ is less than the top mass, ensuring the stop decays as $\tilde{t}_1 \to 
W b \chi_{1}^{0}$. The final number of events at the HL-LHC for the relevant search channel $ bb \ell jj + \met$ is shown in 
Table.~\ref{tab:stst}. The contamination is found to be of the same order as the SM signal. We also note that the other decay mode of 
stop quarks, \textit{viz.}, $t \chi^0_{1}$, will also give rise to $t \bar{t} + \met$ final state, affecting the same channels.
\begin{center}
\begin{table}[htb!]
\centering
\small
\begin{tabular}{|c|c|c|}\hline
SM background & SM $hh$ production & BP3 contamination \\ \hline \hline
1103017.13    & 134.34             & 101.83 \\ \hline
\end{tabular}
\caption{New physics contamination from stop pair production in the $bbWW \to bb \ell jj + \met$ final state.}
\label{tab:stst}
\end{table}
\end{center}
We must stress here that the entire analysis has been performed using boosted decision tree optimisation techniques which has been 
trained using the SM di-Higgs data samples. Hence, the BDT cuts are very efficient in segregating any contamination, \textit{i.e.}, 
non-SM contributions. Now, if a new physics process is still able to contaminate, then it must be very efficient in passing all the 
cuts. This would mean that it must come with a large production cross section or a considerable overlap with the SM kinematic variables, 
so as to contaminate the SM signal. In other words, we can impose stringent bounds on the cross-sections for various BSM scenarios 
discussed above, which can potentially contribute to the di-Higgs signals. The efficiency of the BDT cuts will, of course, depend on 
the particular channel considered. The bound on some BSM physics can be strong from one channel and may not be so strong from the rest. 
It is important to note that there might be two completely different aspects of interpreting our results. The first case would be where 
we are already aware of the presence of new physics (through some other channel). In such situations, we want to ask whether any new 
physics process might contaminate the di-Higgs signal. If so, we will get an idea of how large the cross section will be for such 
processes and prepare our strategy. The second one is similar to our present situation, where we would be still looking for new physics. 
This is a much more complex scenario as we are looking for new physics in various directions. Our purpose in this work is to classify 
di-Higgs searches in multiple channels in a model independent manner so as to extract the best possible information about potential 
contaminating channels. In this case, we can, at best, put bounds on the cross-sections coming from new physics scenarios. This will 
give us an idea if the measurement of the Higgs self-coupling is possible and if yes, then which channel to look out for.

We wish to conclude this section by reiterating our philosophy for the second part of our study with the following 
observations. In the fortunate case that we discover new physics in the near future, for instance discoveries of heavy Higgs boson(s), 
superpartners of quarks, to name a few, then the measurement of $\lambda_{hhh}$ will be affected because of the effects of contamination 
to the SM channels as have been quantified above. For a possible scenario where we have hints of new physics but these are below the 
discovery significance, then also care must be taken to study the effects of contamination which can tell us more about the viability of
such scenarios. A third possible scenario which we did not look for in this present study is the effects of new physics only modifying
$\lambda_{hhh}$. For such possibilities, it might happen that we will see no new particles and the shapes of the kinematic distributions
involving the Higgs pair production can only shed light on new physics.

\section{Summary and outlook}
\label{sec4}
In the first part of this work, we evaluated the prospects of di-Higgs searches 
in numerous well motivated final states. Optimised cut-based analyses were 
performed for the $b\bar{b}\gamma\gamma$ and $b\bar{b}\tau^+\tau^-$ states. We 
followed this up with multivariate analyses using the boosted decision tree (BDT) 
algorithm for the majority of our search channels. The multivariate analyses 
yielded improved signal to background ratio ($S/B$) and the overall statistical 
significance. The $b\bar{b}\gamma\gamma$ final state presented itself as the 
most promising search channel with a statistical significance of 1.46 (1.76) for 
the cut-based (multivariate) analysis. The $b\bar{b}\tau^+\tau^-$ channel was 
looked for in the fully hadronic, semi-leptonic and leptonic sub-states. This 
channel, even upon having a higher yield as compared to its predecessor, is 
marred by much larger backgrounds and our limitation to reconstruct the $\tau$ 
invariant mass precisely. However, upon employing the collinear mass variable for 
reconstructing the Higgs decaying to a pair of $\tau$s, we finally obtain 
statistical significances of 0.65 (0.74), 0.44 (0.49) and 0.07 (0.08) for the 
cut-based (multivariate) analyses in the hadronic, semi-leptonic and leptonic 
modes respectively. The signal to background ratio improves significantly upon 
using the collinear mass technique. The $b\bar{b}WW^*$ state in the leptonic 
final state serves as a clean channel with a moderate $S/B$ and a statistical 
significance of 0.62. This serves as the third most important contribution 
after the $b\bar{b} \gamma \gamma$ and the fully hadronic $b\bar{b} 
\tau^+ \tau^-$ channels. The semi-leptonic final state for $b\bar{b}WW^*$ pales 
in comparison with a much smaller $S/B$ and a statistical significance of 0.13. 
Both the leptonic ($S/B$= 0.40) and semi-leptonic ($S/B$ = 0.11) final states 
for the $W W^*\gamma \gamma$ channel show great potential for higher-energy and 
higher-luminosity colliders. The limitation in design-luminosity at the HL-LHC 
in addition to the smallness of BR($h \to \gamma \gamma$) forbid us from 
utilising these final states while computing the combined significance. We 
conclude the first part of this work upon considering the $SS2\ell$, $3\ell$ 
and $4\ell$ final states emerging from the $h h \to WW^*WW^*$ search channel. 
The tri-leptonic channel yields a statistical significance of $0.20$,
however, with an insignificant $S/B$. One would require a manifold increase in 
the production cross-section in these three channels for them to become 
noteworthy, even in the future colliders. For all channels with less than 5 
signal events, we were unable to define a statistical significance. A combined 
zero-systematics significance of $\sim 2.1\sigma$ was obtained upon combining all the 
statistically significant signals for the HL-LHC analysis at 14 TeV. The quoted 
significance values can get severely diluted, once systematic uncertainties are 
taken into account.

After this we studied the importance of considering varying 
values of the Higgs trilinear coupling and how it affects our conclusions. We
trained the boosted decision trees with the SM case for once and then with 
each of the $\lambda_{hhh}$ samples and found that one can have a difference
in significance because of the difference in the distributions of certain
kinematic variables. We faithfully recover the expected
exclusion on the Higgs trilinear coupling for the HL-LHC, as computed by
ATLAS, upon using a log-likelihood CLs hypothesis for the $\lambda_{SM}$
BDT optimisation. Upon changing the training to a different value of $\lambda$
and also upon choosing a hypothesis different from that of the SM, we obtain
stronger upper limits.

In the final chapter of this work, we analysed some new physics scenarios which 
may potentially contaminate the SM di-Higgs search channels. We used the same 
multivariate training and cut on the BDT variable for the new physics cases as 
have been obtained for the SM non-resonant di-Higgs searches, in order to 
estimate the contaminations. Three major contamination scenarios were studied, 
\textit{viz.}, $hh(+X)$, $hX$ and $X$, $X$ being a set of objects not ensuing 
from the SM-like Higgs, and upper limits on the production cross-section of 
heavy scalar ($H$), pseudoscalar ($A$) and charged Higgs ($H^{\pm}$) bosons were 
obtained. In particular, we derived upper limits on $\sigma(pp\to H \to hh)$, 
$\sigma(pp\to A \to Z h)$, $\sigma(pp\to H \to t\bar{t})$ and $\sigma(pp \to 
H^{+} t\bar{b} \to t\bar{b} (\tau\nu)t\bar{b})$ for the various search channels. 
The $b\bar{b}\gamma\gamma$ channel emerged as the most sensitive search channel, 
with results indicating that for $m_{H} = 500~{\rm GeV}$, a production cross-section 
of $\sigma(pp \to H \to hh)\sim 36~{\rm fb}$ would result in a $2\sigma$-level 
of contamination to the SM search. This is closely followed by the $b\bar{b}\tau^+\tau^-$ 
channel, putting an upper limit of 104 fb for the same resonance mass. The 
limits from the leptonic decay mode of $b\bar{b}WW^*$ also present competitive 
upper limits with $\sigma(p \to H \to hh)$ attaining values of 
$\sim 98~{\rm fb}$ at $m_{H}=500~{\rm GeV}$ for a $2\sigma$-level 
contamination. The upper limits from the remaining decay channels are 
$\sim 5-10$ times weaker. In the resonant $A\to Zh$ search, the $b\bar{b}\gamma\gamma$ 
mode presents the strongest upper limit on the cross-section at 233 fb with 
$m_A=500$ GeV. The $b\bar{b} \tau^+ \tau^-$ mode closely follows with a contaminating 
cross-section of 238 fb for the same mass of the pseudoscalar. The di-leptonic 
final state for the $b\bar{b}WW^*$ channel also imposes upper limit of the same 
order. Next, we derived upper limits on $\sigma(pp \to H \to t\bar{t})$, and the 
results were found to be significantly weaker than the previous scenarios. The 
2$\sigma$ upper limits derived for the charged Higgs production also exhibit similar 
results, with the semi-leptonic $b\bar{b}WW^*$ channel offering the best 
sensitivity with cross-section requirements of the order of 217 fb for $m_{H^+} = 500$ 
GeV, in $H^+\to tb$ mode. The epilogue to this story is provided by the contaminations from various SUSY processes. Here, we had chosen three 
experimentally viable benchmark points, optimised for squark pair production, chargino-neutralino pair production and stop pair 
production, with subsequent cascade decay modes mimicking various di-Higgs final states. Of particular interest is the contribution 
from the $\chi_{2}^{0}-\chi^{\pm}_{1}$ pair production which may significantly contaminate the $SS2\ell$ and $3\ell$ final states in 
the $h h \to 4W$ channel, and the semi-leptonic decay mode of the $b\bar{b}WW^*$ channel, with event yield much higher than the
corresponding SM di-Higgs signal. It would be logical to argue that the presence of such SUSY signatures would lead to a clear and 
strong contamination in these di-Higgs final state searches paving an interesting and complicated road ahead for the search of Higgs 
trilinear coupling.

As seen in this work, the prospects of discovering di-Higgs signals for a SM-like scenario is extremely difficult owing to the smallness
of the production cross-section and the overwhelmingly large backgrounds. However, many of the search channels considered must motivate
the particle physics community to either aim for higher integrated luminosities, beyond 3 ab$^{-1}$ or to build higher energy colliders,
\textit{viz.}, a 28 TeV/33 TeV and ideally 100 TeV machines. Even in our present setup, in all probability, the sensitivities can be 
improved upon having a better handle over the backgrounds by either minimising the uncertainties due to the Monte-Carlo computation 
order or by adopting data driven backgrounds. Besides, there might be certain novel discriminatory variables or certain boosted 
techniques which might help in reducing the backgrounds further. We also learnt from this study that looking for di-Higgs search channels may
in principle be masked by new physics effects. For such scenarios our multivariate optimisation tries the best to separate the 
SM-signal from the new physics effects. However, in certain cases, due to similarities in kinematic distributions with the SM 
counterparts or due to a large cross-section yield, we may have considerable contamination effects. The techniques outlined in this
paper can be easily extended and optimised as searches for the various new physics effects listed above.

\acknowledgments
We thank Amit Chakraborty, Mikael Chala, Christoph Englert, Christopher Lester, Michele Selvaggi and Christian Veelken for helpful discussions at various stages of the work. This work was supported in part by the CNRS LIA-THEP (Theoretical High Energy Physics) and the INFRE-HEPNET (IndoFrench Network on High Energy Physics) of CEFIPRA/IFCPAR (Indo-French Centre for the Promotion of Advanced Research). SB is supported by a Durham Junior Research Fellowship COFUNDed between Durham University and the European Union under grant agreement number 609412. SB acknowledges the hospitality of the Indian Institute of Science, Bangalore, where the project was conceived. The work of BB is supported by the Department of Science and Technology, Government of India, under the Grant Agreement number IFA13-PH-75 (INSPIRE Faculty Award). BB acknowledges the hospitality of LAPTh where parts of this work were carried out.

\bibliographystyle{JHEP}
\bibliography{refs}

\newpage

\appendix
\section{Appendix A}
\label{sec:appendixA}

\begin{table}[htb!]
\centering
\begin{bigcenter}
\scalebox{0.6}{%
\begin{tabular}{|c|c|c|c|c|}

\hline
Process & Signal and Background & \makecell{Generation-level cuts ($\ell=e^\pm,\mu^\pm$)\\ (NA : Not Applied)} & Cross section (fb)  \\ \hline

\multirow{7}{*}{$b \bar{b} \gamma \gamma$} & \multicolumn{1}{l|}{Signal ($hh\to b \bar{b} \gamma \gamma$)} & \multicolumn{1}{l|}{NA} & \multicolumn{1}{l|}{$0.105$}   \\\cline{2-4}
                                           & \multicolumn{1}{l|}{$hb\bar{b}$, $h\to\gamma\gamma$} & \multicolumn{1}{l|}{NA}  & \multicolumn{1}{l|}{$1.32$}   \\\cline{2-4}
                                           & \multicolumn{1}{l|}{$t\bar{t}h$, $h\to\gamma\gamma$} & \multicolumn{1}{l|}{NA}  & \multicolumn{1}{l|}{$1.39$}  \\\cline{2-4}
                                           & \multicolumn{1}{l|}{$Zh$, $h\to\gamma\gamma$, $Z\to b\bar{b}$} & \multicolumn{1}{l|}{NA}  & \multicolumn{1}{l|}{ $0.33$}   \\\cline{2-4}
                                           
                                           & \multicolumn{1}{l|}{$b\bar{b}\gamma\gamma *$} 
                                           & \multicolumn{1}{l|}{ \makecell{$p_{T,j/b/\gamma}>20~\text{GeV}$, $|\eta_j|<5.0$, $|\eta_\gamma|<2.5$, \\ $\Delta R_{\gamma\gamma}>0.4$, $\Delta R_{\gamma j}>0.4$}} 
                                           & \multicolumn{1}{l|}{$348.32$\footnote{Including $b\bar{b}j\gamma$, $c\bar{c}j\gamma$ fake backgrounds, the cross-section is multiplied by a factor $\sim 1.57 \; (2.23)$ for cut based  (BDT) analysis.}}
                                            \\\cline{2-4}

                                           & \multicolumn{1}{l|}{Fake1} 
                                           & \multicolumn{1}{l|}{\makecell{$p_{T,j/b/\gamma}>20~\text{GeV}$, $|\eta_j|<5.0$, $|\eta_{b/\gamma}|<2.5$, $\Delta R_{\gamma\gamma}>0.4$,\\ $\Delta R_{b/b/\gamma/\gamma,b/j/j/b}$\footnote{$\Delta R_{a/b,c/d}$ signifies $\Delta R_{ac}$ and $\Delta R_{bd}$}>0.2, $m_{bb}>50$ GeV}}  
                                           & \multicolumn{1}{l|}{$480.00$\footnote{Cross section for $pp\to b\bar{b}j\gamma$ is $480$ pb and $j\to\gamma$ fake rate is $0.1\%$. Including $c\bar{c}j\gamma$ fake background, the cross-section is multiplied by a factor $\sim 1.14 \; (0.97)$ for cut based (BDT) analysis.}}  
                                            \\\cline{2-4}

                                           & \multicolumn{1}{l|}{Fake2} 
                                           & \multicolumn{1}{l|}{\makecell{$p_{T,j}>10~\text{GeV}$, $p_{T,b}>20~\text{GeV}$, $|\eta_{j/b}|<5.0$, \\  $m_{jj}>50$ GeV, $m_{bb}>50$ GeV}}  
                                           & \multicolumn{1}{l|}{$48.31$\footnote{Cross section for $pp\to b\bar{b}jj$ is $48308.75$ pb and $j\to\gamma$ fake rate is $0.1\%$. The cross-section is multiplied by a factor $\sim 0.88$ for BDT analysis.}}   
                                           \\\hline


\multirow{13}{*}{$b \bar{b} \tau^+ \tau^-$} & \multicolumn{1}{l|}{Signal (hh$\to b \bar{b} \tau^+ \tau^-$)} & \multicolumn{1}{l|}{NA} & \multicolumn{1}{l|}{$2.89$}   \\\cline{2-4}

                                           & \multicolumn{1}{l|}{$t\bar{t}$ hadronic} & \multicolumn{1}{l|}{\makecell{$p_{T,j/b}>20~\text{GeV}$, $p_{T,l}>8~\text{GeV}$, $|\eta_j|<5.0$, \\ $|\eta_{b/\ell}|<3.0$, $\Delta R_{b/j/\ell}>0.2$, $m_{bb}>50$ GeV }}  & \multicolumn{1}{l|}{$168236.00$}  \\\cline{2-4}
                                                                                      
                                           & \multicolumn{1}{l|}{$t\bar{t}$ semi-leptonic} & \multicolumn{1}{l|}{$same$ as $t\bar{t}~full~had$}  & \multicolumn{1}{l|}{$213424.00$}   \\\cline{2-4}
                                                                                      
                                           & \multicolumn{1}{l|}{$t\bar{t}$ leptonic} & \multicolumn{1}{l|}{$same$ as $t\bar{t}~full~had$}  & \multicolumn{1}{l|}{$67629.00$}   \\\cline{2-4}
                                           
                                           & \multicolumn{1}{l|}{$\ell\ell b\bar{b}$} & \multicolumn{1}{l|}{\makecell{$p_{T,b}>20~\text{GeV}$, $|\eta_b|<3.0$, \\ $\Delta R_{\ell b}>0.2$, $m_{bb}>50$ GeV, $m_{\ell\ell}>30$ GeV }}  & \multicolumn{1}{l|}{$8322.30$}   \\\cline{2-4}
                                                                                      
                                            & \multicolumn{1}{l|}{$b\bar{b}h$, $h\to\tau\tau$} & \multicolumn{1}{l|}{\makecell{$p_{T,b}>20~\text{GeV}$, $p_{T,\ell}>10~\text{GeV}$, $|\eta_{b/\ell}|<3.0$, \\ $\Delta R_{\ell b}>0.2$, $m_{bb}>50$ GeV }}  & \multicolumn{1}{l|}{$1.57$}  \\\cline{2-4}

                                           & \multicolumn{1}{l|}{Zh} & \multicolumn{1}{l|}{NA}  & \multicolumn{1}{l|}{$28.21$}   \\\cline{2-4}
                                           & \multicolumn{1}{l|}{$t\bar{t}h$} & \multicolumn{1}{l|}{NA}  & \multicolumn{1}{l|}{$552.00$}   \\\cline{2-4}
                                           & \multicolumn{1}{l|}{$t\bar{t}Z$} & \multicolumn{1}{l|}{NA}  & \multicolumn{1}{l|}{$853.82$}  \\\cline{2-4}
                                           & \multicolumn{1}{l|}{$t\bar{t}W$} & \multicolumn{1}{l|}{NA}  & \multicolumn{1}{l|}{$521.28$}   \\\cline{2-4}
                                           & \multicolumn{1}{l|}{$bbjj$} & \multicolumn{1}{l|}{\makecell{$p_{T,j}>10~\text{GeV}$, $p_{T,b}>20~\text{GeV}$, $|\eta_{j/b}|<5.0$, \\  $m_{jj}>50$ GeV, $m_{bb}>50$ GeV}}  & \multicolumn{1}{l|}{$193.23$\footnote{Cross section for $pp\to b\bar{b}jj$ is $48308.75$ pb and $j\to\tau$ fake rate is $0.2\%$.}}   \\\hline


\multirow{7}{*}{$b \bar{b} WW^*$ } & \multicolumn{1}{l|}{Signal (hh$\to b \bar{b} W^+ W^-\to b\bar{b}l\nu l\nu$)} & \multicolumn{1}{l|}{NA} & \multicolumn{1}{l|}{$1.045$}  \\\cline{2-4}
                                                                       
                                              & \multicolumn{1}{l|}{Signal (hh$\to b \bar{b} W^+ W^-\to b\bar{b}l\nu jj$)} & \multicolumn{1}{l|}{NA}  & \multicolumn{1}{l|}{$9.847$}  \\\cline{2-4}
                                                                                                                                 
                                           & \multicolumn{1}{l|}{$t\bar{t}$ semi-leptonic} & \multicolumn{1}{l|}{\makecell{$p_{T,j/b}>20~\text{GeV}$, $p_{T,l}>8~\text{GeV}$, $|\eta_j|<5.0$, \\ $|\eta_{b/\ell}|<3.0$, $\Delta R_{b/j/\ell}>0.2$, $m_{bb}>50$ GeV }}  & \multicolumn{1}{l|}{$213424.00$}  \\\cline{2-4}
                                                                                      
                                           & \multicolumn{1}{l|}{$t\bar{t}$ leptonic} & \multicolumn{1}{l|}{\makecell{$p_{T,j/b}>20~\text{GeV}$, $p_{T,l}>8~\text{GeV}$, $|\eta_j|<5.0$, \\ $|\eta_{b/\ell}|<3.0$, $\Delta R_{b/j/\ell}>0.2$}}  & \multicolumn{1}{l|}{$67629.00$}  \\\cline{2-4}
                                      & \multicolumn{1}{l|}{$\ell\ell b\bar{b}$} & \multicolumn{1}{l|}{\makecell{$p_{T,b}>20~\text{GeV}$, $|\eta_b|<3.0$, \\ $\Delta R_{\ell b}>0.2$, $m_{bb}>50$ GeV, $m_{\ell\ell}>30$ GeV }}  & \multicolumn{1}{l|}{$8322.30$}   \\\cline{2-4}       
                                                                                 
                                            & \multicolumn{1}{l|}{$Wbb+jets$, $W\to \ell \nu$, $\ell$ also includes $\tau$} & \multicolumn{1}{l|}{\makecell{$p_{T,j/b}>20~\text{GeV}$, $p_{T,\ell}>8~\text{GeV}$, $|\eta_j|<5.0$, $\Delta R_{ll}>0.2$ }}  & \multicolumn{1}{l|}{$38811.80$}  \\\cline{2-4}
                                                                                      
                                           & \multicolumn{1}{l|}{$t\bar{t}h$} & \multicolumn{1}{l|}{NA}  & \multicolumn{1}{l|}{$552.00$}   \\\cline{2-4}
                                           & \multicolumn{1}{l|}{$t\bar{t}Z$} & \multicolumn{1}{l|}{NA}  & \multicolumn{1}{l|}{$853.82$}  \\\cline{2-4}
                                           & \multicolumn{1}{l|}{$t\bar{t}W$} & \multicolumn{1}{l|}{NA}  & \multicolumn{1}{l|}{$521.28$}   \\\hline
                                                        
      \end{tabular}}
       \end{bigcenter}
           \caption{Generation level cuts and cross-sections for the signals and various backgrounds used in the analyses.}

\label{app1:1}
          \end{table}
          
\begin{table}[htb!]
\centering
\begin{bigcenter}
\scalebox{0.6}{%
\begin{tabular}{|c|c|c|c|c|}

\hline
Process & Signal and Background & \makecell{Generation-level cuts ($\ell=e^\pm,\mu^\pm$)\\ (NA : Not Applied)} & Cross section (fb)  \\ \hline


\multirow{6}{*}{$\gamma\gamma WW^*$ } & \multicolumn{1}{l|}{Signal (hh$\to \gamma\gamma W^+ W^-$)} & \multicolumn{1}{l|}{NA} & \multicolumn{1}{l|}{$0.04$}  \\\cline{2-4}                                                                   
                                                                                                                                 
                                           & \multicolumn{1}{l|}{$t\bar{t}h$, $h\to \gamma\gamma$} & \multicolumn{1}{l|}{NA}  & \multicolumn{1}{l|}{$1.39$}   \\\cline{2-4}
                                           
                                           & \multicolumn{1}{l|}{$Zh$ + jets, $h\to\gamma\gamma$} & \multicolumn{1}{l|}{NA}  & \multicolumn{1}{l|}{$2.20$}  \\\cline{2-4}
                                           
                                           & \multicolumn{1}{l|}{$\ell\nu\gamma\gamma$ + jets, $\ell$ also includes $\tau$} & \multicolumn{1}{l|}{\makecell{$p_{T,\gamma/\ell}>10~\text{GeV}$, $|\eta_\gamma/\ell|<2.5$, $\Delta R_{\gamma,\gamma}>0.2$, \\ $\Delta R_{\gamma,\ell}>0.2$, 120 GeV  $<m_{\gamma\gamma}<$ 130 GeV }}  & \multicolumn{1}{l|}{$3.28$}  \\\cline{2-4}   
                                                 
                                           & \multicolumn{1}{l|}{$\ell\ell\gamma\gamma$ + jets, $\ell$ also includes $\tau$} & \multicolumn{1}{l|}{$same$ as $\ell\nu\gamma\gamma$ with $m_{\ell\ell}>$ 20 GeV }  & \multicolumn{1}{l|}{$1.05$}  \\\cline{2-4}            
              
                                           & \multicolumn{1}{l|}{$Wh$ + jets, $h\to\gamma\gamma$} & \multicolumn{1}{l|}{$|\eta_j|<5.0$}  & \multicolumn{1}{l|}{$3.45$}   \\\hline


\multirow{7}{*}{$WW^*WW^*$ } & \multicolumn{1}{l|}{Signal (hh$\to W^+ W^- W^+ W^-$)} & \multicolumn{1}{l|}{NA} & \multicolumn{1}{l|}{$1.81$}  \\\cline{2-4}
                                                                       
                                           & \multicolumn{1}{l|}{$W^\pm W^\pm$ + jets} & \multicolumn{1}{l|}{ $|\eta_j|<5.0$}  & \multicolumn{1}{l|}{$614.75$}  \\\cline{2-4}
                                                                                                                                 
                                           & \multicolumn{1}{l|}{$t\bar{t}$ semi-leptonic} & \multicolumn{1}{l|}{\makecell{$p_{T,j/b}>20~\text{GeV}$, $p_{T,l}>8~\text{GeV}$, $|\eta_j|<5.0$, \\ $|\eta_{b/\ell}|<3.0$, $\Delta R_{b/j/\ell}>0.2$, $m_{bb}>50$ GeV }}  & \multicolumn{1}{l|}{$213424.00$}  \\\cline{2-4}
                                                                                      
                                           & \multicolumn{1}{l|}{$t\bar{t}$ leptonic} & \multicolumn{1}{l|}{\makecell{$p_{T,j/b}>20~\text{GeV}$, $p_{T,l}>8~\text{GeV}$, $|\eta_j|<5.0$, \\ $|\eta_{b/\ell}|<3.0$, $\Delta R_{b/j/\ell}>0.2$ }}  & \multicolumn{1}{l|}{$67629.00$}  \\\cline{2-4}
                                           
                                      & \multicolumn{1}{l|}{$Wh$ + jets} & \multicolumn{1}{l|}{$|\eta_j|<5.0$ }  & \multicolumn{1}{l|}{$1522.00$}   \\\cline{2-4}       
                                                                                 
                                      & \multicolumn{1}{l|}{$Zh$ + jets} & \multicolumn{1}{l|}{$|\eta_j|<5.0$}  & \multicolumn{1}{l|}{$969.00$}   \\\cline{2-4} 
                                      
                                      & \multicolumn{1}{l|}{$WZ$ + jets, $W\to \ell \nu$, $Z\to \ell \ell$, $\ell$ also includes $\tau$} & \multicolumn{1}{l|}{$|\eta_j|<5.0$ }  & \multicolumn{1}{l|}{$1350.19$}   \\\cline{2-4} 

                                      & \multicolumn{1}{l|}{$VVV$} & \multicolumn{1}{l|}{NA}  & \multicolumn{1}{l|}{$255.27$}   \\\cline{2-4} 

                                      & \multicolumn{1}{l|}{$4\ell$, $\ell$ includes $\tau$ also} & \multicolumn{1}{l|}{\makecell{$p_{T,\ell}>8~\text{GeV}$, $|\eta_\ell|<3.0$, \\ $\Delta R_{\ell\ell}>0.2$, $p_{T,\ell}>15~\text{GeV}$ for at least 1 charged $\ell$}}  & \multicolumn{1}{l|}{$124.75$}   \\\cline{2-4}

                                           & \multicolumn{1}{l|}{$t\bar{t}h$} & \multicolumn{1}{l|}{NA}  & \multicolumn{1}{l|}{$552.00$}   \\\cline{2-4}
                                           & \multicolumn{1}{l|}{$t\bar{t}Z$} & \multicolumn{1}{l|}{NA}  & \multicolumn{1}{l|}{$853.82$}  \\\cline{2-4}
                                           & \multicolumn{1}{l|}{$t\bar{t}W$} & \multicolumn{1}{l|}{NA}  & \multicolumn{1}{l|}{$521.28$}   \\\hline

\end{tabular}}
\end{bigcenter}
\caption{Generation level cuts and cross-sections for the signals and various backgrounds used in the analyses.}

\label{app1:2}
\end{table}


\section{Appendix B}
\label{sec:appendixB}
In this section, we will discuss the technique employed in Ref.~\cite{Barr:2013tda} where they construct the $m^{\text{Higgs-bound}}_{\tau\tau}$ variable which is shown to be useful in separating the irreducible backgrounds ensuing 
from the SM $Z$-boson~\cite{Barr:2013tda} from the $h \to \tau^+ \tau^-$ decay. $m^{\text{Higgs-bound}}_{\tau\tau}$ is essentially 
constructed along the lines of the stranverse mass variable ($m_{T2}$)~\cite{Lester:1999tx,Barr:2003rg} but signifies the maximum lower 
bound for an on-shell parent particle decaying into a pair of $\tau$s. In an ideal detector, this observable, by construction, must
sharply fall off at $m_Z$ for the $Z \to \tau \tau$ process. Hence, a suitable cut on this variable should make the region above $m_Z$ 
free from such SM backgrounds. Smearing effects, however, will tamper the sharpness of this fall and hence upon incorporating detector 
effects, the results will be less dramatic. As we shall discuss below, this variable actually significantly reduces our signal
yields along with a reduction in the backgrounds, but ultimately leads to much smaller $S/B$ and smaller significance. We will, hence,
study this observable with caution and keep this at the level of discussion at the end of this subsection.
There are several tools which reconstructs $m_{\tau \tau}$ and now we state the results that we 
obtain upon using the $m_{\tau \tau}^{\textrm{Higgs-bound}}$~\cite{Lester:1999tx,Barr:2003rg} variable. We do not perform a multivariate
analysis for this scenario but perform an optimised cut-based analysis. For the three modes, we find the following to be the most
optimal cut choices:
\begin{itemize}
 \item $\tau_h \tau_h$: $p_{T,bb} > 100$ GeV, $\tau_h \tau_{\ell}$: $p_{T,bb} > 115$ GeV and $\tau_{\ell} \tau_{\ell}$: $p_{T,bb} > 140$ GeV
 \item $\tau_h \tau_h$: $m_{T2} > 110$ GeV, $\tau_h \tau_{\ell}$: $m_{T2} > 130$ GeV and $\tau_{\ell} \tau_{\ell}$: $m_{T2} > 120$ GeV
 \item $\tau_h \tau_h$: $100 \; \textrm{GeV} \; < m_{\tau \tau}^{\textrm{Higgs-bound}} < 165$ GeV, $\tau_h \tau_{\ell}$: 
       $90 \; \textrm{GeV} \; < m_{\tau \tau}^{\textrm{Higgs-bound}} < 150$ GeV and $\tau_{\ell} \tau_{\ell}$: 
       $80 \; \textrm{GeV} \; < m_{\tau \tau}^{\textrm{Higgs-bound}} < 140$ GeV.
\end{itemize}
Upon imposing the cuts, we are left with around 2.93, 10.81 and 10.03 signal and 193.63, 1282.85 and 13315.49 background events for the $\tau_h \tau_h$, 
$\tau_h \tau_{\ell}$ and $\tau_{\ell} \tau_{\ell}$ cases, respectively. We find a considerable reduction in the backgrounds with respect
to the cut-based analysis performed earlier with the $m_{\tau \tau}^{\textrm{vis}}$ variable. However, the signal yield also falls 
sharply. Finally, we find $S/\sqrt{B}$ values of 0.21, 0.30 and 0.09 for the three aforementioned cases, respectively. We do not use 
this variable for a detailed study as the sharpness of this variable reduces upon including smearing and other detector effects.


\end{document}